\documentclass[10pt,journal,compsoc]{IEEEtran}

\ifCLASSOPTIONcompsoc
  \usepackage[nocompress]{cite}
\else
  \usepackage{cite}
\fi

\ifCLASSINFOpdf
\else
\fi

\usepackage{amsmath,amssymb}
\usepackage{cases}
\usepackage{tabularx}
\usepackage{multicol}
\usepackage{graphicx}
\usepackage{float}
\usepackage{amsmath}
\usepackage{amssymb}
\usepackage{array}
\usepackage{color}
\usepackage{xcolor}
\usepackage{multirow}
\usepackage{bbm}
\usepackage{bm}
\usepackage{booktabs}
\usepackage{textcomp}
\usepackage{gensymb}
\newcommand{\specialcell}[2][c]{%
  \begin{tabular}[#1]{@{}c@{}}#2\end{tabular}}
\graphicspath{{figures/}}
\usepackage{url}
\usepackage{tikz}
\usetikzlibrary{decorations.pathreplacing}
\newcolumntype{C}[1]{>{\centering\let\newline\\\arraybackslash\hspace{0pt}}m{#1}}
\newcolumntype{L}[1]{>{\raggedright\let\newline\\\arraybackslash\hspace{0pt}}m{#1}}
\newcolumntype{R}[1]{>{\raggedleft\let\newline\\\arraybackslash\hspace{0pt}}m{#1}}
\newcommand{\etal}{\textit{et al}.}
\newcommand{\ie}{\textit{i}.\textit{e}.}
\newcommand{\eg}{\textit{e}.\textit{g}.}
\newcommand{\mdegr}{360{\degree} }
\DeclareMathOperator*{\argmin}{argmin}
\graphicspath{{figs/}}
\usepackage{algpseudocode}
\usepackage{algorithm}

\begin{document}

\title{Pseudocylindrical Convolutions for Learned Omnidirectional Image Compression}

\author{Mu~Li,
        Kede~Ma,~\IEEEmembership{Member,~IEEE,}
        Jinxing~Li,
        and~David~Zhang,~\IEEEmembership{Life Fellow,~IEEE}
\IEEEcompsocitemizethanks{\IEEEcompsocthanksitem{This project is supported by China Postdoctoral Science Foundation (2020TQ0319, 2020M682034), NSFC Foundation (61906162, 62102339), and  Shenzhen Science and Technology Program (RCBS20200714114910193).}
\IEEEcompsocthanksitem Mu Li is with the School of Data Science, The Chinese University of Hong Kong (Shenzhen), Shenzhen, 518172, China, and also with the School of Information Science and Technology, University of Science and Technology of China, Hefei, 230026, China (e-mail: limuhit@gmail.com).
\IEEEcompsocthanksitem Kede Ma is with the Department
of Computer Science, City University of Hong Kong, Kowloon, Hong Kong (e-mail: kede.ma@cityu.edu.hk).
\IEEEcompsocthanksitem Jinxing Li is with the School of Computer Science and Technology, Harbin Institute of Technology, Shenzhen, 518055, China (e-mail: lijinxing158@gmail.com).
\IEEEcompsocthanksitem David Zhang is with the School of Data Science, The Chinese University of Hong Kong (Shenzhen), and also with the Shenzhen Institute of Artificial Intelligence and Robotics for Society, Shenzhen, 518172, China (e-mail: davidzhang@cuhk.edu.cn).}
}

\markboth{}%
{Shell \MakeLowercase{\textit{et al.}}: Bare Demo of IEEEtran.cls for Computer Society Journals}
 
\IEEEtitleabstractindextext{%
\begin{abstract}
Although equirectangular projection (ERP) is a convenient form to store omnidirectional images (also known as \mdegr images), it is neither equal-area nor conformal, thus not friendly to subsequent visual communication. In the context of image compression, ERP will over-sample and deform things and stuff near the poles, making it difficult for perceptually optimal bit allocation. In conventional \mdegr image compression, techniques such as region-wise packing and tiled representation are introduced to alleviate the over-sampling problem, achieving limited success. In this paper, we make one of the first attempts to learn deep neural networks for omnidirectional image compression. We first describe parametric pseudocylindrical representation as a generalization of common pseudocylindrical map projections. A computationally tractable greedy method is presented to determine the (sub)-optimal configuration of the pseudocylindrical representation in terms of a novel proxy objective for rate-distortion performance. We then propose pseudocylindrical convolutions for \mdegr image compression. Under reasonable constraints on the parametric representation, the pseudocylindrical convolution can be efficiently implemented by standard convolution with the so-called pseudocylindrical padding.
To demonstrate the feasibility of our idea, we implement an end-to-end \mdegr image compression system, consisting of the learned pseudocylindrical representation, an analysis transform, a non-uniform quantizer, a synthesis transform, and an entropy model. Experimental results on $19,790$ omnidirectional images show that our method achieves consistently better rate-distortion performance than the competing methods. Moreover, the visual quality by our method is significantly improved for all images at all bitrates.
\end{abstract}

\begin{IEEEkeywords}
Omnidirectional image compression, pseudocylindrical representation, pseudocylindrical convolution, map projection
\end{IEEEkeywords}}

\maketitle

\IEEEdisplaynontitleabstractindextext

%
\IEEEpeerreviewmaketitle

\IEEEraisesectionheading{\section{Introduction}\label{sec:introduction}}
 
\IEEEPARstart{O}{mnidirectional} images, also referred to as spherical and \mdegr images, provide 360{\degree}$\times$180{\degree} panoramas of natural scenes, and enable free view direction exploration. Recent years have witnessed a dramatic increase in the volume of \mdegr image data being generated. On the one hand, average users have easy access to \mdegr imaging and display devices, and are getting used to play with this format of virtual reality content on a daily basis. On the other hand, there is a trend to capture ultra-high-definition panoramas to provide an excellent immersive experience, pushing the spatial resolution to be exceedingly high (\eg, 8K). 
The increasing need for storing and transmitting the enormous amount of panoramic data calls for novel effective \mdegr image compression methods.

Currently, the prevailing scheme for \mdegr image compression takes a two-step approach. First, select (or create) a map projection~\cite{snyder1987map} with the optimized hyperparameter setting for the sphere-to-plane mapping. Second, pick (or adapt) a standard image codec that is compatible with central-perspective images for compression. In differential geometry, the Theorema Egregium by Gauss states that all planar projections of a sphere will necessarily be distorted \cite{snyder1987map}. Among the three major projection desiderata: equal-area, conformal, and equidistant\footnote{Equal-area, conformal, and equidistant map projections preserve relative scales of things and stuff, local angles, and great-circle distances between points, respectively, on the sphere.}, the most widely used equirectangular projection (ERP) does not satisfy the former two, and thus is not friendly to subsequent visual communication applications. Regional resampling \cite{lee2017omnidirectional,boyce2017hevc,youvalari2016analysis,yu2015content} and adaptive quantization \cite{li2017spherical,liu2018rate,tang2017optimized,xiu2018adaptive} techniques have been proposed to mitigate the sampling problem of ERP. Compression-friendly projections, such as the tiled representation \cite{li2016novel,yu2015content} and hybrid cubemap projection \cite{lin2019efficient,he2018content,hanhart2018360} have also been investigated. In \mdegr content streaming, viewport-based format is often preferred for coding and transmission \cite{ozcinar2017viewport,corbillon2017viewport}. Other projection methods for image display \cite{zelnik2005squaring,chang2013rectangling,kim2017automatic} and 
visual recognition (\eg, icosahedron \cite{kimerling1999comparing} and tangent images \cite{eder2020tangent}) also emerge in the field of computer vision. With many possible projections at hand, it remains unclear which one is the best choice for learned \mdegr image compression in terms of rate-distortion performance, computation and implementation complexity, and compatibility with standard deep learning-based analysis/synthesis transform, and entropy model. 

 \begin{figure*}[htb!]
     \begin{minipage}{1.\linewidth}
     \begin{tikzpicture}
    	\node[inner sep=0pt] (nd1) at (0,0) {\includegraphics[width=1.\linewidth]{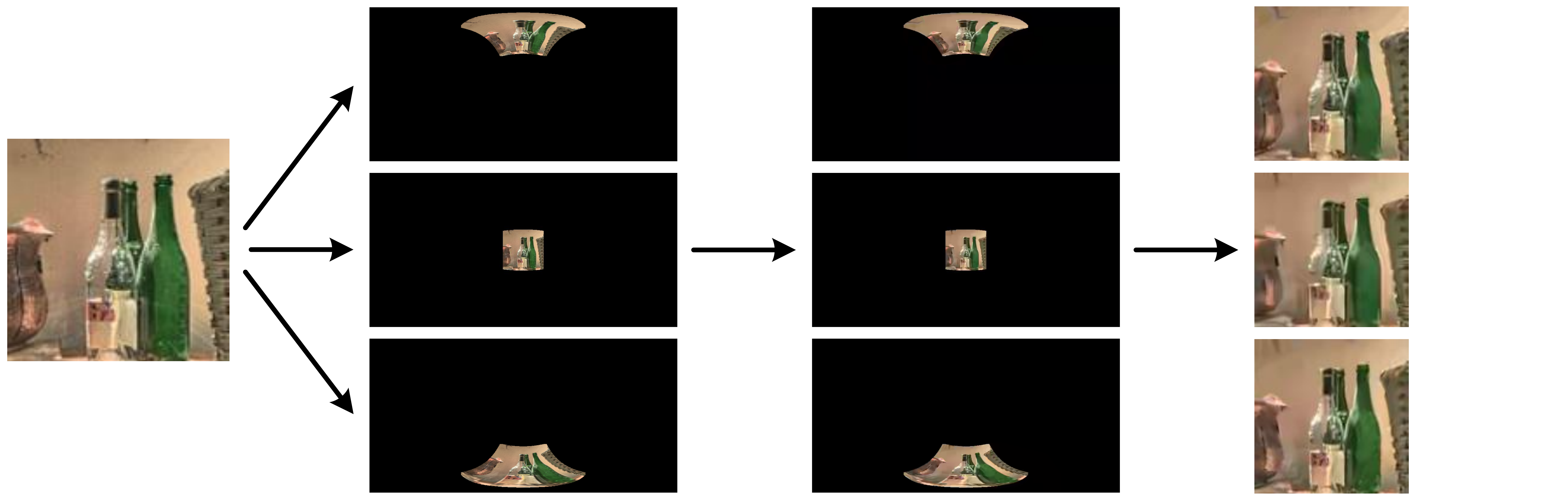}};
		\node[text width=2cm,align=center] (cp2) at (-5.85,1.7){\baselineskip=3pt \scriptsize{ $\theta$ = $\frac{\pi}{3}$} \par};
		\node[text width=2cm,align=center] (cp3) at (-5.7,-0.2){\baselineskip=3pt \scriptsize{ $\theta$ = $0$} \par};
		\node[text width=2cm,align=center] (cp4) at (-5.85,-1.7){\baselineskip=3pt \scriptsize{ $\theta$ = $-\frac{\pi}{3}$} \par};
		\node[text width=2cm,align=center] (cp7) at (-0.5,0.5){\baselineskip=3pt \scriptsize{$\mbox{QP}$ = 42} \par};
		\node[text width=2cm,align=center] (cp10) at (4.55,3.3){\baselineskip=3pt \scriptsize{ERP to Viewport} \par};
		\node[text width=2cm,align=center] (cp10) at (-5.65,3.3){\baselineskip=3pt \scriptsize{Viewport to ERP} \par};
		\node[text width=2cm,align=center] (cp10) at (-0.5,3.3){\baselineskip=3pt \scriptsize{HEVC Intra Codec} \par};
		\node[text width=2cm,align=center] (cp10) at (8.1,2){\baselineskip=9pt \scriptsize{11,112 bytes \\ 28.02 dB} \par};
		\node[text width=2cm,align=center] (cp10) at (8.1,-0.2){\baselineskip=9pt \scriptsize{6,672 bytes \\ 26.78 dB} \par};
		\node[text width=2cm,align=center] (cp10) at (8.1,-2){\baselineskip=9pt \scriptsize{11,952 bytes \\ 28.25 dB} \par};
		\node[text width=2cm,align=center] (cp10) at (-7.6,-1.6){\baselineskip=3pt \scriptsize{Original Patch} \par};
		\node[text width=2cm,align=center] (cp10) at (-3,-3.1){\baselineskip=3pt \scriptsize{Original ERP} \par};
		\node[text width=2cm,align=center] (cp10) at (2.1,-3.1){\baselineskip=3pt \scriptsize{Compressed ERP} \par};
		\node[text width=3cm,align=center] (cp10) at (6.4,-3.1){\baselineskip=3pt \scriptsize{Compressed Patch} \par};
	 \end{tikzpicture}
     \end{minipage}
     \caption{Illustration of the non-uniform sampling problem of ERP. We first project the same image patch to different latitudes of the ERP images (padded with zeros), and compress them by the HEVC intra coding with the identical hyperparameter. The performance is given in the format of  bytes / peak signal-to-noise ratio (PSNR in dB).}
     \label{fig:over_sample}
 \end{figure*}
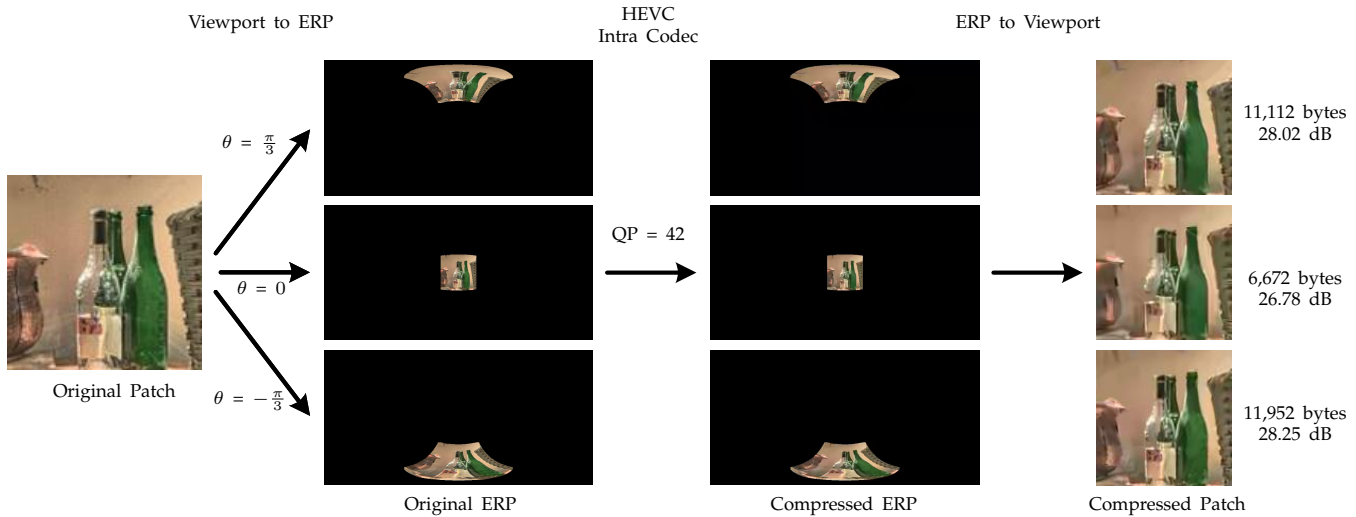

Deep neural networks (DNNs) have been proved effective in many low-level vision tasks, including central-perspective image compression~\cite{balle2016end, theis2017lossy, balle2018variational,minnen2018joint,torfason2018towards,agustsson2019generative,li2020efficient}. Following a transform coding scheme, the raw RGB image is first transformed to a latent code representation, quantized to the discrete code, and final transformed back to the RGB domain, all with DNNs that can be end-to-end optimized with respect to rate-distortion performance. 
Recently, a growing research trend is to enable DNNs for \mdegr computer vision, which is broadly sorted into three categories depending on how they address the sphere-to-plane distortion: spatially adaptive convolution \cite{zioulis2018omnidepth,tateno2018distortion}, knowledge distillation \cite{su2017learning,su2019kernel}, and reparameterization \cite{cohen2018spherical,esteves2018learning,perraudin2019deepsphere,jiang2019spherical}. Nevertheless, it is highly nontrivial to directly adapt these techniques to learned \mdegr image compression. This is because these methods typically require to modify convolution filters, and cannot benefit from years of sophisticated code optimization of standard convolution. As a result, compressing a high-resolution omnidirectional image would be painfully slow. Moreover, non-uniform sampling (especially over-sampling at high latitudes) is a more urgent issue in \mdegr image compression than geometric deformation (see Fig. \ref{fig:over_sample}), because the latter can be handled by adopting a perceptual image quality metric as the learning objective \cite{sui2021perceptual}. From this perspective, reparameterization methods that directly work with spherical signals and are independent of projection methods seem to be more appropriate for rate reduction. But reparameterization comes with its own problem apart from computational complexity: the spherical representation is orderless, which may hinder the context-based entropy modeling for accurate rate estimation \cite{minnen2018joint,li2020efficient}.

In this paper, we take initial steps towards learned omnidirectional image compression based on DNNs. Our main contributions are three-fold.
\begin{itemize}
    \item We describe parametric pseudocylindrical representation as a generalization of common pseudocylindrical map projections and the tiled representation by Yu \etal \cite{yu2015content}. We propose a computationally tractable greedy algorithm to determine the (sub)-optimal parameter configuration in terms of the rate-distortion performance, estimated by a novel proxy objective. Interestingly, we find that the optimized representation does not correspond to pseudocylindrical projections with the equal-area property (\eg, sinusoidal projection). Empirically, the rate-distortion performance will benefit from slight over-sampling at mid-latitudes. 
    
    \item We propose pseudocylindrical convolutions that work seamlessly with the parametric pseudocylindrical representation for \mdegr image compression. Under reasonable constraints on the representation (\ie, the tiled representation), the pseudocylindrical convolution can be efficiently implemented by standard convolution with pseudocylindrical padding. In particular, given the current tile, we pad the latitudinal side with adjacent tiles resized to the same width, and pad the longitudinal side circularly to respect the spherical structure. The manipulation on feature representation instead of convolution leads to a significant advantage of our approach: we are able to  transfer the large zoo of DNN-based compression methods for central-perspective images to omnidirectional images. 
    
    \item We build an end-to-end \mdegr image compression system, which is composed of the optimized pseudocylindrical representation,  an analysis transform, a non-uniform quantizer, a synthesis transform, and  a context-based entropy model. Extensive experiments show that our method outperforms compression standards and DNN-based methods for central-perspective images with region-wise packing (RWP). More importantly, the visual quality of the compressed images is much better for all images at all bitrates.
\end{itemize}

\section{Related Work}
In this section, we provide a short overview of learned image compression methods for planar images and  standards (and tricks) for compressing omnidirectional images.  Relevant techniques for \mdegr computer vision will also be briefly summarized.

\subsection{Learned Planar Image Compression}
Learned image compression learns to trade off the rate and distortion, in which DNNs are commonly used to build the analysis transform (\ie, encoder) and the synthesis transform (\ie, decoder), and to model the rate of the codes.

For rate estimation, the discrete entropy serves as a general choice, which requires keeping track of the joint probability of the discrete codes that varies with
changes in the network parameters. Side information in the form of hyper-prior \cite{balle2018variational} and code context \cite{minnen2018joint,li2020efficient} can be introduced to boost the accuracy of entropy modeling. Ball{\'e}~\etal~\cite{balle2016end} adopted a parametric piece-wise probability distribution function for codes of the same channel; they~\cite{balle2018variational} later assumed a univariate Gaussian distribution for codes of the same spatial location, whose mean and variance are estimated using a hyper-prior. Minnen~\etal~\cite{minnen2018joint} modeled the code distribution with a mixture of Gaussians (MoG). A $5 \times 5$ code context was adopted as an auto-regressive prior to better predict the MoG parameters. Similarly, a MoG distribution was used in~\cite{theis2017lossy}. Li~\etal~\cite{li2020efficient} proposed a context-based DNN for efficient entropy modeling. 

Besides precise estimation of the code rate using the discrete entropy, various upper-bounds (\eg, the number and dimension of codes) have been derived.
Toderici~\etal~\cite{toderici2015variable,toderici2016full} proposed a progressive compression scheme, in which the rate was controlled by the number of iterations. Johnston~\etal~\cite{johnston2017improved} took a step further, and  presented a content-adaptive bit allocation strategy. Ripple~\etal~\cite{rippel2017real} implemented  pyramid networks for the encoder and the decoder, with an adaptive code length regularization for rate control. In a similar spirit, Li~\etal~\cite{li2017learning} described a spatially adaptive bit allocation scheme, where the rate was estimated as the total number of codes allocated to different regions. They~\cite{li2020learning} further designed better relaxation strategies for learning optimal bit allocation.

For distortion quantification, conventional metrics, including mean squared error (MSE), peak signal-to-noise ratio (PSNR), structural similarity (SSIM)~\cite{wang2004image}, and multi-scale SSIM (MS-SSIM)~\cite{wang2003multiscale}, were employed to evaluate the ``perceptual'' distance between the original and compressed images in learned image compression. Other metrics were also incorporated  for special considerations. For instance, to boost the visual quality of low-bitrate images, the  adversarial loss~\cite{goodfellow2014generative} 
was introduced in~\cite{rippel2017real, agustsson2019generative}. Torfason~\etal~\cite{torfason2018towards} suggested to utilize task-dependent losses (\eg, the classification and segmentation error) for task-driven image compression. 

\subsection{\mdegr Image Compression}
Most \mdegr image compression methods were designed on  top of widely used compression standards such as HEVC \cite{ghaznavi2017comparison,boyce2017hevc,sullivan2012overview} for central-perspective content. Thus sphere-to-plane projections are inevitable for compatibility purposes~\cite{xu2020state}.
One popular branch of work is to introduce practical tricks  to tackle the non-uniform sampling problem, such as regional resampling and adaptive quantization. Budagavi~\etal~\cite{budagavi2015360} adopted Gaussian blurring to smooth high-latitude regions of the ERP image to ease subsequent compression. Regional down-sampling~\cite{lee2017omnidirectional,boyce2017hevc,youvalari2016analysis,yu2015content} partitions
the ERP image into several regions according to the latitude, and resamples and assembles them into a new image of reduced size for compression. Of particular interest, RWP, which repacks the ERP image by reducing the sizes of  polar regions, is adopted in HEVC when dealing with \mdegr content~\cite{boyce2017hevc}. 
Close to our work, Yu~\etal~\cite{yu2015content} introduced the tile representation for \mdegr images, and suggested to optimize the height and width of each tile for the sampling rate and bitrate.
Adaptive quantization~\cite{li2017spherical,liu2018rate,tang2017optimized,liu2017novel,xiu2018adaptive}, on the other hand, adjusts quantization parameters (QPs) for different regions in ERP with respect to spherical ``perceptual'' metrics such as S-PSNR~\cite{yu2015framework} and WS-PSNR~\cite{sun2016ahg8}.

Apart from ERP, cubemap projection is also commonly seen in \mdegr compression. Su~\etal~\cite{su2018learning} learned to rotate the \mdegr images to boost the coding performance. Other variants of cubemap formats, such as hybrid equi-angular cubemap projection (HEC)~\cite{lin2019efficient}, hybrid cubemap projection (HCP)~\cite{he2018content}, and hybrid angular cubemap projection (HAP)~\cite{hanhart2018360}, were investigated for uniform and content adaptive sampling. In addition, viewport-based~\cite{zare2017virtual,ghaznavi2017comparison,ozcinar2017viewport,ozcinar2019visual} and saliency-aware  methods~\cite{hadizadeh2013saliency,luz2017saliency,sitzmann2018saliency} were proposed to spend most of the bits on coding the viewports of interest, when streaming \mdegr content. 

Taking inspirations from Yu~\etal~\cite{yu2015content}, we propose parametric pseudocylindrical representation for learned \mdegr image compression. The optimal parameter configuration is determined by a greedy algorithm optimized for a proxy rate-distortion objective. With reasonable constraints, the parametric representation supports an efficient implementation of the proposed pseudocylindrical convolutions.

\subsection{\mdegr Computer Vision}
Recently, there has been a surge of interest to develop DNNs for \mdegr computer vision with four main types of techniques. The first is spatially adaptive convolution. Designed around ERP, the most straightforward implementation is to expand the receptive field  horizontally via rectangular filters or dilated convolutions \cite{zioulis2018omnidepth}. A more advanced version is to design distortion-aware and deformable convolution kernels \cite{tateno2018distortion}, in combination with spiral spherical sampling \cite{saff1997distributing}. This type of approach is less scalable to deeper networks, which is necessary to achieve satisfactory rate-distortion performance in learned image compression. The second is knowledge distillation, with the goal of training DNNs on ERP images to predict the responses of a target model on viewport images. As one of the first attempts to learn ``spherical convolution'' for \mdegr vision, Su and Grauman \cite{su2017learning} tied the kernel weights along each row of the ERP image to accommodate the over-sampling issue. Due to the incorporation of secondary DNNs, this type of approach is often computationally expensive, whose performance may also be limited as the sphere-to-plane distortion is not explicitly modeled. The third is a family of reparameterization methods that are rooted in spherical harmonics and spectral analysis. These often define spherical convolution mathematically to seek rotational equivariance and invariance for dense and global prediction problems. Cohen \etal \cite{cohen2018spherical} defined spherical correlation with an efficient implementation based on the generalized fast Fourier transform \cite{kostelec2008ffts}. Concurrently, Esteves \etal \cite{esteves2018learning} defined spherical convolution \cite{driscoll1994computing} as a specific case of group convolution \cite{weinstein1996groupoids}, which admits a spectral domain implementation. To avoid the computational cost of the spherical Fourier transform, 
Perraudin \etal \cite{perraudin2019deepsphere} relied on the hierarchical equal area isolatitude pixelization (HEALPix) to formulate graph convolutions for cosmological applications. However, it is only practical to apply the method to part of the sphere. Jiang \etal \cite{jiang2019spherical} reparameterized convolution filters as linear combinations of first-order and second-order differential operators with learnable weights. Despite being mathematically appealing, rotational equivariance and invariance are less relevant to \mdegr image compression, and may cause inconvenience in context-based entropy modeling because of the orderless nature of the spherical representation.

The above-mentioned methods typically require modifying or re-designing the convolution operation, which are generally computationally expensive. Moreover, they may not enable the desired transferability of existing DNNs for central-perspective images, which enjoy years of research into optimal architecture design and efficient convolution implementation. In contrast, the proposed pseudocylindrical convolution resolves the over-sampling problem and can be efficiently implemented by standard 
convolution with pseudocylindrical padding.

\section{Proposed Method}\label{sec:lat_constraint}
In this section, we first introduce parametric pseudocylindrical representation, and describe a greedy algorithm to determine the parameter configuration for a proxy rate-distortion objective. We then propose pseudocylindrical  convolutions as the main building block for our learned \mdegr image compression system.

\subsection{Parametric Pseudocylindrical Representation}
From the practical standpoint, we start with a \mdegr image $\bm{x}\in\mathbb{R}^{H\times W}$ stored in ERP format, where $H$ and $W$ are the maximum numbers of samples in each column and row, respectively. The plane-to-sphere coordinate conversion can be calculated by 
\begin{align}
\theta_i &= \left(0.5-\frac{i+0.5}{H}\right) \times \pi, \quad i = \{0,\ldots, H-1\},\label{eq:erptheta}\\
\phi_j &= \left(\frac{j+0.5}{W}-0.5\right)\times 2\pi, \quad j = \{0,\ldots, W-1\},\label{eq:erpphi}
\end{align}
where $\theta$ and $\phi$  index the latitude and the longitude, respectively. Bilinear interpolation is used as the optional resampling filter if necessary. As a generalization of ERP, the proposed representation is also defined over a 2D grid $\Omega = \{0,\ldots, H-1\}\times \{0,\ldots, W-1\}$, and is parameterized by $\{W_i\}_{i=0}^{H-1}$, where $W_i \in\{1, \ldots, W\}$ is the width of the $i$-th row (with the starting point fixed to zero). By varying $W_i$, our representation offers a precise control over the sampling density of each row. For visualization purposes, we may reparameterize the representation by $\{W_i,S_i\}_{i=0}^{H-1}$, where $S_i$ denotes the starting point:
\begin{align}
    S_i = \lfloor (W-W_i)/2 \rfloor.
\end{align}
We refer to this data structure as parametric pseudocylindrical representation, since it generalizes several pseudocylindrical map projections. Specifically:
\begin{itemize}
\item Choosing $W_i=W$ and Eqs. \eqref{eq:erptheta} and \eqref{eq:erpphi} as the plane-to-sphere mapping yields the standard ERP;
\item Choosing $W_i = \cos(\theta_i)W$ and  replacing Eq. \eqref{eq:erpphi} to 
\begin{align}\label{eq:sine}
\phi_j &= \left(\frac{j-S_i+0.5}{W_i}-0.5\right)\times 2\pi,
\end{align}
for $j = \{S_i,\ldots, S_i+ W_i-1\}$
as the longitude mapping yields sinusoidal projection (see Fig. \ref{fig:pseudo} (b));

\item Choosing $W_i = (2\cos(2\theta_i/3)-1)W$, where
\begin{align}
\theta_i&= 3 \arcsin\left(0.5-\frac{i+0.5}{H}\right)
\end{align}
is the latitude mapping and Eq. \eqref{eq:sine} is the longitude mapping 
yields the Craster parabolic projection. This is used in the objective quality metric -  CPP-PSNR \cite{zakharchenko2016ahg8}.

\begin{figure}
\begin{minipage}{1\linewidth}
\begin{minipage}{0.46\linewidth}
\begin{tikzpicture}
    	\node[inner sep=0pt] (nd1) at (0,0) {\includegraphics[width=0.9\linewidth]{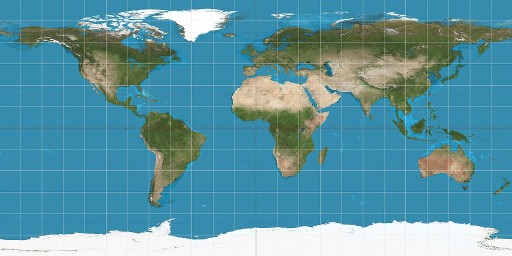}};
		\draw [-stealth, line width=0.3mm] (-1.9,1.) -- (1.9,1.);
		\draw [-stealth, line width=0.3mm] (-1.9,1.) -- (-1.9,-1.);
		\node[text width=0.1cm,align=center] (cp10) at (0,1.25){\baselineskip=3pt \scriptsize{$\phi$} \par};
		\node[text width=0.1cm,align=center] (cp10) at (-1.8,1.25){\baselineskip=3pt \scriptsize{$-\pi$} \par};
		\node[text width=0.1cm,align=center] (cp10) at (1.8,1.25){\baselineskip=3pt \scriptsize{$\pi$} \par};
		\node[text width=0.1cm,align=center] (cp10) at (-2.1,0){\baselineskip=3pt \scriptsize{$\theta$} \par};
		\node[text width=0.1cm,align=center] (cp10) at (-2.2,0.9){\baselineskip=3pt \scriptsize{$\frac{\pi}{2}$} \par};
		\node[text width=0.1cm,align=center] (cp10) at (-2.35,-0.8){\baselineskip=3pt \scriptsize{$-\frac{\pi}{2}$} \par};
\end{tikzpicture}
\end{minipage}
\begin{minipage}{0.46\linewidth}
\begin{tikzpicture}
    	\node[inner sep=0pt] (nd1) at (0,0) {\includegraphics[width=0.9\linewidth]{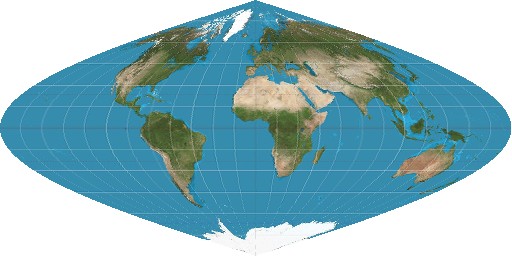}};
		\draw [-stealth, line width=0.3mm] (-1.9,1.) -- (1.9,1.);
		\draw [-stealth, line width=0.3mm] (-1.9,1.) -- (-1.9,-1.);
		\node[text width=0.1cm,align=center] (cp10) at (0,1.25){\baselineskip=3pt \scriptsize{$j$} \par};
		\node[text width=1.6cm,align=center] (cp10) at (-1.8,1.25){\baselineskip=3pt \scriptsize{$0$} \par};
		\node[text width=0.8cm,align=center] (cp10) at (1.5,1.25){\baselineskip=3pt \scriptsize{$W-1$} \par};
		\node[text width=0.1cm,align=center] (cp10) at (-2.15,0.2){\baselineskip=3pt \scriptsize{$i$} \par};
		\node[text width=0.8cm,align=center,rotate=90] (cp10) at (-2.15,0.9){\baselineskip=3pt \scriptsize{$0$} \par};
		\node[text width=0.8cm,align=center,rotate=90] (cp10) at (-2.15,-0.6){\baselineskip=3pt \scriptsize{$H-1$} \par};
\end{tikzpicture}
\end{minipage}
\end{minipage}

\begin{minipage}{1\linewidth}
\begin{minipage}{0.54\linewidth}
\center{\scriptsize{(a)}}
\end{minipage}
\begin{minipage}{0.01\linewidth}
\hspace{0.2cm}
\end{minipage}
\begin{minipage}{0.42\linewidth}
\center{\scriptsize{(b)}}
\end{minipage}
\end{minipage}

\hfill\vline\hfill

\begin{minipage}{1\linewidth}
\begin{minipage}{0.06\linewidth}
\hspace{0.3cm}
\end{minipage}
\begin{minipage}{0.4\linewidth}
\includegraphics[width=1\linewidth]{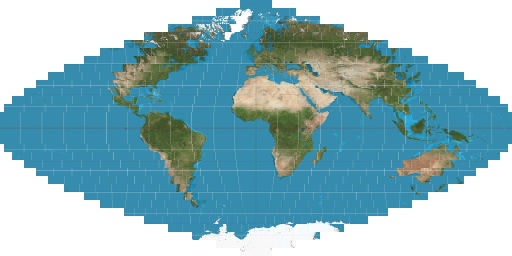}
\end{minipage}
\begin{minipage}{0.08\linewidth}
\hspace{0.3cm}
\end{minipage}
\begin{minipage}{0.4\linewidth}
\includegraphics[width=1\linewidth]{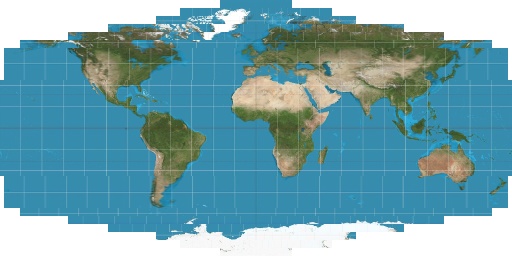}
\end{minipage}
\end{minipage}

\begin{minipage}{1\linewidth}
\begin{minipage}{0.06\linewidth}
\hspace{0.3cm}
\end{minipage}
\begin{minipage}{0.4\linewidth}
\center{\scriptsize{(c)}}
\end{minipage}
\begin{minipage}{0.09\linewidth}
\hspace{0.3cm}
\end{minipage}
\begin{minipage}{0.4\linewidth}
\center{\scriptsize{(d)}}
\end{minipage}
\end{minipage}

\caption{Comparison of different omnidirectional image representations. (a) ERP. (b) Sinusoidal projection. (c) Tiled sinusoidal representation. (d) Optimized pseudocylindrical representation.}\label{fig:pseudo}
\end{figure}
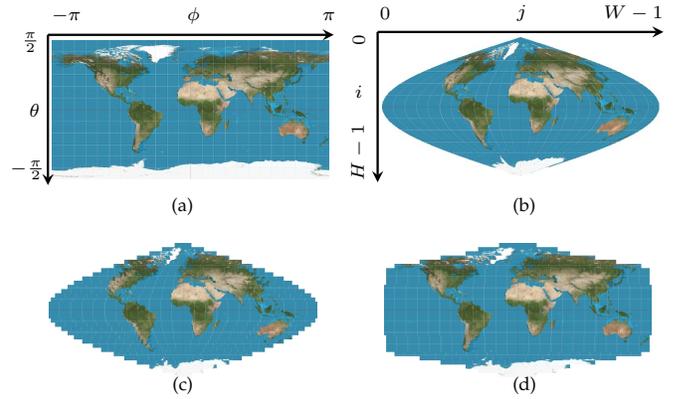








\item Another special case of interest arises when setting adjacent rows to the same width, leading to the tiled representation proposed by Yu \etal~\cite{yu2015content}, which plays a crucial role in accelerating pseudocylindrical convolutions, as will be immediately clear. 
\end{itemize}

With different combinations of the width configuration and the plane-to-sphere mapping, our pseudocylindrical representation not only includes a broad class of pseudocylindrical map projections as special cases but also opens the door of novel data structures that may be more suitable for \mdegr image compression. \textit{Without loss of generality, in the remainder of the paper, we use Eqs. \eqref{eq:erptheta} and \eqref{eq:sine} for the plane-to-sphere coordinate conversion, and assume $S_i = 0$.}

\subsection{Pseudocylindrical Convolutions} \label{sec:irregular_rep}
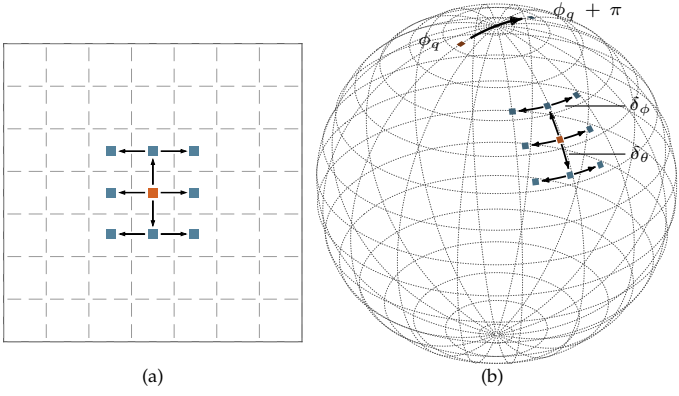
\begin{figure}[tb!]
     \begin{minipage}{1\linewidth}
     \begin{minipage}{0.45\linewidth}
     \vspace{0.3cm}
     \begin{tikzpicture}
    	\node[inner sep=0pt] (nd1) at (0,0) {\includegraphics[width=1.\linewidth]{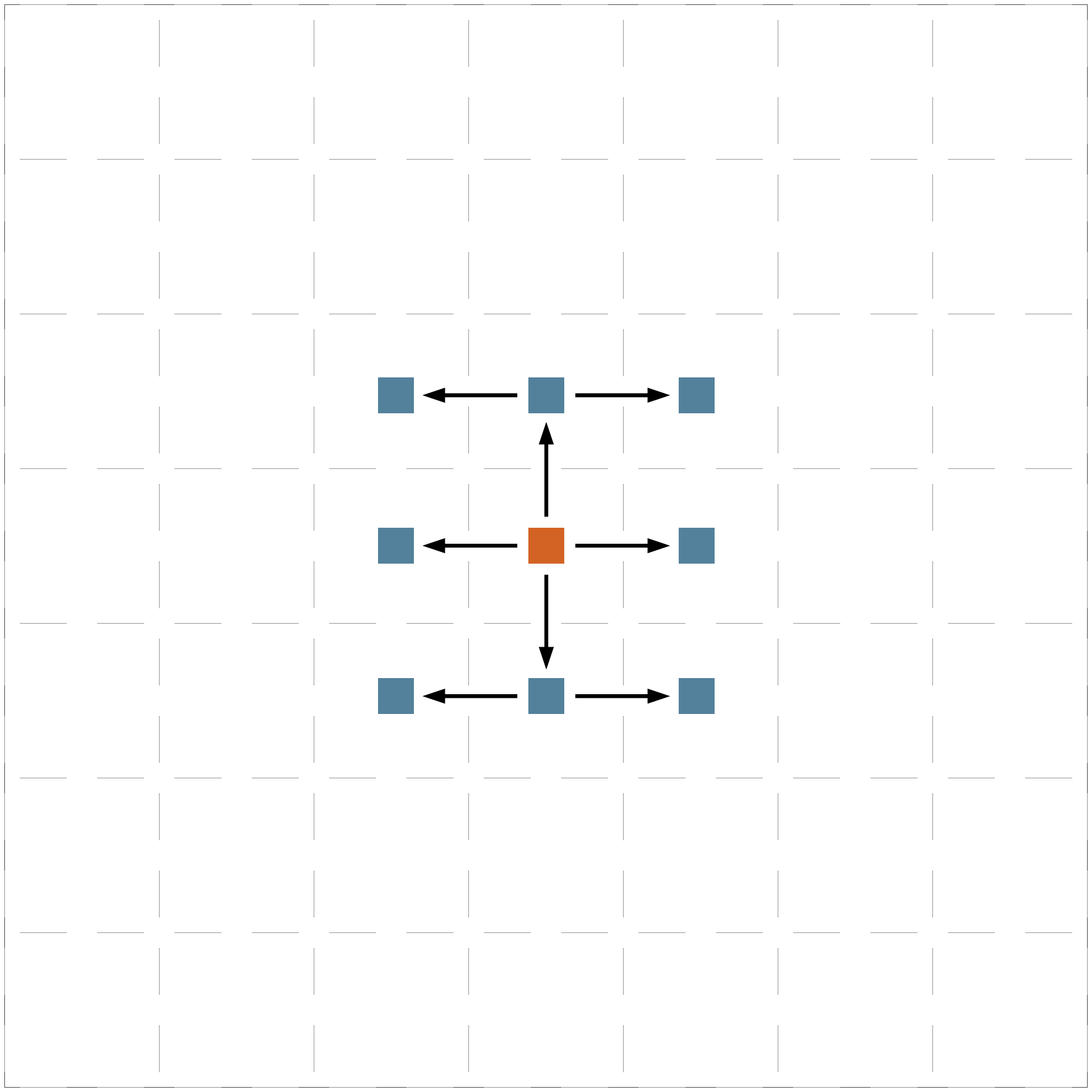}};
	 \end{tikzpicture}
     \end{minipage}
     \begin{minipage}{0.55\linewidth}
     \begin{tikzpicture}
    	\node[inner sep=0pt] (nd1) at (0,0) {\includegraphics[width=1.\linewidth]{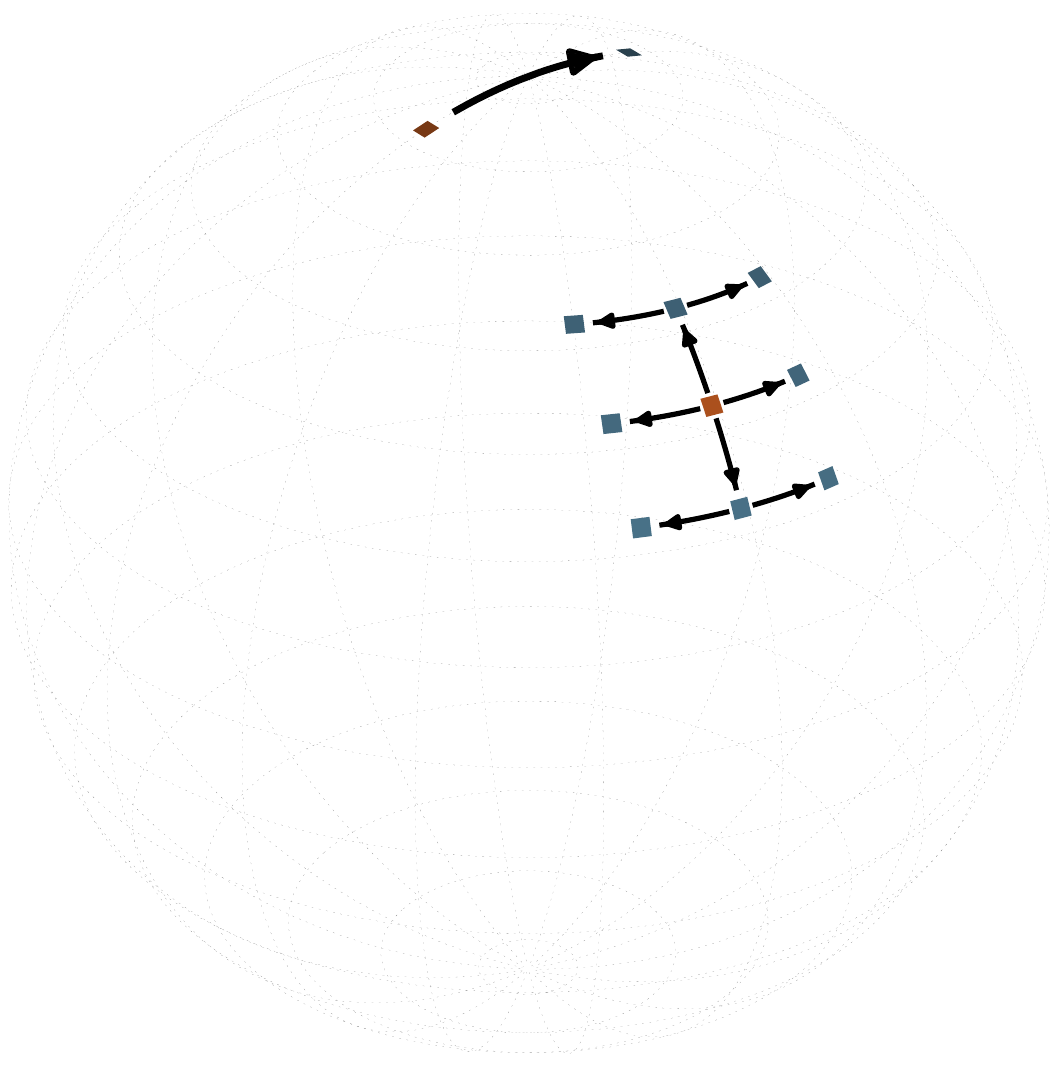}};
		\draw [line width=0.15mm] (0.9,1.04) -- (1.7,1.04);
		\node[text width=3cm,align=center] (cp10) at (1.9,1.04){\baselineskip=3pt \scriptsize{$\delta_{\phi}$} \par};
		\draw [line width=0.15mm] (0.95,0.4) -- (1.7,0.4);
		\node[text width=3cm,align=center] (cp11) at (1.9,0.4){\baselineskip=3pt \scriptsize{$\delta_{\theta}$} \par};
		\node[text width=3cm,align=center] (cp11) at (-0.9,1.9){\baselineskip=3pt \scriptsize{$\phi_q$} \par};
		\node[text width=3cm,align=center] (cp11) at (1.2,2.3){\baselineskip=3pt \scriptsize{$\phi_q  + \pi$} \par};
	 \end{tikzpicture}
     \end{minipage}
     \end{minipage}
     \begin{minipage}{1\linewidth}
     \begin{minipage}{0.45\linewidth}
     \center{\scriptsize{(a)}}
     \end{minipage}
     \begin{minipage}{0.55\linewidth}
     \center{\scriptsize{(b)}}
     \end{minipage}
     \end{minipage}
     
     \caption{Illustration of the neighborhood in the (a) planar and (b) spherical representation.}
     \label{fig:neighbor}
 \end{figure}

Based on the  pseudocylindrical representation, we define the pseudocylindrical convolution operation\footnote{In fact, nearly all DNNs implement cross-correlation instead of convolution. Here we assume the convolution filter has already been reflected around the center.} by first specifying a neighboring grid:
\begin{align}
\mathcal{N} = \{(i,j)| i,j\in\{-K,\ldots, K\}\},
\end{align}
where $2K+1$ is the spread of the convolution kernel. For a central-perspective image $\bm x$, it is straightforward to define the neighbors using the Manhattan distance (see Fig. \ref{fig:neighbor}). The response $\bm y$ of the convolution on $\bm x$ at the position $(p,q)$ is  computed by
\begin{align}\label{eq:standardconv}
\bm y(p,q) = \sum_{(i,j) \in \mathcal{N}} \bm w(i,j)\bm x(p_i, q_j),
\end{align}
where $\bm w$ denotes the convolution filter, $p_i=p+i$, and $q_j=q+j$. 
To convolve the filter over the pseudocylindrical representation, we also need to define the neighbors, with the goal of approaching the uniform sampling density over the sphere. Relying on a variant of Manhattan distance, we start from $(\theta_p, \phi_q)$ corresponding to $(p,q)\in \Omega$, and retrieve the neighbor at $(i,j)\in \mathcal{N}$ by first moving $i\delta_\theta$ along the longitude and then $j\delta_\phi$ along the latitude\footnote{We assume a unit sphere.}, where $\delta_\theta = \pi/H$ and $\delta_\phi = 2\pi \cos(\theta_p)/ W_p$. Positive (negative) values of $i$ head towards the north (south) pole. Similarly, positive (negative) values of $j$ mean anticlockwise (clockwise) movement from a bird's-eye view. We obtain the neighbors on the pseudocylindrical representation via the sphere-to-plane projection: 
\begin{align}
    p_i &= p + i,\label{eq:pi}\\
    q_j &= \frac{W_{p_i}}{W_p}(q+0.5)-0.5+\frac{\cos\theta_p}{\cos\theta_{p_i}}\frac{W_{p_i}}{W_p}j \nonumber\\
&= \frac{W_{p_i}}{W_p}\left(q+\frac{\cos\theta_p}{\cos\theta_{p_i}}j+0.5\right) - 0.5.\label{eq:loc}
\end{align}
We assume circular boundary condition, and give a careful treatment of the boundary handling near the two poles (\ie, $p_i<0$ and $p_i > H$):
\begin{align}
p_i &= (-1-p_i) \mbox{ mod } H, \nonumber\\
q_j &= (q_j + 0.5 W_{p_i}) \mbox{ mod } W_{p_i}. \label{eq:loc_pole}
\end{align}
Fig.~\ref{fig:neighbor} (b) shows an example of the adjustment of $q_j$ over the sphere when $p_i<0$ (\ie, crossing the north pole). For fractional $q_j$, we compute $\bm x(p_i,q_j)$ via linear interpolation:
\begin{align}\label{eq:interp}
\bm x(p_i,q_j) = \sum_{k\in\mathcal{N}}\bm b(q_j,k)\bm x(p_i, \lfloor q_j \rfloor + k),
\end{align}
where $\bm b$ is the linear kernel. Last, the pseudocylindrical convolution is computed by plugging Eqs. \eqref{eq:pi}, \eqref{eq:loc}, and \eqref{eq:interp} into Eq. \eqref{eq:standardconv}.

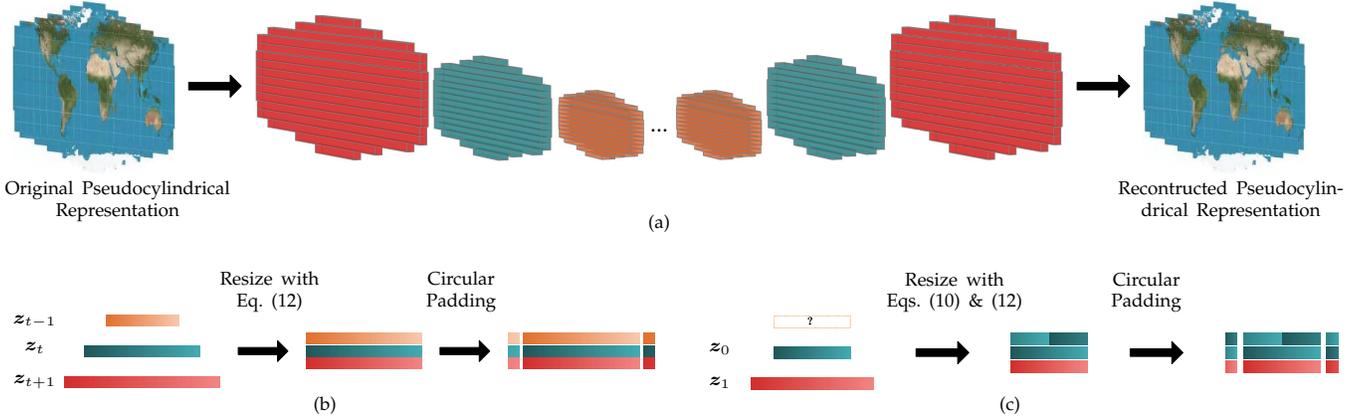
\begin{figure*}[htb!]
     \begin{minipage}{1.\linewidth}
     \begin{tikzpicture}
    	\node[inner sep=0pt] (nd1) at (0,0) {\includegraphics[width=0.95\linewidth]{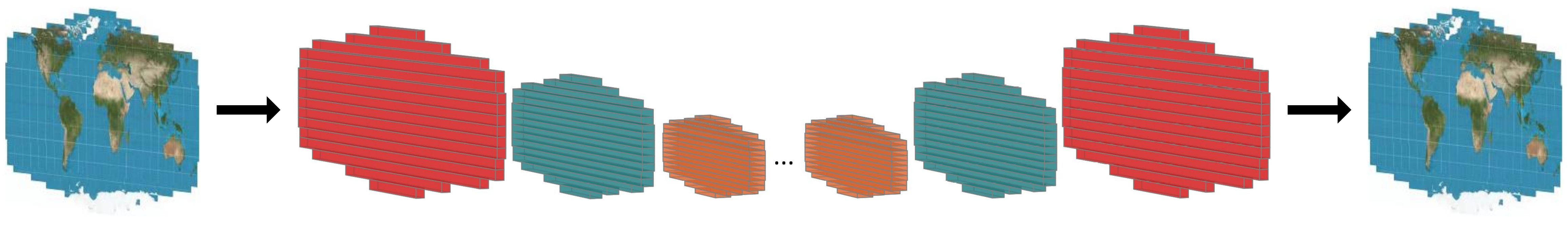}};
		\node[text width=3cm,align=center] (cp10) at (-7.2,-1.5){\baselineskip=3pt \scriptsize{Original Pseudocylindrical Representation} \par};
		\node[text width=4cm,align=center] (cp10) at (7.6,-1.5){\baselineskip=3pt \scriptsize{Recontructed Pseudocylindrical Representation} \par};
		\node[text width=4cm,align=center] (cp10) at (0,-1.8){\baselineskip=3pt \scriptsize{(a)} \par};
	 \end{tikzpicture}
     \end{minipage}

     \vspace{0.3cm}

     \begin{minipage}{1\linewidth}
     \begin{minipage}{0.48\linewidth}
     \begin{minipage}{1.\linewidth}
     \begin{tikzpicture}
    	\node[inner sep=0pt] (nd1) at (0,0) {\includegraphics[width=0.9\linewidth]{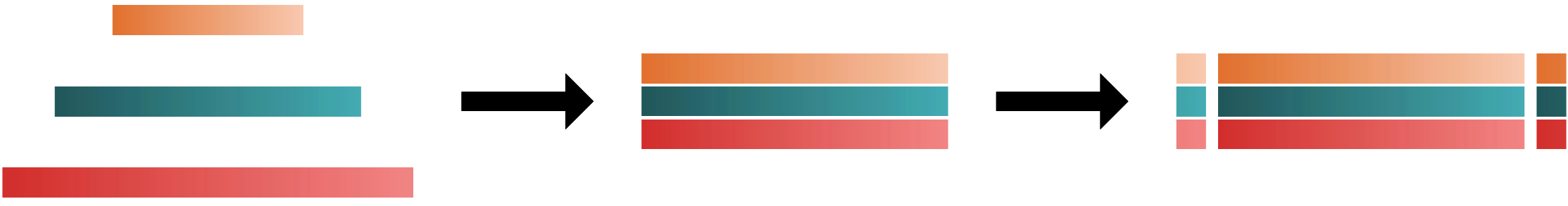}};
		\node[text width=0.8cm,align=center] (cp10) at (-4.32,0.4){\baselineskip=3pt \scriptsize{$\bm z_{t-1}$} \par};
		\node[text width=0.8cm,align=center] (cp10) at (-4.32,0.05){\baselineskip=3pt \scriptsize{$\bm z_t$} \par};
		\node[text width=0.8cm,align=center] (cp10) at (-4.32,-0.4){\baselineskip=3pt \scriptsize{$\bm z_{t+1}$} \par};
		
		\node[text width=2cm,align=center] (cp10) at (-1.2,0.8){\baselineskip=3pt \scriptsize{Resize with\\ Eq.~\eqref{eq:loc_new}} \par};
		\node[text width=2cm,align=center] (cp10) at (1.35,0.8){\baselineskip=3pt \scriptsize{Circular \\ Padding} \par};
	 \end{tikzpicture}
     \end{minipage}
     \end{minipage}
     \begin{minipage}{0.01\linewidth}
     \hspace{0.1cm}
     \end{minipage}
     \begin{minipage}{0.48\linewidth}
     \begin{tikzpicture}
     \node[inner sep=0pt] (nd1) at (0,0) {\includegraphics[width=0.9\linewidth]{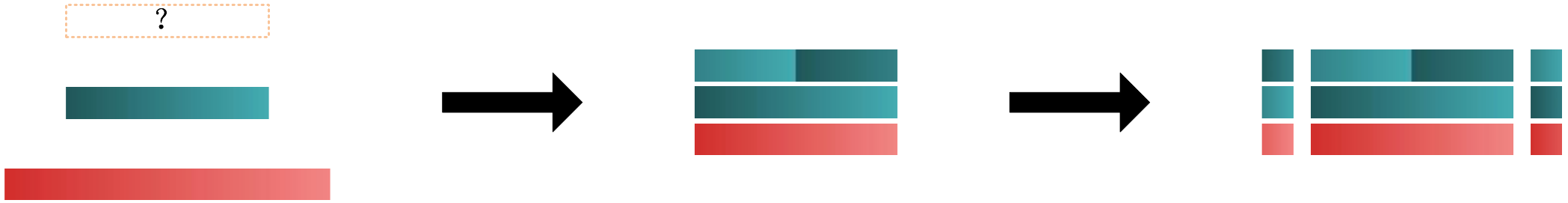}};
		\node[text width=0.8cm,align=center] (cp10) at (-4.32,0.05){\baselineskip=3pt \scriptsize{$\bm z_0$} \par};
		\node[text width=0.8cm,align=center] (cp10) at (-4.32,-0.4){\baselineskip=3pt \scriptsize{$\bm z_{1}$} \par};
		\node[text width=2cm,align=center] (cp10) at (-1.2,0.8){\baselineskip=3pt \scriptsize{Resize with\\ Eqs.~\eqref{eq:loc_pole} \& \eqref{eq:loc_new}} \par};
		\node[text width=2cm,align=center] (cp10) at (1.35,0.8){\baselineskip=3pt \scriptsize{Circular \\ Padding} \par};
     \end{tikzpicture}
     \end{minipage}
     \end{minipage}
     \begin{minipage}{1\linewidth}
     \begin{minipage}{0.48\linewidth}
     \begin{minipage}{1.\linewidth}
    	\centering{\scriptsize{(b)}}
     \end{minipage}
     \end{minipage}
     \begin{minipage}{0.01\linewidth}
     \hspace{0.1cm}
     \end{minipage}
     \begin{minipage}{0.48\linewidth}
     	\centering{\scriptsize{(c)}}
     \end{minipage}
     \end{minipage}
     \caption{Illustration of the pseudocylindrical representation and pseudocylindrical padding. (a) Intermediate pseudocylindrical representations in the DNN. (b) Pseudocylindrical padding. (c) Pseudocylindrical padding for the pole tile.}
     \label{fig:pseudo_conv}
 \end{figure*}

We take a close look at the computational complexity of the pseudocylindrical convolution, which mainly comes from three parts: neighbor search, linear interpolation, and inner product. Searching one neighbor requires calling the transcendental function, $\cos(\cdot)$, twice with four multiplications and three additions.  For linear interpolation, one
modulo operation and one addition are needed to create
the bilinear kernel, and two multiplications and one addition are used to compute interpolated value. For a kernel size of $(2K+1)\times(2K+1)$, we need $28K^2+14K$ and $20K^2+10K$ operations for neighbor search and linear interpolation, respectively,  but only $8K^2+8K+1$ operations for inner product.   

To reduce the computational complexity of neighbor search, we make a mild approximation to Eq.~\eqref{eq:loc}:
\begin{align}\label{eq:loc_new}
    q_j \approx \frac{W_{p_i}}{W_p}(q+j+0.5) - 0.5,
\end{align}
where we assume $\cos(\theta_p) \approx \cos(\theta_{p_i})$. This is reasonably true because $\theta_p$ and $\theta_{p_i}$ correspond to adjacent rows, and are very close provided that $H$ is large. From Eq. \eqref{eq:loc_new}, it is clear $q_j$ = $(q+k)_{j-k}$, meaning that adjacent samples in a row are neighbors of one another with no computation. Moreover, searching for neighbors in an adjacent row amounts to scaling it to the width of the current row\footnote{More precisely, we first shift the adjacent row by half pixel, and scale it to the width of the current row, and shift it back by half pixel.}. Furthermore, we  perform circular padding for $q_j<0$ and $q_j\geq W_p$ with $q_j = q_j \mbox{ mod } W_p$. We refer to this process as \textit{pseudocylindrical padding} (see Fig.~\ref{fig:pseudo_conv} (b) and (c)). For a row with width $W_p$, we greatly reduce the computation\footnote{$(2K+1)2K$ arises from the circular boundary handling along the longitude.} from $(48K^2+24K)W_p$ to $20KW_p + (2K+1)2K$, where $W_p \gg K$.  

We may further simplify Eq. \eqref{eq:loc_new} to \eqref{eq:standardconv} by enforcing $W_p = W_{p_i}$. By doing so, the $p$-th and $(p+i)$-th  rows  become neighborhood of each other with no computation. The neighboring rows with the same width can be viewed as a \textit{tile}, and the pseudocylindrical representation reduces gracefully to the tiled representation \cite{yu2015content} (see Fig. \ref{fig:pseudo} (d)). Pseudocylindrical padding occurs only 
at the boundaries of each tile. In summary, the tiled representation $\bm z$ of $\bm x$ is composed of $\{\bm z_t\}_{t=0}^{T-1}$, where $\bm z_t \in\mathbb{R}^{H_t \times W_t}$ is the $t$-th tile. The set of free parameters are $\{T, \{H_t\}_{t=0}^{T-1}, \{W_t\}_{t=0}^{T-1}\}$.
With these two steps of simplifications,  the pseudocylindrical convolution can be implemented in parallel on $\{\bm z_t\}$ by standard convolution with pseudocylindrical padding. 
On one hand, this offers the opportunity to build DNN-based compression methods for omnidirectional images upon those for central-perspective images with minimal modifications. 
On the other hand, this enables fast implement of the proposed pseudocylindrical convolution. For a tile $\bm z_t$, the computation operations used for pseudocylindrical padding are $20KW_t + (2K+H_t)2K$, which are much smaller than the operations for convolution, \ie, $(8K^2+8K+1) W_t H_t$, when $H_t \gg 1$. In such case, our pseudocylindrical convolution should achieve nearly the same running speed as the standard convolution with zero padding.    

\subsection{Pseudocylindrical Representation Optimization}\label{sec:irregular_learn}
In general, different \mdegr images may have different parameter configurations of the pseudocylindrical representation for optimal compression performance, which depend on the image content. The corresponding combinatorial optimization problem can be formulated as 
\begin{equation}\label{eq:gop}
\begin{aligned}
\min_{T, \{H_t\},\{W_t\}} \quad & \mathrm{RD}(\bm x, \texttt{compress}_{\bm \alpha}(\bm x); T, \{H_t\},\{W_t\})\\
\textrm{s.t.} \quad & T\in\{0,\ldots, H-1\},\\
 & H_t \in\{0,\ldots, H-1\}, \;  t\in \{0,\ldots, T-1\}, \\
  & \sum_t H_t = H-1,  \\
  & W_t \in\{0,\ldots, W-1\}, \;  t\in \{0,\ldots, T-1\}, \\
\end{aligned}
\end{equation}
where $\bm x$ is the given \mdegr image, $\mathrm{RD}(\cdot)$ is a quantitative measure for rate-distortion performance, and $\texttt{compress}_{\bm \alpha}(\cdot)$ is a generic compression method with a learnable parameter vector $\bm \alpha$. As noted by Yu \etal \cite{yu2015content},  Problem \eqref{eq:gop} is essentially a  multiple-dimensional, multiple-choice knapsack problem, which prohibits exhaustive search when $H$ and $W$ are large. We choose to simplify the problem in the follow ways.

\begin{enumerate}
    \item Treat $T$ and $\{H_t\}_{t=1}^{T-1}$ as hyperparameters, and pre-set them, where $H_t = H_0$ and $T = H/ H_0$. The general guideline is to set $H_0$ large enough to enjoy the fast computation of standard convolution, while making Eq.~\eqref{eq:loc_new} approximately hold.
    \item Quantize the width to $L$ levels, where $L\ll W$. Thus enumeration of the possible widths is performed in a much reduced search space.
    \item Enforce the symmetry of the pseudocylindrical representation with respect to the equator, which further halves the search space.
    \item Discourage oversampling at higher latitudes by adding the constraint $W_t \le W_{t'}$ for $t \le t'$ and $t,t'\in \{0,\ldots, T_{\mathrm{half}}\}$, where $T_{\mathrm{half}} =  \lfloor (T-1)/2\rfloor - 1$ (see the coordinate system in Fig. \ref{fig:pseudo}). 
    \item Solve the problem at the dataset level instead of the image level for two reasons. First, even with the above simplifications, it still takes quite some time to obtain the sub-optimal configuration in practice. Second, the content-dependent configuration may render the training of DNN-based \mdegr image compression unstable.  
\end{enumerate}

\begin{figure}[!tbp]
\centering
\includegraphics[width=1.0\linewidth]{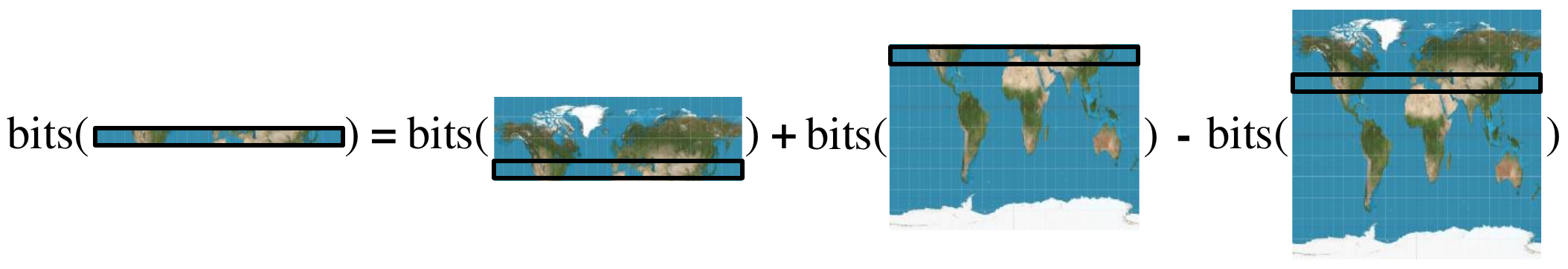}
\caption{Estimation of the bits used to code a single tile.}
\label{fig:size}
\end{figure}

Putting together, Problem~\eqref{eq:gop} is simplified to 
\begin{equation}\label{eq:sop}
\begin{aligned}
\min_{\{W_t\}} \quad & \frac{1}{\vert\mathcal{D}\vert}\sum_{\bm x\in \mathcal{D}}\mathrm{RD}(\bm x, \texttt{compress}_{\bm \alpha}(\bm x); \{W_t\})\\
\textrm{s.t.} \quad 
 & W_t \in\{\overline{W}_0,\ldots, \overline{W}_{L-1}\},\;  t\in \{0,\ldots, T-1\}, \\
  & W_t = W_{T-1-t},\; t\in \{0,\ldots, T_{\mathrm{half}}\},  \\
  & W_t \le W_{t'},\; \textrm{for } t \le t' \textrm{ and }  t,t'\in \{0,\ldots, T_{\mathrm{half}}\}, \\
\end{aligned}
\end{equation}
where $\overline{W}_t=(t+1)\left\lfloor \frac{W}{L} \right\rfloor$. Now, an exhaustive search which evaluates
all possible configurations may be feasible. Alternatively, we propose a divide-and-conquer greedy solver to Problem \eqref{eq:sop} in case $L$ and $T$ are still large. We first initialize all $\{W_t\}$ to $\overline{W}_{T-1}$. From the pole to the equator, we enumerate the possible widths of the current tile while holding higher-latitude tiles to the estimated widths and lower-latitude tiles to the initialized widths. This procedure is repeated until all tiles are visited, which is summarized in Algorithm \ref{alg1}.

\begin{algorithm}
\caption{Greedy Estimation  of the Pseudocylindrical Representation Parameters}\label{alg1}
\hspace*{\algorithmicindent} \textbf{Input:} ERP image set $\mathcal{D}=\{\bm x^{(0)}, \ldots, \bm x^{(\vert\mathcal{D}\vert-1)}\}$, and the quantized width set $\{\overline{W}_0, \ldots, \overline{W}_{L-1}\}$.\\
\hspace*{\algorithmicindent}  \textbf{Output:} The optimized parameter set $\{W_t^\star\}$.
\begin{algorithmic}[1]

\For{$t \gets 0 \mbox{ to } T-1$}
\State $W^\star_t \gets \overline{W}_{L-1}$; \Comment{Initialization}
\EndFor

\For{$t \gets 0 \mbox{ to } T_{\mathrm{half}}$}
\State $V_{\mathrm{best}} \gets \infty, \;\;\; W_{\mathrm{best}} \gets 0$;
\If{$t = 0$}
	\State $\mathrm{start}\_\mathrm{width} \gets \overline{W}_0$;
\Else
	\State $\mathrm{start}\_\mathrm{width} \gets W_{t-1}^\star$;
\EndIf
\For{$W^\star_t \gets \mathrm{start}\_\mathrm{width} \mbox{ to } \overline{W}_{L-1}$}
\State $W^\star_{T-t-1} \gets W^\star_t$;
\State $ V_\mathrm{temp} \gets \mathbb{E}_{\bm x \in \mathcal{D}}\mathrm{RD}(\bm x, \texttt{compress}_{\bm \alpha}(\bm x); \{W^\star_t\})$;
\If{$V_\mathrm{temp} < V_\mathrm{best}$}
\State	$V_\mathrm{best} \gets V_\mathrm{temp}, \;\; W_\mathrm{best} \gets W_t$;
\EndIf
\EndFor
\State $W^\star_{t} \gets W_\mathrm{best}, \;\; W^\star_{T-t-1} \gets W_\mathrm{best}$;
\EndFor
\end{algorithmic}
\end{algorithm}

It remains to instantiate the objective function in Problem \eqref{eq:sop}, which quantifies the rate-distortion trade-off given a particular parameter configuration $\{W_t\}$. To obtain an accurate estimate, it is preferable but  impractical to optimize a set of DNN-based image compression models (\ie, optimize $\bm \alpha$ in $\texttt{compress}_{\bm \alpha}(\cdot)$) for each configuration. A workaround is to apply  existing codecs on each tile and ``sum'' the results. In our implementation, we use JPEG2000 as the off-the-shelf compression method. As context is crucial in image compression for  bitrate reduction, a na\"{i}ve compression of the tile without considering adjacent ones would lead to an inaccurate parameter estimation of our representation that admits pseudocylindrical padding. To alleviate this issue, we introduce a proxy rate-distortion objective. For the rate estimation of the $t$-th tile, $\bm z_t$, we resize the width of $\bm x$ to $W_t$, and crop two subimages such that the intersection is $\bm z_t$ and the union is the resized image. The rate of $\bm z_t$ is calculated as the difference between the number of bits of the subimages and the resized image. This process is better illustrated in Fig.~\ref{fig:size}.

For the distortion, we suggest to use viewport-based objective quality metrics, which correctly reflect how humans view  \mdegr images. According to~\cite{sui2021perceptual}, viewport-based metrics deliver so far the best quality prediction performance on \mdegr images. In our implementation, we use MSE as the base quality metric. We sample $14$ viewports by first mapping the pseudocylindrical representation back to the unit sphere followed by rectilinear projections\footnote{The centers of the $14$ viewports correspond to  $(0,-\frac{\pi}{2})$, $(0,0)$, $(0,\frac{\pi}{2})$, $(0,\pi)$, $(-\frac{\pi}{4},-\frac{\pi}{2})$, $(-\frac{\pi}{4},0)$, $(-\frac{\pi}{4},\frac{\pi}{2})$, $(-\frac{\pi}{4},\pi)$, $(\frac{\pi}{4},-\frac{\pi}{2})$, $(\frac{\pi}{4},0)$, $(\frac{\pi}{4},\frac{\pi}{2})$, $(\frac{\pi}{4},\pi)$, $(\frac{\pi}{2},0)$ and $(-\frac{\pi}{2},0)$, respectively, on the unit sphere.}. Each viewport is a $H_v \times W_v$ rectangle, where $H_v = \lceil\frac{H}{3}\rceil$ and $ W_v = \lceil \frac{W}{4} \rceil$, with a field of view (FoV) of $\frac{\pi}{3}\times\frac{\pi}{2}$. Together, they cover all spherical content. 
%

By varying the QP values of JPEG2000, we produce a rate-distortion curve for each image $\bm{x}\in\mathcal{D}$, and we average all curves\footnote{We average the rates and the distortion scores separately over images compressed with identical QP values.} as the rate-distortion performance of the current parameter configuration of the proposed pseudocylindrical representation.  We compare two curves using the Bjontegaard delta bitrate saving (BD-BR) metric \cite{bjontegaard2001calculation} to identify a better configuration.

\begin{figure}[!tbp]
\centering

\includegraphics[width=0.8\linewidth]{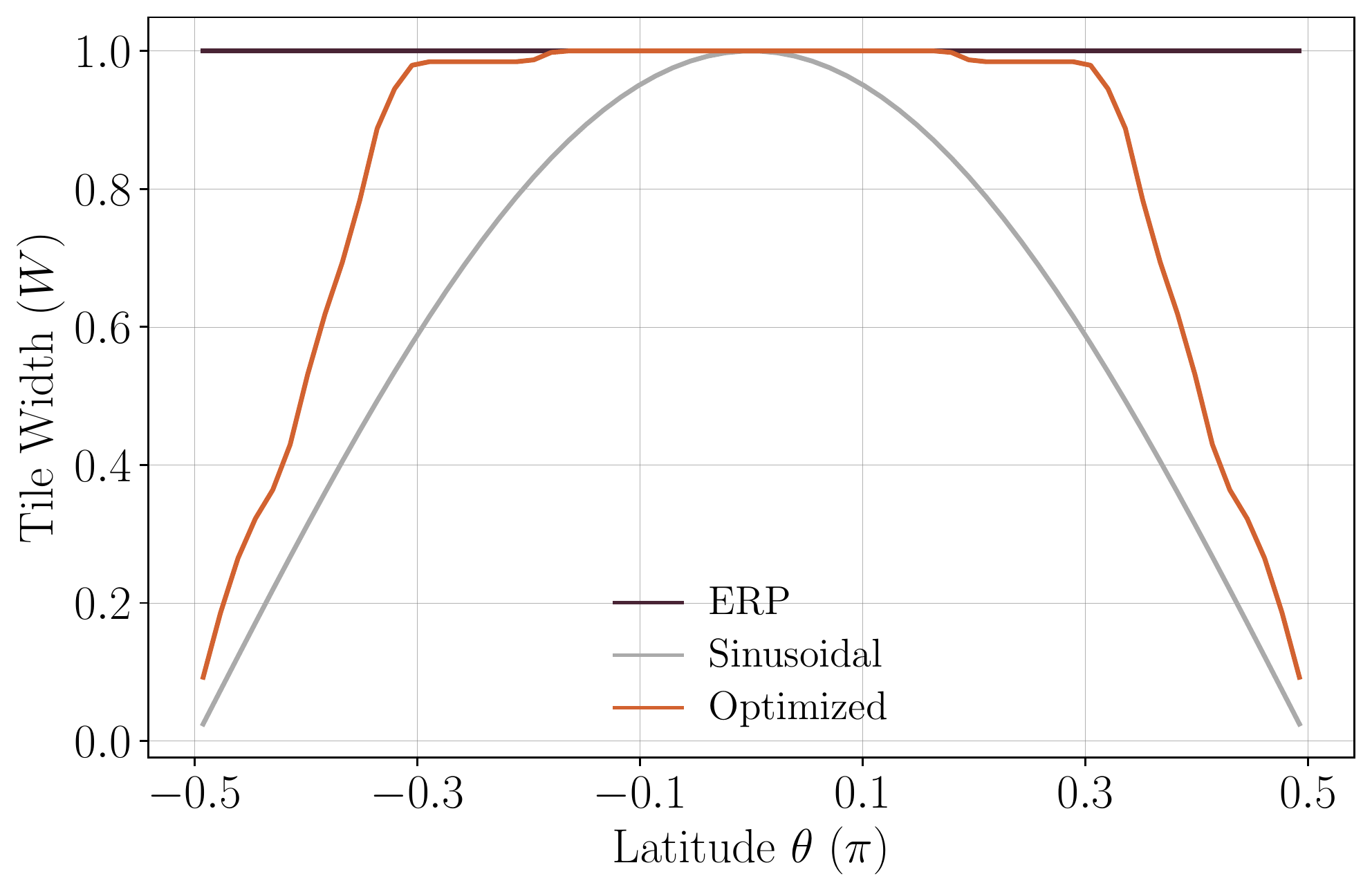}

\caption{The optimized width configuration of the pseudocylindrical representation in comparison to ERP and the sinusoidal projection.}\label{fig:slice_param}
\end{figure}

\begin{figure*}[!tbp]
\centering
\scriptsize

\begin{minipage}{1\linewidth}
\centering
\begin{minipage}{0.25\linewidth}

\begin{tikzpicture}
\node (nd1) at (0,1.3) {\includegraphics[width=0.95\linewidth]{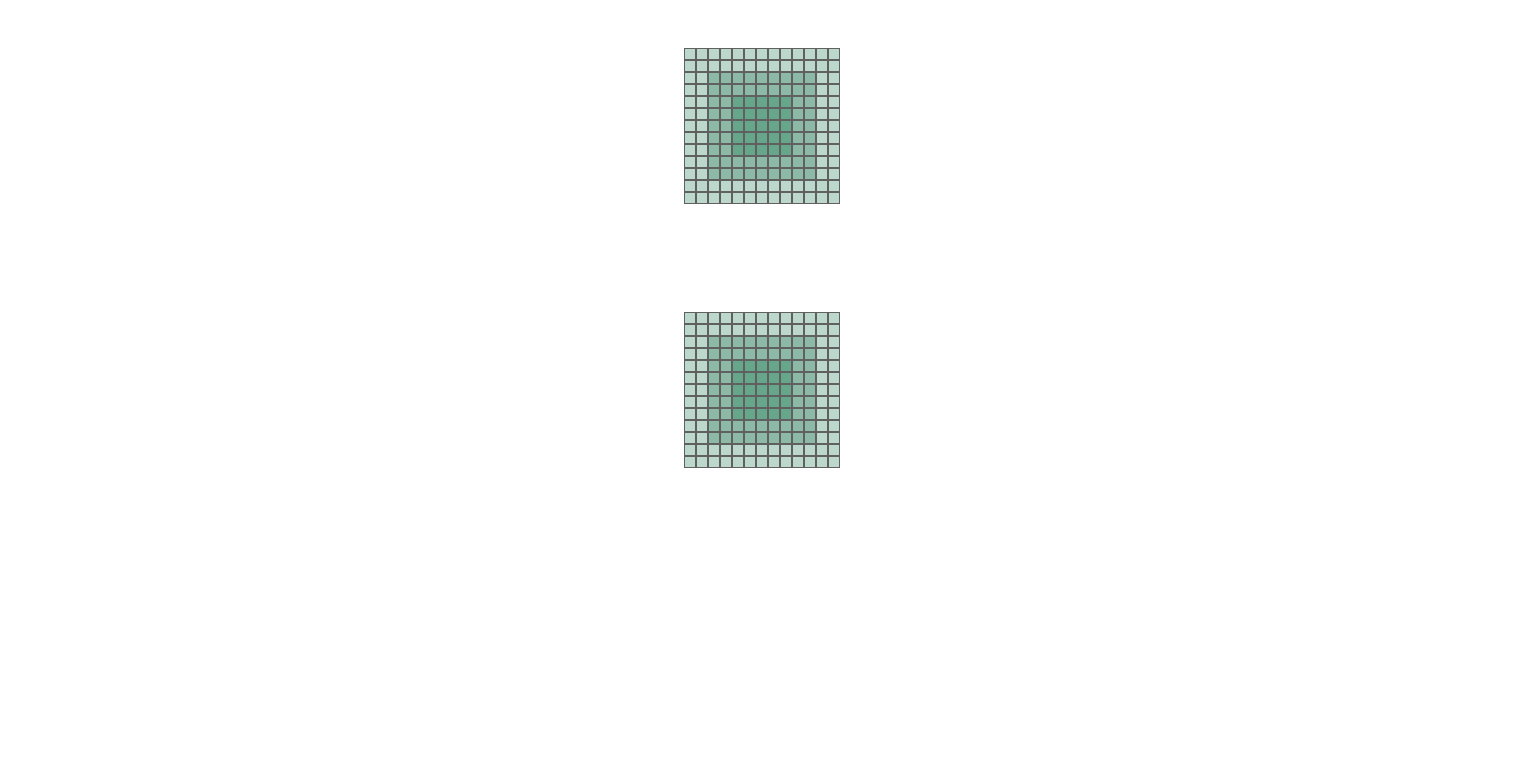}};
\node (nd2) at (0,-1.3) {\includegraphics[width=0.95\linewidth]{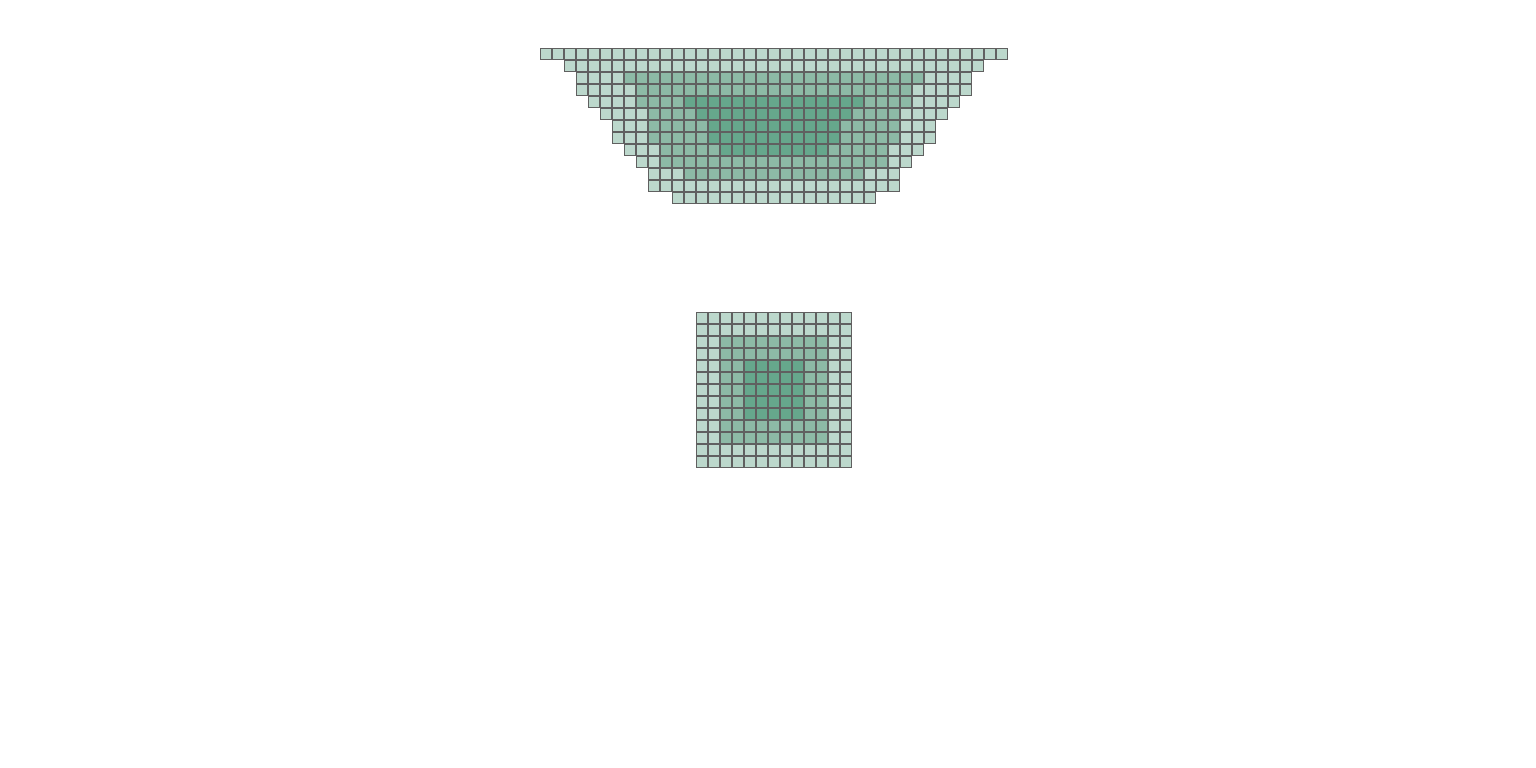}};
\draw[draw=lightgray] (-2.2,-2.41) rectangle ++(4.4,2.22);
\draw[draw=lightgray] (-2.2,0.19) rectangle ++(4.4,2.22);
\draw [line width=0.25mm,densely dotted] (0.32,2) -- (0.8,2); 
\node[text width=2cm,align=center] (t1) at (1.5,2) {\baselineskip=3pt \scriptsize{$\theta=0.36\pi$} \par};
\draw [line width=0.25mm,densely dotted] (0.32,1.35) -- (0.8,1.35); 
\node[text width=2cm,align=center] (t1) at (1.25,1.35) {\baselineskip=3pt \scriptsize{$\theta=0$} \par};
\draw [line width=0.25mm,densely dotted] (0.52,-0.6) -- (0.8,-0.6); 
\node[text width=2cm,align=center] (t1) at (1.5,-0.6) {\baselineskip=3pt \scriptsize{$\theta=0.36\pi$} \par};
\draw [line width=0.25mm,densely dotted] (0.32,-1.25) -- (0.8,-1.25); 
\node[text width=2cm,align=center] (t1) at (1.25,-1.25) {\baselineskip=3pt \scriptsize{$\theta=0$} \par};
\node[text width=2cm,align=center] (cp7) at (0,0.7) {\baselineskip=3pt \scriptsize{Standard Convolution} \par};
\node[text width=2cm,align=center] (cp7) at (0,-1.9) {\baselineskip=3pt \scriptsize{Pseudocylindrical Convolution} \par};
    	
\end{tikzpicture}

\centering{(a)}
\end{minipage}
\begin{minipage}{0.47\linewidth}
\begin{minipage}{0.48\linewidth}
\centering
\scriptsize{Standard Convolution}
\end{minipage}
\begin{minipage}{0.48\linewidth}
\centering
\scriptsize{\hspace{-0.5cm} Pseudocylindrical Convolution}
\end{minipage}

\vspace{0.05cm}

\begin{minipage}{0.48\linewidth}
\includegraphics[width=1\linewidth]{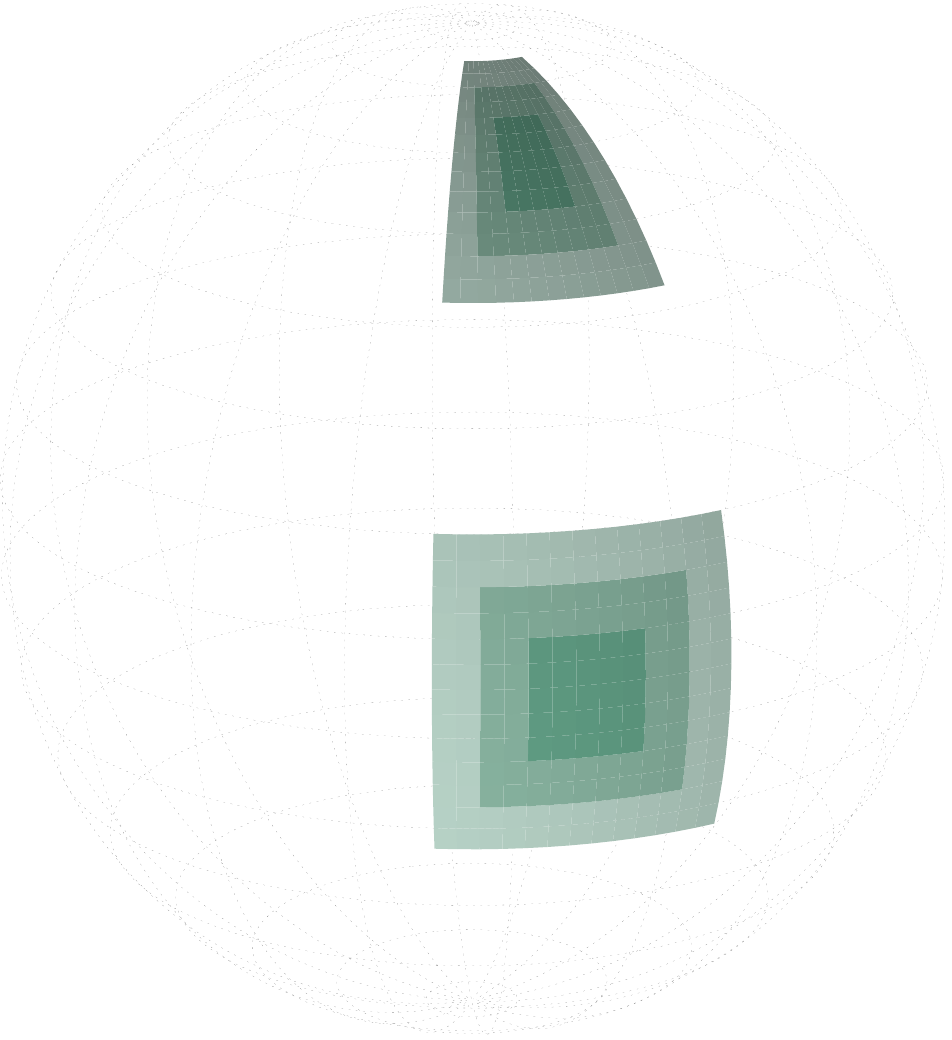}
\end{minipage}
\begin{minipage}{0.48\linewidth}
\includegraphics[width=1\linewidth]{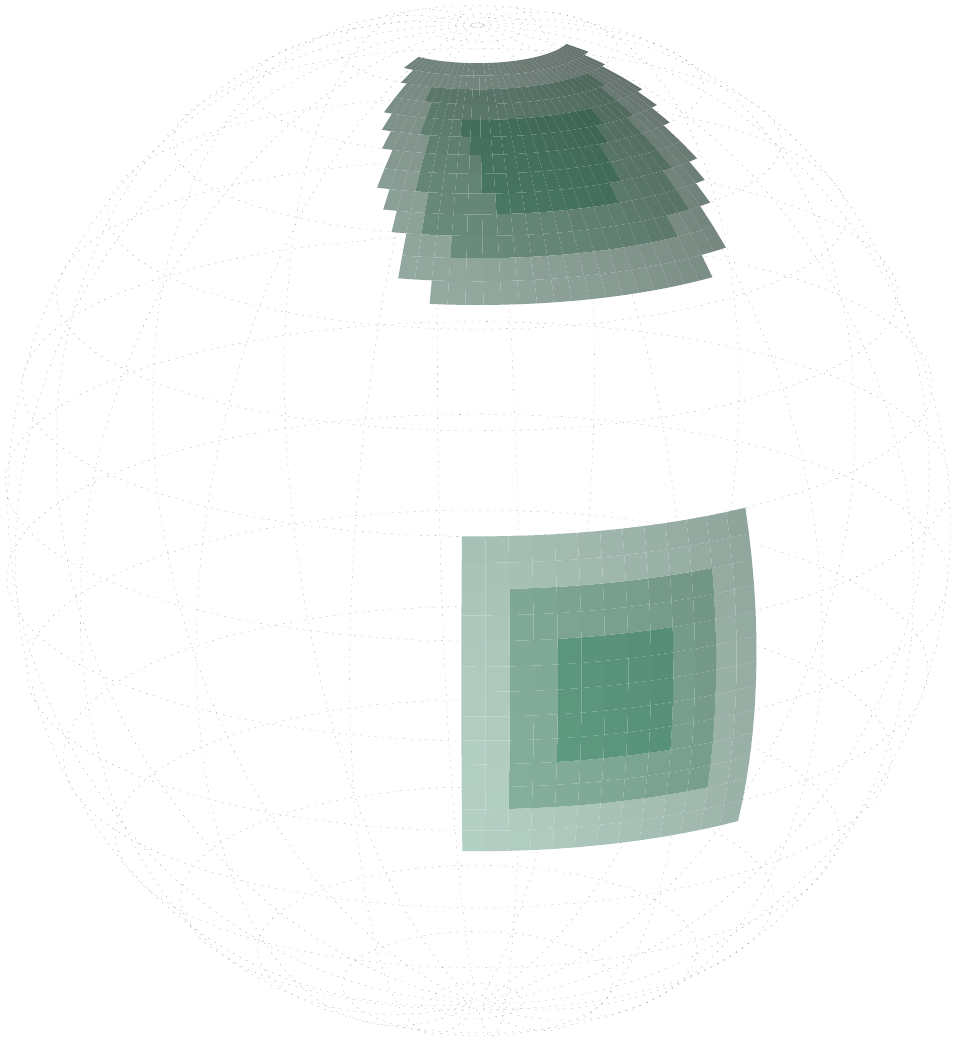}
\end{minipage}

\vspace{0.1cm}

\centering{(b)}
\end{minipage}
\begin{minipage}{0.25\linewidth}
\centering
\begin{minipage}{0.48\linewidth}
\centering
\scriptsize{Standard Convolution}
\end{minipage}
\begin{minipage}{0.48\linewidth}
\centering
\scriptsize{Pseudocylindrical Convolution}
\end{minipage}

\vspace{0.1cm}

\begin{minipage}{0.48\linewidth}
\begin{tikzpicture}
\node (nd1) at (0,0) {\includegraphics[width=1.\linewidth]{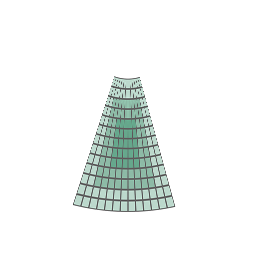}}; 
\draw[draw=lightgray] (-1.1,-1.1) rectangle ++(2.2,2.2);
\node[text width=2cm,align=center] (t1) at (0,0.8) {\baselineskip=3pt \scriptsize{$\theta=0.36\pi$} \par};
\end{tikzpicture}
\end{minipage}
\begin{minipage}{0.48\linewidth}
\begin{tikzpicture}
\node (nd1) at (0,0) {\includegraphics[width=1.\linewidth]{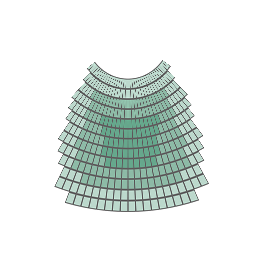}};
\draw[draw=lightgray] (-1.1,-1.1) rectangle ++(2.2,2.2); 
\node[text width=2cm,align=center] (t1) at (0,0.8) {\baselineskip=3pt \scriptsize{$\theta=0.36\pi$} \par};
\end{tikzpicture}
\end{minipage}

\vspace{0.1cm}

\begin{minipage}{0.48\linewidth}
\begin{tikzpicture}
\node (nd1) at (0,0) {\includegraphics[width=1.\linewidth]{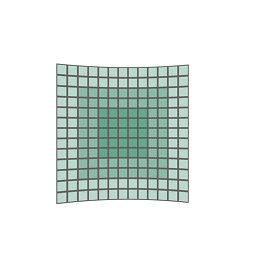}}; 
\draw[draw=lightgray] (-1.1,-1.1) rectangle ++(2.2,2.2);
\node[text width=2cm,align=center] (t1) at (0,0.8) {\baselineskip=3pt \scriptsize{$\theta=0$} \par};
\end{tikzpicture}
\end{minipage}
\begin{minipage}{0.48\linewidth}
\begin{tikzpicture}
\node (nd1) at (0,0) {\includegraphics[width=1.\linewidth]{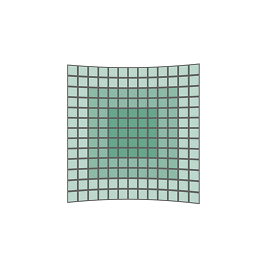}}; 
\draw[draw=lightgray] (-1.1,-1.1) rectangle ++(2.2,2.2);
\node[text width=2cm,align=center] (t1) at (0,0.8) {\baselineskip=3pt \scriptsize{$\theta=0$} \par};
\end{tikzpicture}
\end{minipage}

\vspace{0.1cm}

(c)

\end{minipage}
\end{minipage}

\caption{Illustration of the receptive field of the standard and pseudocylindrical  convolution.  (a) ERP domain. (b) Spherical domain. (c) Viewport domain.}\label{fig:spherical_conv}

\end{figure*}
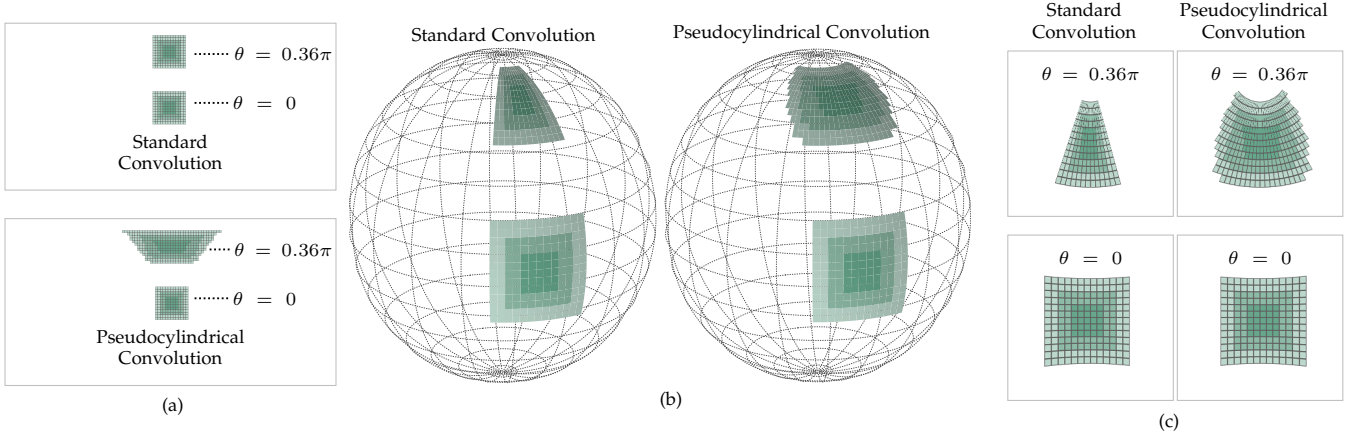

\begin{figure*}
\centering
\begin{tikzpicture}
    	\node[inner sep=0pt] (nd1) at (0,0) {\includegraphics[width=0.9\linewidth]{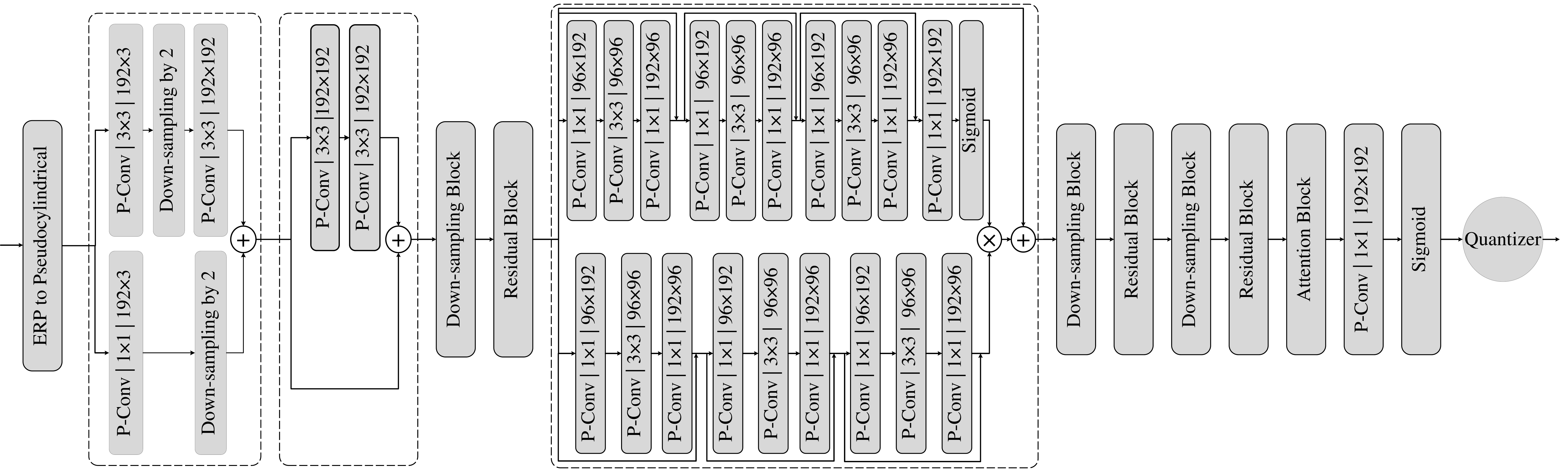}};
		\node[text width=0.1cm,align=center] (cp10) at (-8.4,-0.1){\baselineskip=3pt \scriptsize{$\bm x$} \par};
		\node[text width=1.8cm,align=center] (cp11) at (-6.3,-2.8){\baselineskip=3pt \scriptsize{Down-sampling \\ Block} \par};
		\node[text width=1.8cm,align=center] (cp12) at (-4.6,-2.8){\baselineskip=3pt \scriptsize{Residual \\ Block} \par};
		\node[text width=1.8cm,align=center] (cp12) at (0,-2.7){\baselineskip=3pt \scriptsize{Attention Block} \par};
		\node[text width=0.1cm,align=center] (cp14) at (8.3,-0.05){\baselineskip=3pt \scriptsize{$\bar{\bm c}$} \par};
		\node[text width=0.1cm,align=center] (cp14) at (6.95,0.15){\baselineskip=3pt \scriptsize{$\bm c$} \par};
\end{tikzpicture}

\vspace{-0.2cm}

\caption{Analysis transform $g_a$. P-Conv: proposed pseudocylindrical convolution with filter support ($S\times S$) and number of channels (output$\times$input).}\label{fig:frame_analysis}
\end{figure*}

\begin{figure}
\centering
\begin{tikzpicture}
\node[inner sep=0pt] (nd1) at (0,0) {\includegraphics[width=0.95\linewidth]{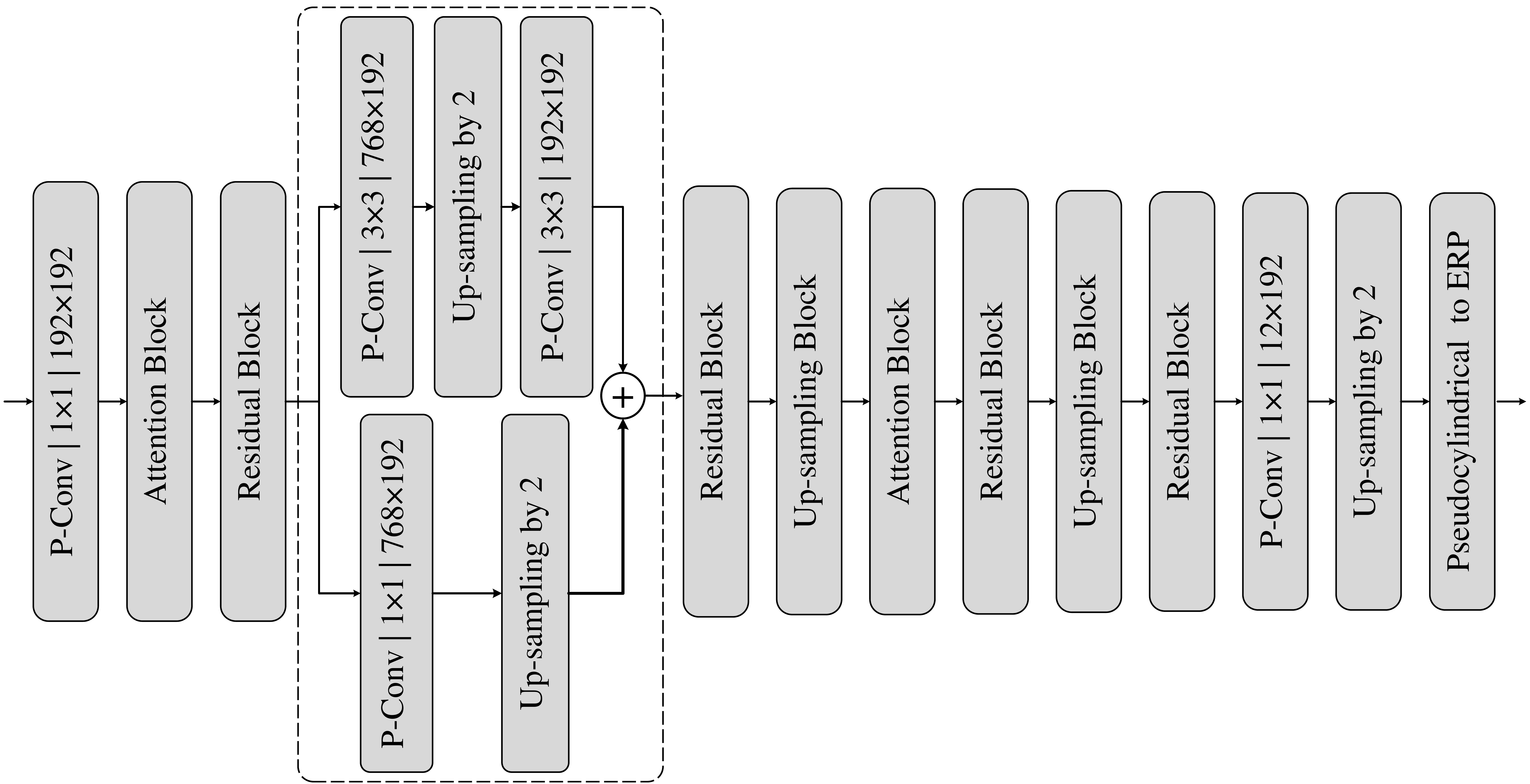}};
		\node[text width=0.1cm,align=center] (cp10) at (-4.4,-0.05){\baselineskip=3pt \scriptsize{$\bar{\bm c}$} \par};
		\node[text width=2.4cm,align=center] (cp11) at (-1.5,-2.5){\baselineskip=3pt \scriptsize{Up-sampling Block} \par};
		\node[text width=0.1cm,align=center] (cp14) at (4.4,-0.05){\baselineskip=3pt \scriptsize{$\hat{\bm x}$} \par};
\end{tikzpicture}

\vspace{-0.2cm}

\caption{Synthesis transform $g_s$. P-Conv: proposed pseudocylindrical convolution with filter support ($S\times S$) and number of channels (output$\times$input).}\label{fig:frame_synthesis}
\end{figure}

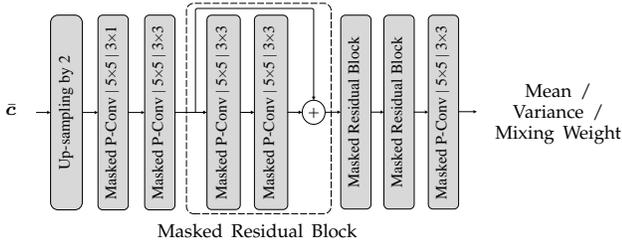
\begin{figure}
\centering
\begin{tikzpicture}
    	\node[inner sep=0pt] (nd1) at (0,0) {\includegraphics[width=0.68\linewidth]{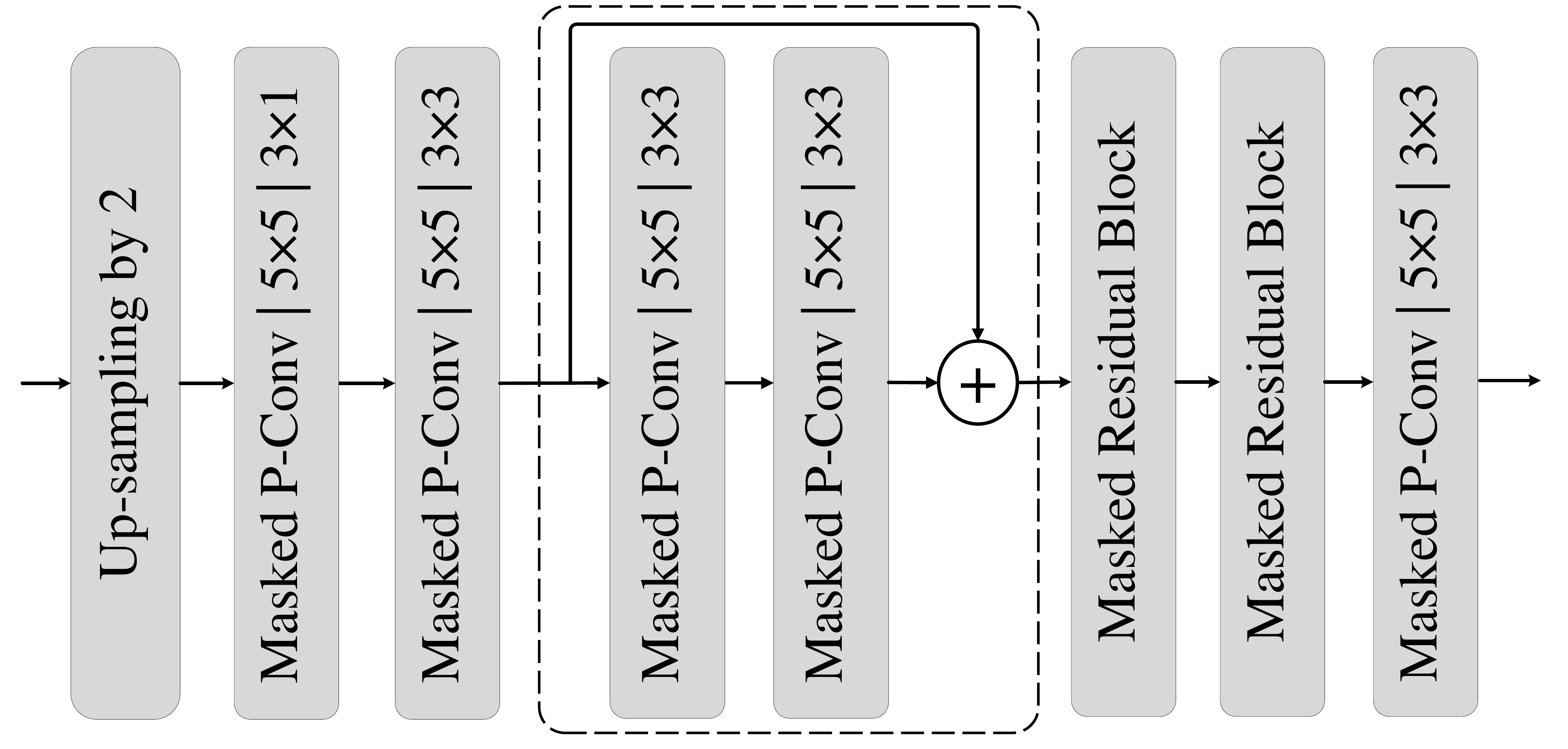}};
		\node[text width=0.1cm,align=center] (cp10) at (-3.3,0){\baselineskip=3pt \scriptsize{$\bar{\bm c}$} \par};
		\node[text width=4cm,align=center] (cp11) at (-0,-1.6){\baselineskip=3pt \scriptsize{Masked Residual Block} \par};
		\node[text width=1.8cm,align=center] (cp14) at (4,-0.05){\baselineskip=3pt \scriptsize{Mean / Variance / Mixing Weight} \par};
\end{tikzpicture}

\vspace{-0.2cm}

\caption{Entropy network $g_e$. Masked P-Conv: masked pseudocylindrical convolution adapted from \cite{li2020efficient}.}\label{fig:frame_entropy}
\end{figure}

We conclude this section by illustrating two interesting properties of our pseudocylindrical representation and convolution. First,  Fig. \ref{fig:pseudo} shows the optimized pseudocylindrical configuration in comparison to sinusoidal projection and its tiled version, where we set $H=512$, $W=1024$, $H_0=32$, and $L=64$. The key observation is that the optimized configuration does not completely resolve the over-sampling problem in ERP (see also Fig. \ref{fig:slice_param}). Surprisingly, to achieve better compression performance, the mid-latitude regions are still over-sampled than the sinusoidal tiles, which is also supported by the ablation experiments in Sec. \ref{sec:abl}. Second, we build a six-layer DNN with $3\times 3$ pseudocylindrical convolutions, and visualize the receptive field at different latitudes in Fig. \ref{fig:spherical_conv}. With the optimized structure as input, the proposed pseudocylindrical convolution is able to produce similar receptive fields for different locations on the sphere, which are latitude-adaptive in the ERP  domain. Although the receptive field is slightly deformed, it should not be a problem in \mdegr image compression because the kernel weights can be learned to adapt to such geometric distortions (if necessary) by optimizing viewport-based perceptual quality metrics.

\section{Learned \mdegr Image Compression System}
In this section, we design a learned \mdegr image compression system based on the proposed pseudocylindrical representation and convolution. At a high level, our system consists of an analysis network $g_a$, a non-uniform quantizer $g_q$, a synthesis network $g_s$, and a context-based  entropy network $g_e$, which are jointly optimized for rate-distortion performance.

\subsection{Analysis, Synthesis, and Entropy Networks}
The analysis transform $g_a$ takes the ERP image $\bm x$ as input and maps it to the proposed pseudocylindrical representation $\bm z$, based on which the code representation $\bm c$ is produced by the network.  $g_a$ is made of ERP to pseudocylindrical representation transform, four down-sampling blocks, four residual blocks, two attention blocks, a back-end pseudocylindrical convolution, and a sigmoid layer, whose computational graph with detailed specifications is shown in Fig~\ref{fig:frame_analysis}. The down-sampling block processes and down-samples the pseudocylindrical feature maps by a factor of two. The residual block~\cite{He_2016_CVPR} has two convolution layers with a skip connection, following each down-sampling block. A simplified attention block~\cite{cheng2020learned} is added right after the second and fourth residual block to increase the model capacity and expand the receptive field. A final convolution layer with $C$ filters followed by a sigmoid activation is used to produce the code representation $\bm c$ with a desired operating range of $[0,1]$.

The synthesis transform $g_s$ (see Fig.~\ref{fig:frame_synthesis}) is a mirror of the analysis transform where the down-sampling blocks are replaced by the up-sampling blocks. Instead of performing deconvolution for upsampling, we expand the feature representation by a factor of four in the channel dimension and reshape it such that the height and width grow by a factor of two~\cite{toderici2016full,shi_2016_real}. 
$g_s$ ends with a pseudocylindrical representation to ERP transform to reconstruct the ERP from the proposed pseudocylindrical representation.
Generalized divisive normalization (GDN) and inverse GDN as bio-inspired  nonlinearities~\cite{balle2016end} are separately adopted after the last convolution of the down-sampling and up-sampling blocks. For other convolution layers, unless stated otherwise, the parametric rectified linear unit (ReLU) is used as the nonlinear activation function.

As for the entropy network $g_e$ (see Fig.~\ref{fig:frame_entropy}), we model the probability distribution of the quantized code $\bar{\bm c}$ as a MoG, whose accuracy can be improved by considering the code context, also known as the auto-regressive prior. Thus we employ the group context\footnote{When performing pseudocylindrical convolution, the specified code order following \cite{li2020efficient} should be respected, which requires careful treatment of linear interpolation and pseudocylindrical padding.} proposed by Li \etal~\cite{li2020efficient}, and train three context-based DNNs to predict the  mean, variance, and mixing weights, respectively. Each DNN comprises a front-end up-sampling layer and two masked pseudocylindrical convolutions~\cite{li2020efficient}, followed by three masked residual blocks and a back-end masked pseudocylindrical convolution. No activation function is added as the output layer in the mean branch; ReLU activation is added in the variance branch to ensure that the output is nonnegative; the softmax activation is added in the mixing weight branch to produce a probability vector as output.

\subsection{Quantizer}
The quantizer $g_q$ is parameterized by $\bm \omega = \{\omega_{k,0},\ldots,\omega_{k,L_q-1}\}$, where $k=0,\ldots, C-1$ is the channel index, and $L_q$ is the number of quantization centers in each channel. We compute the $l$-th quantization center for the $k$-th channel by
\begin{align}
\Omega_{k,l} = \sum_{l'=0}^{l} \exp(\omega_{k,l'}).
\end{align}
The code $c_{k,i,j}$, where $i,j$ are spatial indices, is quantized to its nearest quantization center:
\begin{align}
\bar{c}_{k,i,j} = g_q(c_{k,i,j}) = \argmin_{\{ \Omega_{k,l}\}}\Vert c_{k,i,j} - \Omega_{k,l} \Vert^2_2.
	\label{eq:mapping}
\end{align}
$g_q$ has zero gradients almost everywhere, which hinders training via back-propagation. Inspired by~\cite{courbariaux2016binarized,theis2017lossy}, we approximate $g_q$ by the identify function $\hat{g}_q(c_{k,i,j}) = c_{k,i,j}$. That is, we use $g_d$ and $\hat{g}_d$ in the forward and backward passes, respectively. The parameters $\bm \omega$ can be optimized by minimizing the quantization error, \ie, $\| \bm c - \bar{\bm c} \|^2_2$, on the fly.

\subsection{Rate-Distortion Objective}
As with general data compression, our objective function is a weighted sum of the rate and distortion: 
\begin{align}\label{eq:rd}
\ell = \mathbb{E}_{\bm x \in \mathcal{D}} \Bigg[\ell_{r}\Big(g_q\big(g_a(\bm x)\big)\Big) + \lambda \ell_{d}\bigg(\bm x, g_s\Big(g_q\big(g_a(\bm x)\big)\Big)\bigg)\Bigg],
\end{align}
where $\lambda$ is the trade-off parameter, and $\mathcal{D}$ is the training set. The rate of the quantized code $\bar{\bm c}$ is computed by
\begin{align}
\ell_{r}(\bar{\bm c}) = -\mathbb{E}_{\bar{\bm c}}\left[ \log p(\bar{\bm c})\right] = -\mathbb{E}\left[\sum_{k,i,j} \log p(\bar{c}_{k,i,j})\right],
\end{align}
where we omit the conditional dependency to make the notation uncluttered. The discretized
probability, $p(\bar{c}_{k,i,j})$, can be computed by integrating the continuous MoG with three components:
\begin{align}
p(\bar{c}_{k,i,j}) = \int_{\frac{\Omega_{k,l-1}+\Omega_{k,l}}{2}}^{\frac{\Omega_{k,l+1}+\Omega_{k,l}}{2}} \sum_{m=0}^{2} \pi_{k,i,j}^m \mathcal{N} \left(\xi;\mu_{k,i,j}^m,\left(\sigma_{k,i,j}^m\right)^2\right)d\xi,
\end{align}
where we assume $\Omega_{k,l}=\bar{c}_{k,i,j}$, and $\pi_{k,i,j}^m$, $\mu_{k,i,j}^m$, and $(\sigma_{k,i,j}^m)^2$ are the mixing weight, mean and variance of $m$-th Gaussian component for $\bar{c}_{k,i,j}$, respectively.

As previously discussed in Sec.~\ref{sec:irregular_learn}, we adopt viewport-based MSE and structure similarity (SSIM) as the quality measures, which are denoted by VMSE and VSSIM. It is noteworthy that VSSIM is a quality metric, and we need to convert it into a loss function:
\begin{equation}
\ell_{\mathrm{VSSIM}}(\bm x, \hat{\bm x}) = 1 - \mathrm{VSSIM}(\bm x,\hat{\bm x}),
\end{equation}
where $\hat{\bm x}$ is the reconstructed image of $\bm x$ by our model.

\section{Experiments}
\begin{figure}[!tbp]
\centering

\includegraphics[width=1\linewidth]{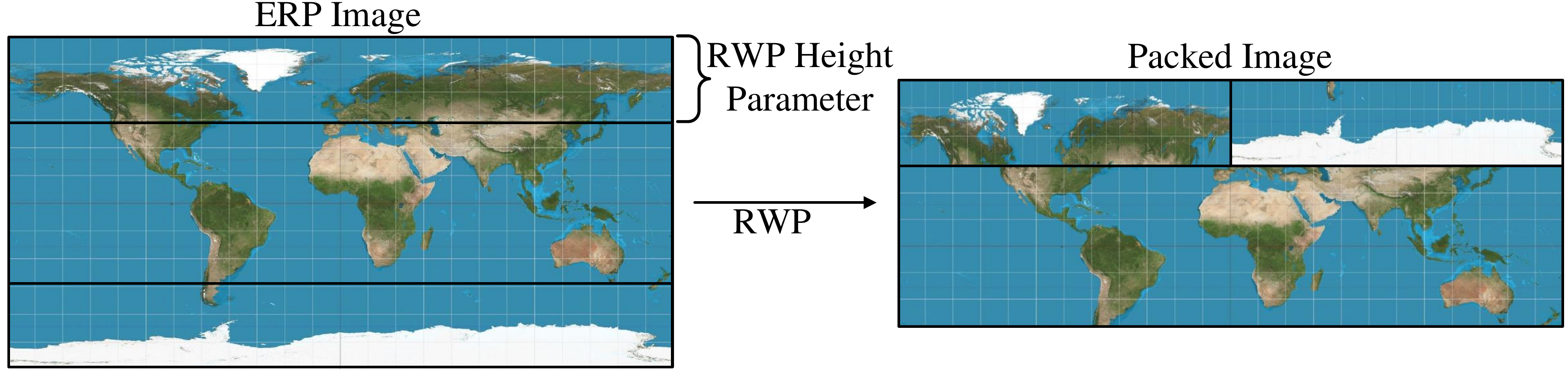}

\caption{Region-wise packing (RWP) for the ERP image.}\label{fig:rwp}
\end{figure}

In this section, we first describe the experimental setup, and then augment existing codecs by RWP with the height parameter optimized for rate-distortion performance (see Fig. \ref{fig:rwp}). We compare our method to the augmented codecs in terms of quantitative metrics and visual quality. Last, we conduct comprehensive ablation studies to single out the contributions of the proposed techniques. The codes and the trained models will be available at: \url{https://github.com/limuhit/pseudocylindrical_convolution}.

\subsection{Experimental Setup}
We collect $19,790$ ERP images from Flickr that carry Creative Commons licenses and save them losslessly. All images are down-sampled to the size of $512 \times 1024$ to further counteract potential compression artifacts. We split the dataset into the training set with $19,590$ images and the test set with $200$ images.  

\begin{figure*}
\begin{minipage}{0.33\linewidth}
\includegraphics[width=1.0\linewidth]{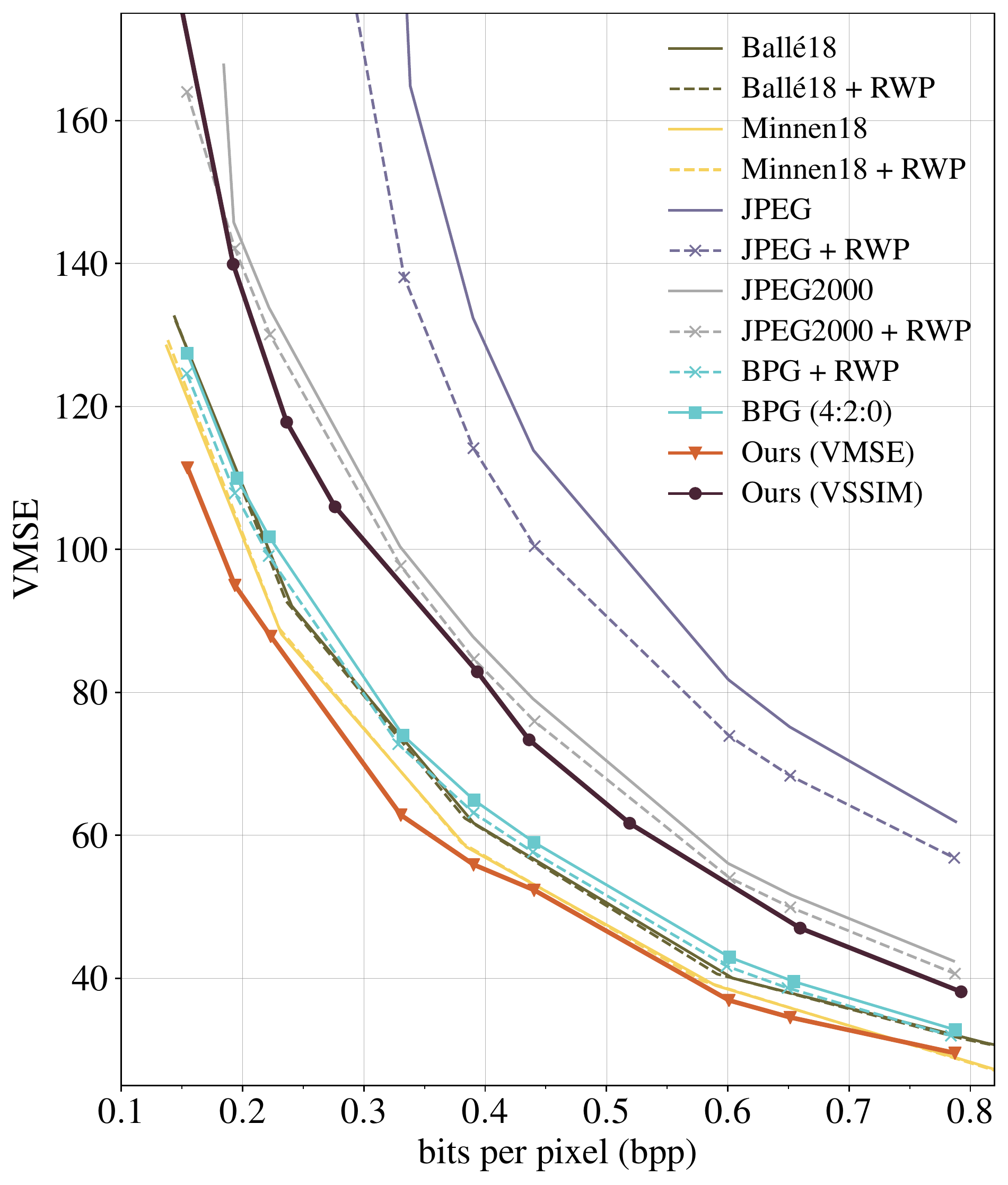}
\centering{\scriptsize{(a)}}
\end{minipage}
\begin{minipage}{0.33\linewidth}
\includegraphics[width=1.0\linewidth]{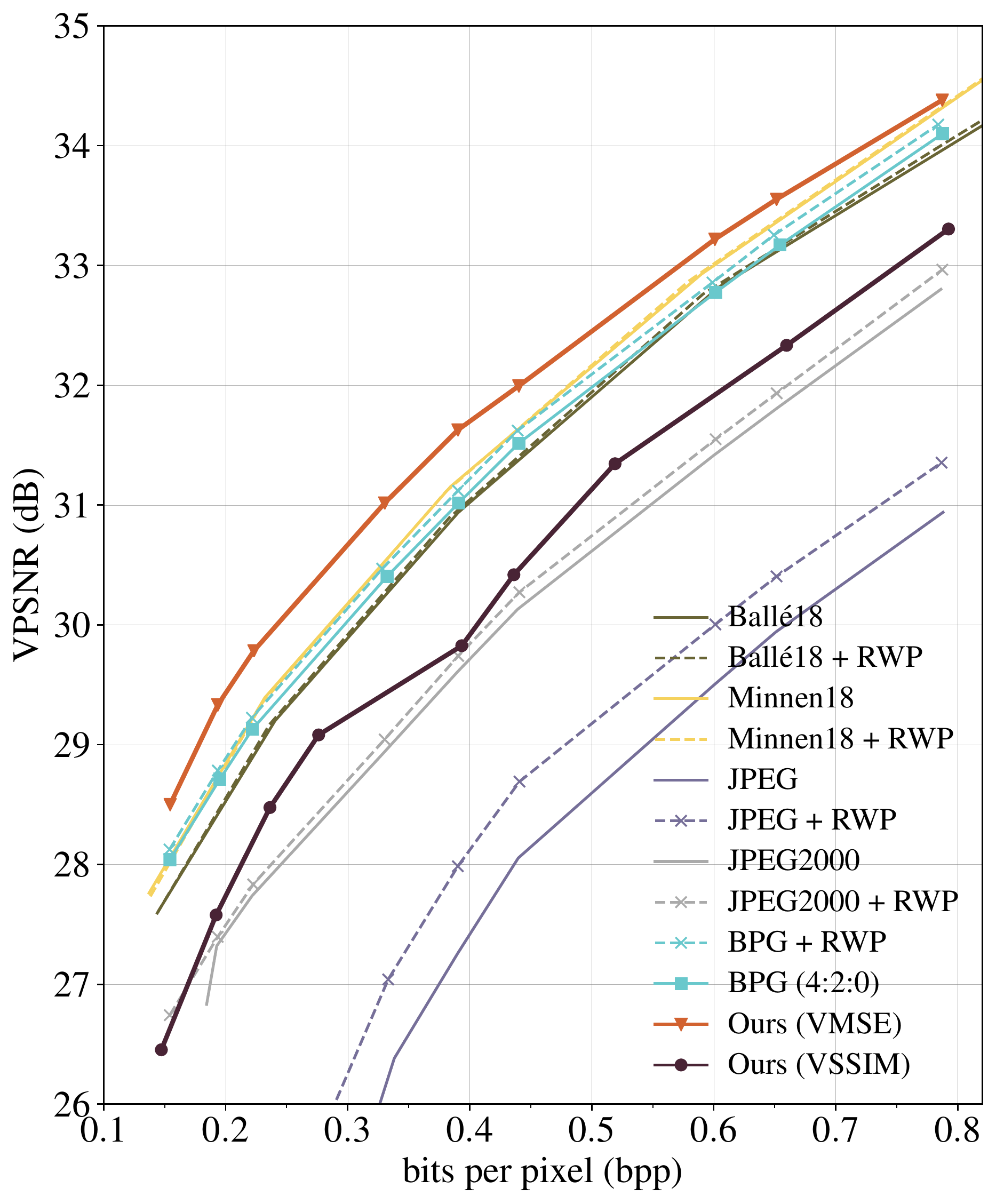}
\centering{\scriptsize{(b)}}
\end{minipage}
\begin{minipage}{0.33\linewidth}
\includegraphics[width=1.0\linewidth]{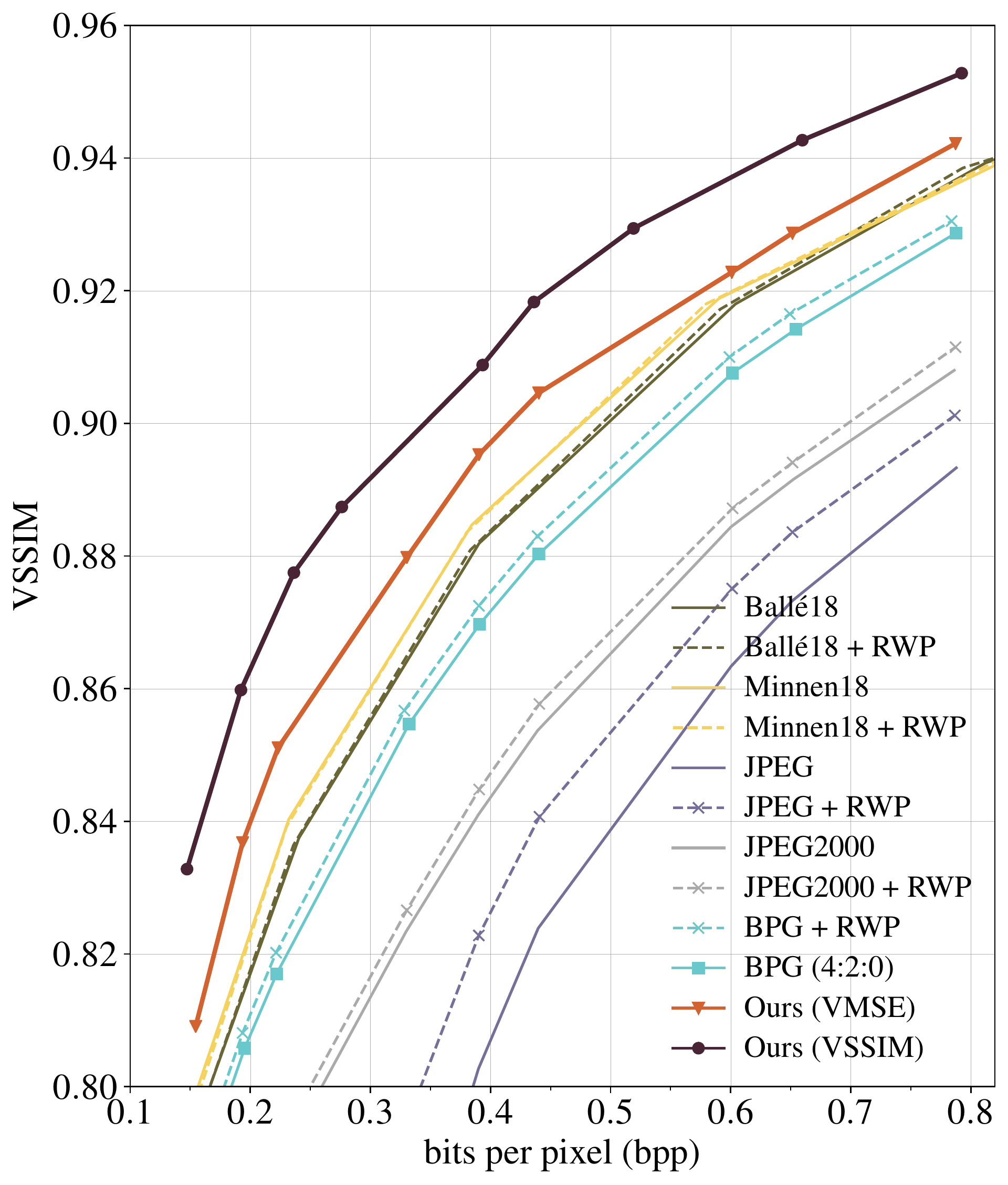}
\centering{\scriptsize{(c)}}
\end{minipage}
\caption{Rate-distortion curves of different compression methods. (a) VMSE. (b) VPSNR. (c) VSSIM.}\label{fig:rd}
\end{figure*}

In main experiments, we set the height of each tile to $H_0=32$ such that it can be down-sampled by $16$ times in the analysis transform. Correspondingly, the number of tiles in our pseudocylindrical representation is $16$. The quantization level $L$ is set to $64$ in Problem \eqref{eq:gop}. The trade-off parameter $\lambda$ in Eq. \eqref{eq:rd} is in the range of $[1/16, 1/3]$.  For the number of channels of the latent code, we choose $C=192, 112, 56$ for high-bitrate, mid-bitrate, and low-bitrate models, respectively. The number of quantization levels adopted by $g_q$ is $L_q=8$.

Stochastic optimization is carried out by minimizing Eq.~\eqref{eq:rd} using the Adam method~\cite{kingma2014adam}  with a learning rate of $10^{-4}$. We decay the learning rate by a factor of $10$ whenever the training plateaus. 
We leverage the favorable transferability  of the proposed method, and pre-train it  using a large set of  central-perspective images with standard convolution (\ie, no pseudocylindrical padding). We then fine-tune the entire method with pseudocylindrical convolutions on the full-size \mdegr images in the training set.

Compression methods are commonly evaluated from two aspects - rate and distortion. In this paper, we adopt the bits per pixel (bpp), calculated as the total number of bits used to code the image divided by the number of pixels, for the rate, and  VMSE, viewport-based PSNR (VPSNR) and VSSIM for the distortion. Then, we are able to draw the rate-distortion (RD) curve, and compute BD-BR and Bjontegaard delta distortion (BD-distortion)~\cite{bjontegaard2001calculation}.  BD-BR in the unit of percentage (\%) calculates the average bitrate saving between two rate-distortion curves, while BD-distortion calculates the average distortion improvement between two curves. A negative BD-BR value and a positive BD-distortion value represent that the test method is better than the anchor method.

\subsection{Optimization of RWP for Existing Codecs}
To the best of our knowledge, there are no learned compression methods specifically designed  for \mdegr images. Thus, we choose to augment five codecs for central-perspective images with the RWP strategy~\cite{boyce2017hevc} - JPEG~\cite{wallace1992jpeg}, JPEG2000~\cite{skodras2001jpeg}, BPG~\cite{bellard2016bpg} (\ie, HEVC intra coding), Minnen18~\cite{minnen2018joint}, and Ball{\'e}18~\cite{balle2018variational}. As shown in Fig.~\ref{fig:rwp}, RWP partitions the images into three parts with a parameter to control  the height of the north (and south) pole region. The assembled image is used as the input for compression. RWP can relieve the non-uniform sampling problem to a certain extent. However, the simple split-and-merge operation will break the image continuity and context, which may have a negative impact on compression performance.

We optimize over a set of the height parameter of RWP, $\{8,16,24,32,40,48,56,64,72,80,100\}$, for each of the three compression standards (JPEG, JPEG2000, and BPG) in terms of BD-VPSNR and BD-BR metrics using the original method without RWP as anchor. 
We find that RWP indeed improves the compression performance for all three codecs.  For example, RWP helps improve $0.59$ dB and saves $10.2\%$ bits for JPEG. It turns out the best heights for JPEG, JPEG2000, and BPG are $64$, $40$, and $48$, respectively. As for DNN-based methods, we choose the height to be $48$, which offers satisfactory performance for all three compression standards.

\subsection{Quantitative Evaluation}

We compare our methods optimized for VMSE and VSSIM, with the five chosen codes and their augmented versions with optimized RWP. Both the JPEG and JPEG2000 are based on the internal implementations in OpenCV 4.2. As for BPG,  we adopt the official codec from \url{https://bellard.org/bpg/} with the 4:2:0 chroma format. For the two DNN-based methods, we test on codes released by the respective authors. 

\begin{table}[t]
 \scriptsize
   \centering
   \caption{Performance comparison of different compression methods in terms of BD-VPSNR, BD-VSSIM, and BD-BR}\label{tab:NS}
   \begin{tabular}{L{2.0cm} R{1cm} R{1cm} R{1cm} R{1cm}}
     \toprule
      \multirow{3}{*}{Method} & \multicolumn{2}{c}{\centering{RATE-VPSNR}} & \multicolumn{2}{c}{\centering{RATE-VSSIM}}\\
      \cmidrule{2-3} \cmidrule{4-5}
       & \multicolumn{1}{c}{\specialcell{BD-VPSNR \\ (dB)}} & \multicolumn{1}{c}{BD-BR (\%)} & \multicolumn{1}{c}{BD-VSSIM} & \multicolumn{1}{c}{BD-BR (\%)}\\
     \midrule
JPEG & $-3.504$ & $140.08$ & $-0.057$ & $79.92$\\
JPEG + RWP & $-3.080$ & $112.04$ & $-0.048$ & $59.34$\\
JPEG2000 & $-1.385$ & $43.86$ & $-0.028$ & $33.67$\\
JPEG2000 + RWP & $-1.300$ & $40.85$ & $-0.025$ & $31.34$\\
BPG & $0.000$ & $0.00$ & $0.000$ & $0.00$\\
BPG + RWP & $0.092$ & $-2.44$ & $0.003$ & $-3.04$\\
Ball{\'e}18 & $-0.110$ & $2.90$ & $0.012$ & $-12.41$\\
Ball{\'e}18 + RWP & $-0.077$ & $2.02$ & $0.012$ & $-13.02$\\
Minnen18 & $\mathbf{0.187}$ & $\mathbf{-4.73}$ & $0.016$ & $-16.99$\\
Minnen18 + RWP & $0.183$ & $-4.62$ & $0.016$ & $-16.68$\\
Ours (VMSE) & $\mathbf{0.547}$ & $\mathbf{-13.97}$ & $\mathbf{0.025}$ & $\mathbf{-25.86}$\\
Ours (VSSIM) & $-0.958$ & $27.99$ & $\mathbf{0.043}$ & $\mathbf{-41.84}$\\
     \bottomrule
   \end{tabular}
\end{table}

\begin{figure*}
\begin{minipage}{1.\linewidth}
\begin{minipage}{0.08\linewidth}\centering{\scriptsize{($\theta$,$\phi$)}}\vspace{0.1cm}\end{minipage}
\begin{minipage}{0.125\linewidth}\centering{\scriptsize{($0$,$-\frac{\pi}{2}$)}}\vspace{0.1cm}\end{minipage}
\begin{minipage}{0.125\linewidth}\centering{\scriptsize{($0$,$0$)}}\vspace{0.1cm}\end{minipage}
\begin{minipage}{0.125\linewidth}\centering{\scriptsize{($0$,$\frac{\pi}{2}$)}}\vspace{0.1cm}\end{minipage}
\begin{minipage}{0.125\linewidth}\centering{\scriptsize{($0$,$\pi$)}}\vspace{0.1cm}\end{minipage}
\begin{minipage}{0.125\linewidth}\centering{\scriptsize{($\frac{\pi}{4}$,$-\frac{\pi}{2}$)}}\vspace{0.1cm}\end{minipage}
\begin{minipage}{0.125\linewidth}\centering{\scriptsize{($\frac{\pi}{4}$,$0$)}}\vspace{0.1cm}\end{minipage}
\begin{minipage}{0.125\linewidth}\centering{\scriptsize{($\frac{\pi}{4}$,$\frac{\pi}{2}$)}}\vspace{0.1cm}\end{minipage}
\end{minipage}
\begin{minipage}{1.\linewidth}
\begin{minipage}{0.08\linewidth}
\centering{\scriptsize{JPEG-RWP}}
\end{minipage}
\begin{minipage}{0.125\linewidth}
\includegraphics[width=1.0\linewidth]{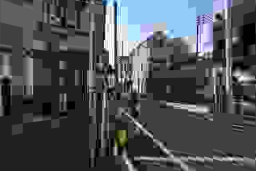}

\vspace{-0.1cm}

\centering{\scriptsize{20.80 / 0.590}}
\vspace{0.1cm}
\end{minipage}
\begin{minipage}{0.125\linewidth}
\includegraphics[width=1.0\linewidth]{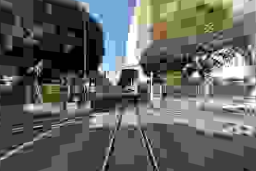}

\vspace{-0.1cm}

\centering{\scriptsize{19.69 / 0.521}}
\vspace{0.1cm}
\end{minipage}
\begin{minipage}{0.125\linewidth}
\includegraphics[width=1.0\linewidth]{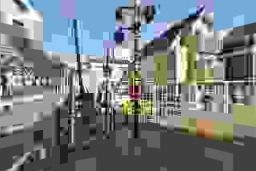}

\vspace{-0.1cm}

\centering{\scriptsize{18.88 / 0.542}}
\vspace{0.1cm}
\end{minipage}
\begin{minipage}{0.125\linewidth}
\includegraphics[width=1.0\linewidth]{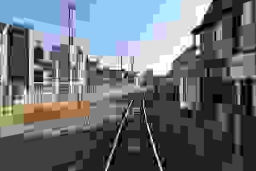}

\vspace{-0.1cm}

\centering{\scriptsize{20.70 / 0.556}}
\vspace{0.1cm}
\end{minipage}
\begin{minipage}{0.125\linewidth}
\includegraphics[width=1.0\linewidth]{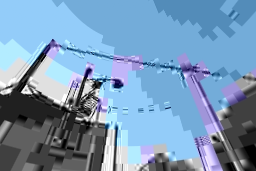}

\vspace{-0.1cm}

\centering{\scriptsize{20.84 / 0.686}}
\vspace{0.1cm}
\end{minipage}
\begin{minipage}{0.125\linewidth}
\includegraphics[width=1.0\linewidth]{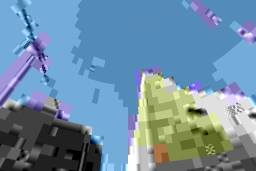}

\vspace{-0.1cm}

\centering{\scriptsize{20.37 / 0.617}}
\vspace{0.1cm}
\end{minipage}
\begin{minipage}{0.125\linewidth}
\includegraphics[width=1.0\linewidth]{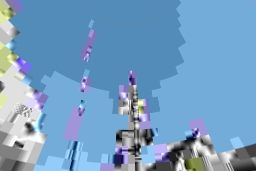}

\vspace{-0.1cm}

\centering{\scriptsize{20.99 / 0.733}}
\vspace{0.1cm}
\end{minipage}
\end{minipage}

\begin{minipage}{1.\linewidth}
\begin{minipage}{0.08\linewidth}
\centering{\scriptsize{JPEG2000-RWP}}
\end{minipage}
\begin{minipage}{0.125\linewidth}
\includegraphics[width=1.0\linewidth]{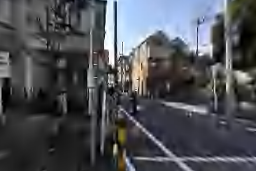}

\vspace{-0.1cm}

\centering{\scriptsize{23.21 / 0.677}}
\vspace{0.1cm}
\end{minipage}
\begin{minipage}{0.125\linewidth}
\includegraphics[width=1.0\linewidth]{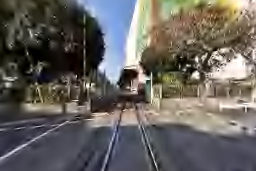}

\vspace{-0.1cm}

\centering{\scriptsize{22.74 / 0.677}}
\vspace{0.1cm}
\end{minipage}
\begin{minipage}{0.125\linewidth}
\includegraphics[width=1.0\linewidth]{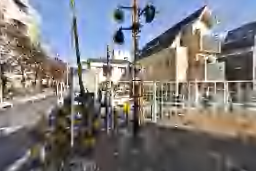}

\vspace{-0.1cm}

\centering{\scriptsize{22.19 / 0.689}}
\vspace{0.1cm}
\end{minipage}
\begin{minipage}{0.125\linewidth}
\includegraphics[width=1.0\linewidth]{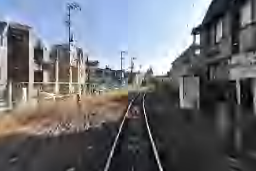}

\vspace{-0.1cm}

\centering{\scriptsize{23.75 / 0.674}}
\vspace{0.1cm}
\end{minipage}
\begin{minipage}{0.125\linewidth}
\includegraphics[width=1.0\linewidth]{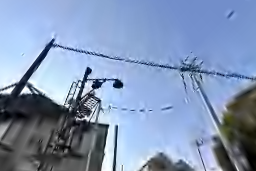}

\vspace{-0.1cm}

\centering{\scriptsize{25.16 / 0.799}}
\vspace{0.1cm}
\end{minipage}
\begin{minipage}{0.125\linewidth}
\includegraphics[width=1.0\linewidth]{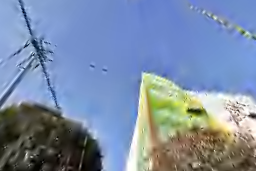}

\vspace{-0.1cm}

\centering{\scriptsize{24.13 / 0.743}}
\vspace{0.1cm}
\end{minipage}
\begin{minipage}{0.125\linewidth}
\includegraphics[width=1.0\linewidth]{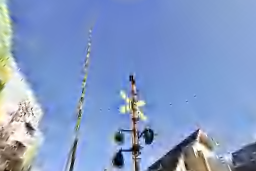}

\vspace{-0.1cm}

\centering{\scriptsize{25.84 / 0.833}}
\vspace{0.1cm}
\end{minipage}
\end{minipage}

\begin{minipage}{1.\linewidth}
\begin{minipage}{0.08\linewidth}
\centering{\scriptsize{BPG-RWP}}
\end{minipage}
\begin{minipage}{0.125\linewidth}
\includegraphics[width=1.0\linewidth]{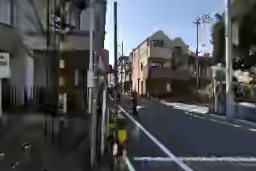}

\vspace{-0.1cm}

\centering{\scriptsize{25.90 / 0.790}}
\vspace{0.1cm}
\end{minipage}
\begin{minipage}{0.125\linewidth}
\includegraphics[width=1.0\linewidth]{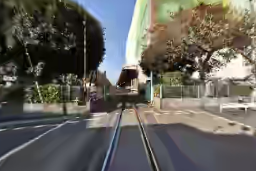}

\vspace{-0.1cm}

\centering{\scriptsize{25.21 / 0.776}}
\vspace{0.1cm}
\end{minipage}
\begin{minipage}{0.125\linewidth}
\includegraphics[width=1.0\linewidth]{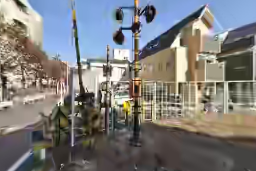}

\vspace{-0.1cm}

\centering{\scriptsize{24.19 / 0.755}}
\vspace{0.1cm}
\end{minipage}
\begin{minipage}{0.125\linewidth}
\includegraphics[width=1.0\linewidth]{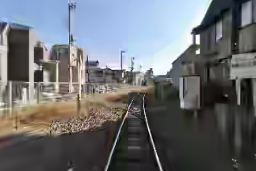}

\vspace{-0.1cm}

\centering{\scriptsize{25.66 / 0.735}}
\vspace{0.1cm}
\end{minipage}
\begin{minipage}{0.125\linewidth}
\includegraphics[width=1.0\linewidth]{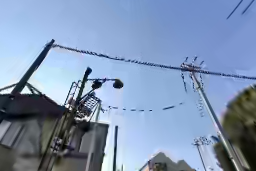}

\vspace{-0.1cm}

\centering{\scriptsize{27.32 / 0.859}}
\vspace{0.1cm}
\end{minipage}
\begin{minipage}{0.125\linewidth}
\includegraphics[width=1.0\linewidth]{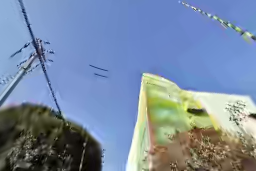}

\vspace{-0.1cm}

\centering{\scriptsize{26.15 / 0.813}}
\vspace{0.1cm}
\end{minipage}
\begin{minipage}{0.125\linewidth}
\includegraphics[width=1.0\linewidth]{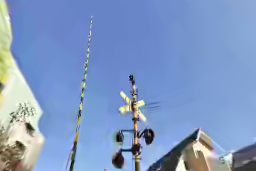}

\vspace{-0.1cm}

\centering{\scriptsize{27.94 / 0.865}}
\vspace{0.1cm}
\end{minipage}
\end{minipage}

\begin{minipage}{1.\linewidth}
\begin{minipage}{0.08\linewidth}
\centering{\scriptsize{Minnen18}}
\end{minipage}
\begin{minipage}{0.125\linewidth}
\includegraphics[width=1.0\linewidth]{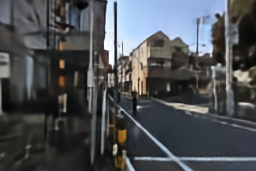}

\vspace{-0.1cm}

\centering{\scriptsize{25.77 / 0.793}}
\vspace{0.1cm}
\end{minipage}
\begin{minipage}{0.125\linewidth}
\includegraphics[width=1.0\linewidth]{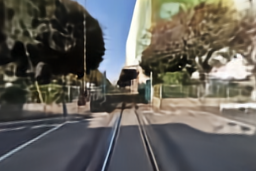}

\vspace{-0.1cm}

\centering{\scriptsize{24.31 / 0.753}}
\vspace{0.1cm}
\end{minipage}
\begin{minipage}{0.125\linewidth}
\includegraphics[width=1.0\linewidth]{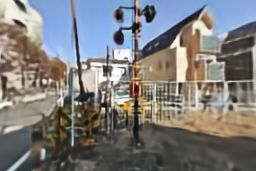}

\vspace{-0.1cm}

\centering{\scriptsize{23.22 / 0.741}}
\vspace{0.1cm}
\end{minipage}
\begin{minipage}{0.125\linewidth}
\includegraphics[width=1.0\linewidth]{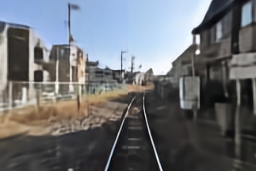}

\vspace{-0.1cm}

\centering{\scriptsize{25.23 / 0.738}}
\vspace{0.1cm}
\end{minipage}
\begin{minipage}{0.125\linewidth}
\includegraphics[width=1.0\linewidth]{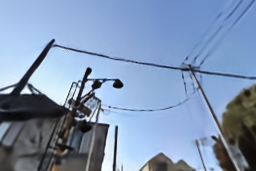}

\vspace{-0.1cm}

\centering{\scriptsize{26.75 / 0.862}}
\vspace{0.1cm}
\end{minipage}
\begin{minipage}{0.125\linewidth}
\includegraphics[width=1.0\linewidth]{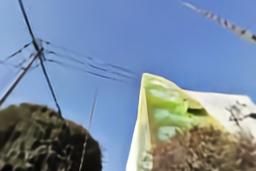}

\vspace{-0.1cm}

\centering{\scriptsize{25.79 / 0.807}}
\vspace{0.1cm}
\end{minipage}
\begin{minipage}{0.125\linewidth}
\includegraphics[width=1.0\linewidth]{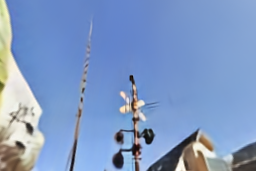}

\vspace{-0.1cm}

\centering{\scriptsize{27.30 / 0.858}}
\vspace{0.1cm}
\end{minipage}
\end{minipage}

\begin{minipage}{1.\linewidth}
\begin{minipage}{0.08\linewidth}
\centering{\scriptsize{Ours (VSSIM)}}
\end{minipage}
\begin{minipage}{0.125\linewidth}
\includegraphics[width=1.0\linewidth]{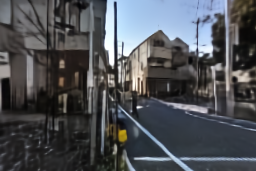}

\vspace{-0.1cm}

\centering{\scriptsize{24.19 / 0.814}}
\vspace{0.1cm}
\end{minipage}
\begin{minipage}{0.125\linewidth}
\includegraphics[width=1.0\linewidth]{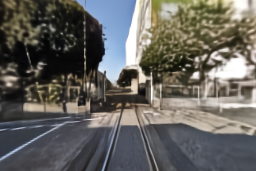}

\vspace{-0.1cm}

\centering{\scriptsize{22.42 / 0.761}}
\vspace{0.1cm}
\end{minipage}
\begin{minipage}{0.125\linewidth}
\includegraphics[width=1.0\linewidth]{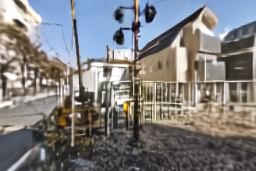}

\vspace{-0.1cm}

\centering{\scriptsize{21.42 / 0.755}}
\vspace{0.1cm}
\end{minipage}
\begin{minipage}{0.125\linewidth}
\includegraphics[width=1.0\linewidth]{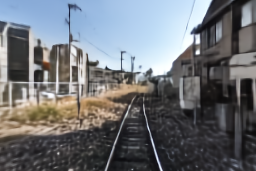}

\vspace{-0.1cm}

\centering{\scriptsize{23.62 / 0.781}}
\vspace{0.1cm}
\end{minipage}
\begin{minipage}{0.125\linewidth}
\includegraphics[width=1.0\linewidth]{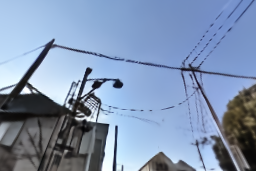}

\vspace{-0.1cm}

\centering{\scriptsize{25.08 / 0.884}}
\vspace{0.1cm}
\end{minipage}
\begin{minipage}{0.125\linewidth}
\includegraphics[width=1.0\linewidth]{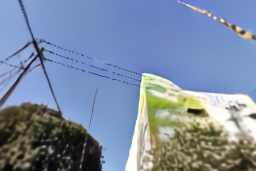}

\vspace{-0.1cm}

\centering{\scriptsize{23.94 / 0.831}}
\vspace{0.1cm}
\end{minipage}
\begin{minipage}{0.125\linewidth}
\includegraphics[width=1.0\linewidth]{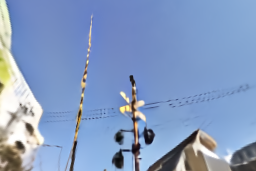}

\vspace{-0.1cm}

\centering{\scriptsize{25.50 / 0.880}}
\vspace{0.1cm}
\end{minipage}
\end{minipage}

\begin{minipage}{1.\linewidth}
\begin{minipage}{0.08\linewidth}
\centering{\scriptsize{Original}}
\end{minipage}
\begin{minipage}{0.125\linewidth}
\includegraphics[width=1.0\linewidth]{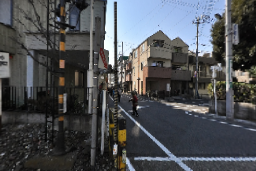}
 \end{minipage}
\begin{minipage}{0.125\linewidth}
\includegraphics[width=1.0\linewidth]{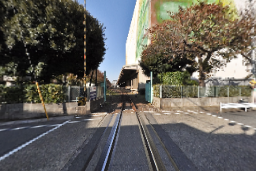}
 \end{minipage}
\begin{minipage}{0.125\linewidth}
\includegraphics[width=1.0\linewidth]{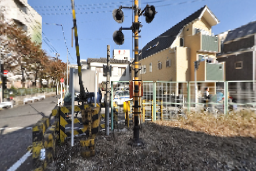}
 \end{minipage}
\begin{minipage}{0.125\linewidth}
\includegraphics[width=1.0\linewidth]{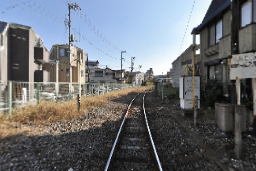}
 \end{minipage}
\begin{minipage}{0.125\linewidth}
\includegraphics[width=1.0\linewidth]{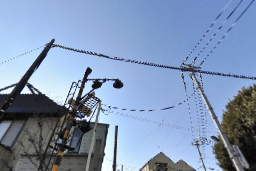}
 \end{minipage}
\begin{minipage}{0.125\linewidth}
\includegraphics[width=1.0\linewidth]{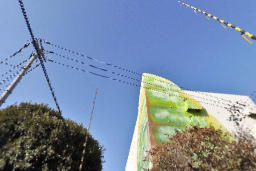}
 \end{minipage}
\begin{minipage}{0.125\linewidth}
\includegraphics[width=1.0\linewidth]{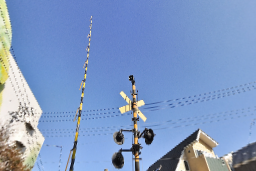}
 \end{minipage}

\hfill\vspace{0.1cm}\hfill

\end{minipage}

\vspace{0.15cm}

\begin{minipage}{1.\linewidth}
\begin{minipage}{0.08\linewidth}\centering{\scriptsize{($\theta$,$\phi$)}}\vspace{0.1cm}\end{minipage}
\begin{minipage}{0.125\linewidth}\centering{\scriptsize{($\frac{\pi}{4}$,$\pi$)}}\vspace{0.1cm}\end{minipage}
\begin{minipage}{0.125\linewidth}\centering{\scriptsize{($-\frac{\pi}{4}$,$-\frac{\pi}{2}$)}}\vspace{0.1cm}\end{minipage}
\begin{minipage}{0.125\linewidth}\centering{\scriptsize{($-\frac{\pi}{4}$,$0$)}}\vspace{0.1cm}\end{minipage}
\begin{minipage}{0.125\linewidth}\centering{\scriptsize{($-\frac{\pi}{4}$,$\frac{\pi}{2}$)}}\vspace{0.1cm}\end{minipage}
\begin{minipage}{0.125\linewidth}\centering{\scriptsize{($-\frac{\pi}{4}$,$\pi$)}}\vspace{0.1cm}\end{minipage}
\begin{minipage}{0.125\linewidth}\centering{\scriptsize{($\frac{\pi}{2}$,$0$)}}\vspace{0.1cm}\end{minipage}
\begin{minipage}{0.125\linewidth}\centering{\scriptsize{($-\frac{\pi}{2}$,$0$)}}\vspace{0.1cm}\end{minipage}
\end{minipage}
\begin{minipage}{1.\linewidth}
\begin{minipage}{0.08\linewidth}
\centering{\scriptsize{JPEG-RWP}}
\end{minipage}
\begin{minipage}{0.125\linewidth}
\includegraphics[width=1.0\linewidth]{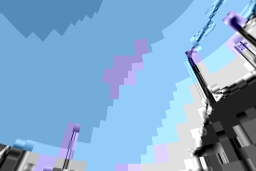}

\vspace{-0.1cm}

\centering{\scriptsize{22.33 / 0.794}}
\vspace{0.1cm}
\end{minipage}
\begin{minipage}{0.125\linewidth}
\includegraphics[width=1.0\linewidth]{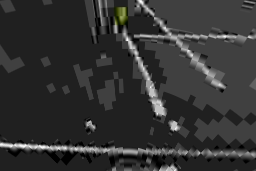}

\vspace{-0.1cm}

\centering{\scriptsize{22.47 / 0.530}}
\vspace{0.1cm}
\end{minipage}
\begin{minipage}{0.125\linewidth}
\includegraphics[width=1.0\linewidth]{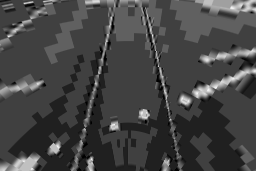}

\vspace{-0.1cm}

\centering{\scriptsize{22.33 / 0.554}}
\vspace{0.1cm}
\end{minipage}
\begin{minipage}{0.125\linewidth}
\includegraphics[width=1.0\linewidth]{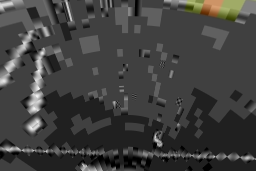}

\vspace{-0.1cm}

\centering{\scriptsize{21.12 / 0.407}}
\vspace{0.1cm}
\end{minipage}
\begin{minipage}{0.125\linewidth}
\includegraphics[width=1.0\linewidth]{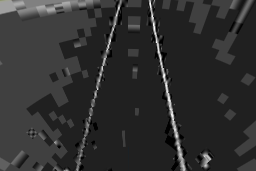}

\vspace{-0.1cm}

\centering{\scriptsize{22.22 / 0.394}}
\vspace{0.1cm}
\end{minipage}
\begin{minipage}{0.125\linewidth}
\includegraphics[width=1.0\linewidth]{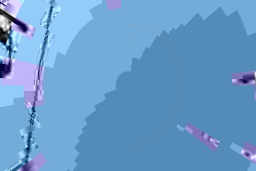}

\vspace{-0.1cm}

\centering{\scriptsize{22.23 / 0.848}}
\vspace{0.1cm}
\end{minipage}
\begin{minipage}{0.125\linewidth}
\includegraphics[width=1.0\linewidth]{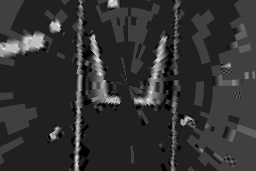}

\vspace{-0.1cm}

\centering{\scriptsize{21.69 / 0.470}}
\vspace{0.1cm}
\end{minipage}
\end{minipage}

\begin{minipage}{1.\linewidth}
\begin{minipage}{0.08\linewidth}
\centering{\scriptsize{JPEG2000-RWP}}
\end{minipage}
\begin{minipage}{0.125\linewidth}
\includegraphics[width=1.0\linewidth]{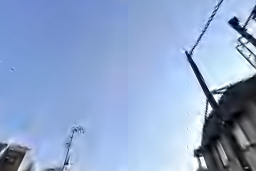}

\vspace{-0.1cm}

\centering{\scriptsize{28.69 / 0.902}}
\vspace{0.1cm}
\end{minipage}
\begin{minipage}{0.125\linewidth}
\includegraphics[width=1.0\linewidth]{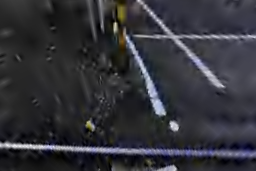}

\vspace{-0.1cm}

\centering{\scriptsize{25.00 / 0.610}}
\vspace{0.1cm}
\end{minipage}
\begin{minipage}{0.125\linewidth}
\includegraphics[width=1.0\linewidth]{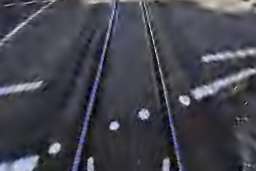}

\vspace{-0.1cm}

\centering{\scriptsize{25.52 / 0.657}}
\vspace{0.1cm}
\end{minipage}
\begin{minipage}{0.125\linewidth}
\includegraphics[width=1.0\linewidth]{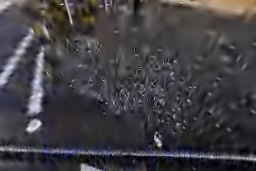}

\vspace{-0.1cm}

\centering{\scriptsize{23.11 / 0.529}}
\vspace{0.1cm}
\end{minipage}
\begin{minipage}{0.125\linewidth}
\includegraphics[width=1.0\linewidth]{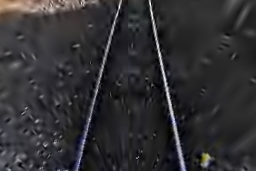}

\vspace{-0.1cm}

\centering{\scriptsize{23.94 / 0.482}}
\vspace{0.1cm}
\end{minipage}
\begin{minipage}{0.125\linewidth}
\includegraphics[width=1.0\linewidth]{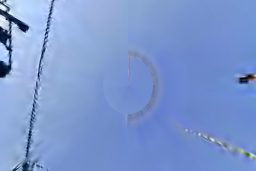}

\vspace{-0.1cm}

\centering{\scriptsize{29.75 / 0.917}}
\vspace{0.1cm}
\end{minipage}
\begin{minipage}{0.125\linewidth}
\includegraphics[width=1.0\linewidth]{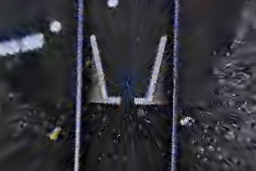}

\vspace{-0.1cm}

\centering{\scriptsize{24.31 / 0.596}}
\vspace{0.1cm}
\end{minipage}
\end{minipage}

\begin{minipage}{1.\linewidth}
\begin{minipage}{0.08\linewidth}
\centering{\scriptsize{BPG-RWP}}
\end{minipage}
\begin{minipage}{0.125\linewidth}
\includegraphics[width=1.0\linewidth]{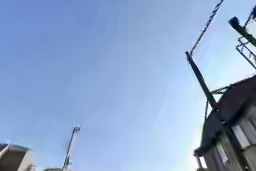}

\vspace{-0.1cm}

\centering{\scriptsize{30.63 / 0.916}}
\vspace{0.1cm}
\end{minipage}
\begin{minipage}{0.125\linewidth}
\includegraphics[width=1.0\linewidth]{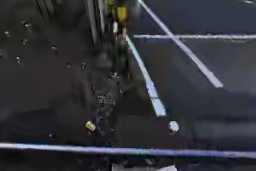}

\vspace{-0.1cm}

\centering{\scriptsize{26.63 / 0.658}}
\vspace{0.1cm}
\end{minipage}
\begin{minipage}{0.125\linewidth}
\includegraphics[width=1.0\linewidth]{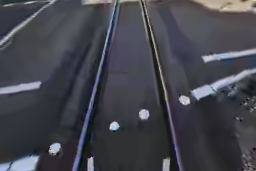}

\vspace{-0.1cm}

\centering{\scriptsize{27.21 / 0.695}}
\vspace{0.1cm}
\end{minipage}
\begin{minipage}{0.125\linewidth}
\includegraphics[width=1.0\linewidth]{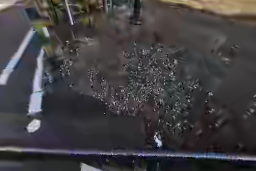}

\vspace{-0.1cm}

\centering{\scriptsize{24.73 / 0.611}}
\vspace{0.1cm}
\end{minipage}
\begin{minipage}{0.125\linewidth}
\includegraphics[width=1.0\linewidth]{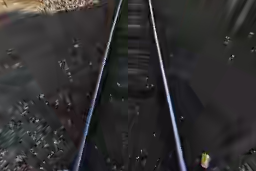}

\vspace{-0.1cm}

\centering{\scriptsize{25.19 / 0.523}}
\vspace{0.1cm}
\end{minipage}
\begin{minipage}{0.125\linewidth}
\includegraphics[width=1.0\linewidth]{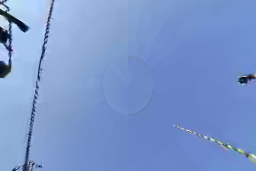}

\vspace{-0.1cm}

\centering{\scriptsize{32.34 / 0.946}}
\vspace{0.1cm}
\end{minipage}
\begin{minipage}{0.125\linewidth}
\includegraphics[width=1.0\linewidth]{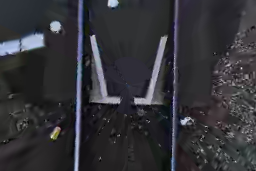}

\vspace{-0.1cm}

\centering{\scriptsize{25.94 / 0.658}}
\vspace{0.1cm}
\end{minipage}
\end{minipage}

\begin{minipage}{1.\linewidth}
\begin{minipage}{0.08\linewidth}
\centering{\scriptsize{Minnen18}}
\end{minipage}
\begin{minipage}{0.125\linewidth}
\includegraphics[width=1.0\linewidth]{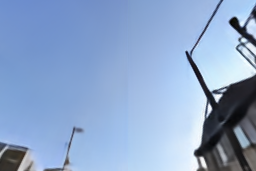}

\vspace{-0.1cm}

\centering{\scriptsize{29.94 / 0.923}}
\vspace{0.1cm}
\end{minipage}
\begin{minipage}{0.125\linewidth}
\includegraphics[width=1.0\linewidth]{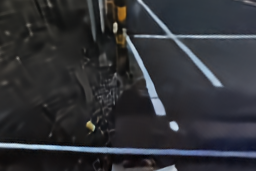}

\vspace{-0.1cm}

\centering{\scriptsize{27.36 / 0.688}}
\vspace{0.1cm}
\end{minipage}
\begin{minipage}{0.125\linewidth}
\includegraphics[width=1.0\linewidth]{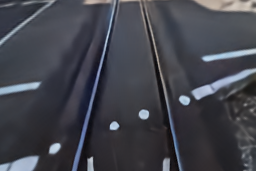}

\vspace{-0.1cm}

\centering{\scriptsize{28.01 / 0.712}}
\vspace{0.1cm}
\end{minipage}
\begin{minipage}{0.125\linewidth}
\includegraphics[width=1.0\linewidth]{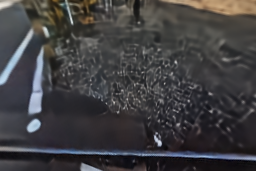}

\vspace{-0.1cm}

\centering{\scriptsize{25.27 / 0.651}}
\vspace{0.1cm}
\end{minipage}
\begin{minipage}{0.125\linewidth}
\includegraphics[width=1.0\linewidth]{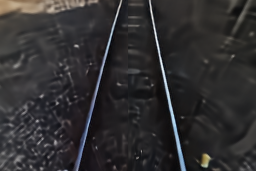}

\vspace{-0.1cm}

\centering{\scriptsize{25.62 / 0.572}}
\vspace{0.1cm}
\end{minipage}
\begin{minipage}{0.125\linewidth}
\includegraphics[width=1.0\linewidth]{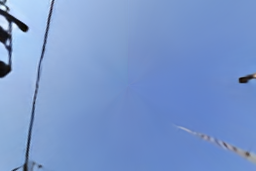}

\vspace{-0.1cm}

\centering{\scriptsize{30.87 / 0.944}}
\vspace{0.1cm}
\end{minipage}
\begin{minipage}{0.125\linewidth}
\includegraphics[width=1.0\linewidth]{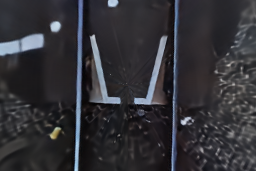}

\vspace{-0.1cm}

\centering{\scriptsize{26.56 / 0.693}}
\vspace{0.1cm}
\end{minipage}
\end{minipage}

\begin{minipage}{1.\linewidth}
\begin{minipage}{0.08\linewidth}
\centering{\scriptsize{Ours (VSSIM)}}
\end{minipage}
\begin{minipage}{0.125\linewidth}
\includegraphics[width=1.0\linewidth]{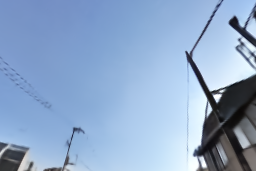}

\vspace{-0.1cm}

\centering{\scriptsize{27.73 / 0.932}}
\vspace{0.1cm}
\end{minipage}
\begin{minipage}{0.125\linewidth}
\includegraphics[width=1.0\linewidth]{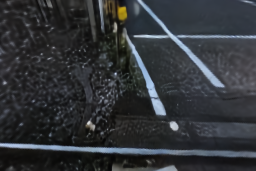}

\vspace{-0.1cm}

\centering{\scriptsize{26.71 / 0.780}}
\vspace{0.1cm}
\end{minipage}
\begin{minipage}{0.125\linewidth}
\includegraphics[width=1.0\linewidth]{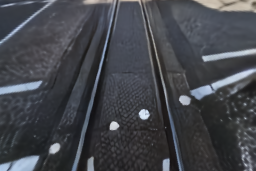}

\vspace{-0.1cm}

\centering{\scriptsize{27.39 / 0.797}}
\vspace{0.1cm}
\end{minipage}
\begin{minipage}{0.125\linewidth}
\includegraphics[width=1.0\linewidth]{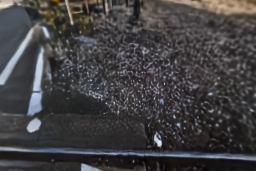}

\vspace{-0.1cm}

\centering{\scriptsize{24.38 / 0.725}}
\vspace{0.1cm}
\end{minipage}
\begin{minipage}{0.125\linewidth}
\includegraphics[width=1.0\linewidth]{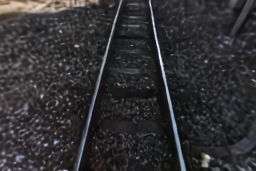}

\vspace{-0.1cm}

\centering{\scriptsize{25.12 / 0.704}}
\vspace{0.1cm}
\end{minipage}
\begin{minipage}{0.125\linewidth}
\includegraphics[width=1.0\linewidth]{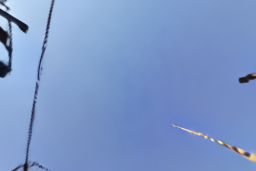}

\vspace{-0.1cm}

\centering{\scriptsize{28.99 / 0.950}}
\vspace{0.1cm}
\end{minipage}
\begin{minipage}{0.125\linewidth}
\includegraphics[width=1.0\linewidth]{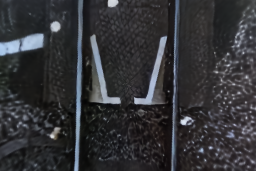}

\vspace{-0.1cm}

\centering{\scriptsize{25.77 / 0.773}}
\vspace{0.1cm}
\end{minipage}
\end{minipage}

\begin{minipage}{1.\linewidth}
\begin{minipage}{0.08\linewidth}
\centering{\scriptsize{Original}}
\end{minipage}
\begin{minipage}{0.125\linewidth}
\includegraphics[width=1.0\linewidth]{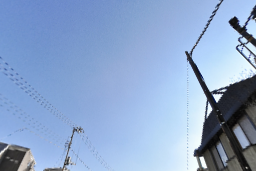}
 \end{minipage}
\begin{minipage}{0.125\linewidth}
\includegraphics[width=1.0\linewidth]{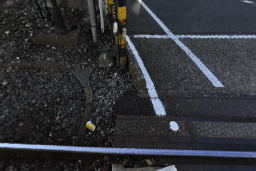}
 \end{minipage}
\begin{minipage}{0.125\linewidth}
\includegraphics[width=1.0\linewidth]{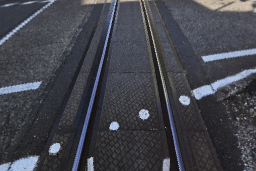}
 \end{minipage}
\begin{minipage}{0.125\linewidth}
\includegraphics[width=1.0\linewidth]{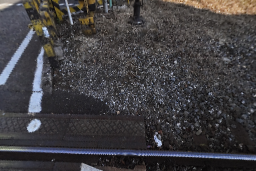}
 \end{minipage}
\begin{minipage}{0.125\linewidth}
\includegraphics[width=1.0\linewidth]{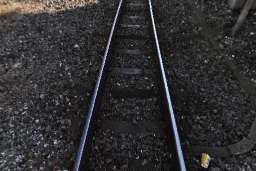}
 \end{minipage}
\begin{minipage}{0.125\linewidth}
\includegraphics[width=1.0\linewidth]{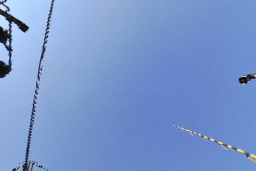}
 \end{minipage}
\begin{minipage}{0.125\linewidth}
\includegraphics[width=1.0\linewidth]{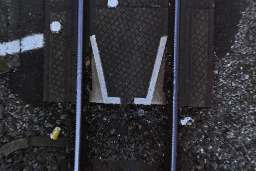}
 \end{minipage}

\hfill\vspace{0.05cm}\hfill

\end{minipage}

\caption{Visual quality comparison of JPEG-RWP, JPEG2000-RWP, BPG-RWP, Minnen18, and our VSSIM-optimized method using $14$ viewports indexed by $(\theta,\phi)$. We quantify the distortion in the form of PSNR (dB) / SSIM under each viewport. The bitrates of the ERP image produced by JPEG-RWP, JPEG2000-RWP, BPG-RWP, Minnen18 and ours are separately $0.165$ bpp, $0.171$ bpp, $0.188$ bpp, $0.169$ bpp, and $0.161$ bpp.}\label{fig:visual1}
\end{figure*}

Fig.~\ref{fig:rd} draws the rate-distortion curves, where we find that our VMSE-optimized method clearly outperforms BPG(-RWP),  Minnen18(-RWP)  Ball{\'e}18(-RWP), and overwhelms JPEG(-RWP) and JPEG2000(-RWP) under VMSE and VPSNR. 
Similar observations can be drawn for the proposed method under VSSIM. Our VMSE-optimized method ranks second in terms of VSSIM, outperforming the competing methods by a clear margin. It is interesting to note that all DNN-based compression methods optimized for VMSE achieve  better VSSIM performance compared with the three compression standards. As SSIM is widely regarded as a more perceptual quality metric, it is expected DNN-based compression methods to deliver better visual quality. Table \ref{tab:NS} lists the BD-BR and BD-distortion metrics computed from the rate-distortion curves in Fig.~\ref{fig:rd}, where BPG is the anchor method. We observe that our VMSE-optimized method improves VPSNR by $0.547$ dB and  saves $13.97\%$ bitrate on average.   Our VSSIM-optimized method increases VSSIM by $0.043$ and saves $41.84\%$ bitrate on average. One caveat should be pointed out: although RWP is beneficial to JPEG, JPEG2000, and BPG in compressing \mdegr images, it has no clear contribution to DNN-based image compression methods. For example, RWP boosts Ball{\'e}18 by $0.033$ dB, but hurts Minnen18 by $0.004$ dB. This result arises because RWP breaks image context and creates image discontinuity, to which the learned compression methods fail to adapt.

\begin{figure}[!tbp]
\begin{minipage}{0.49\linewidth}
\includegraphics[width=1.0\linewidth]{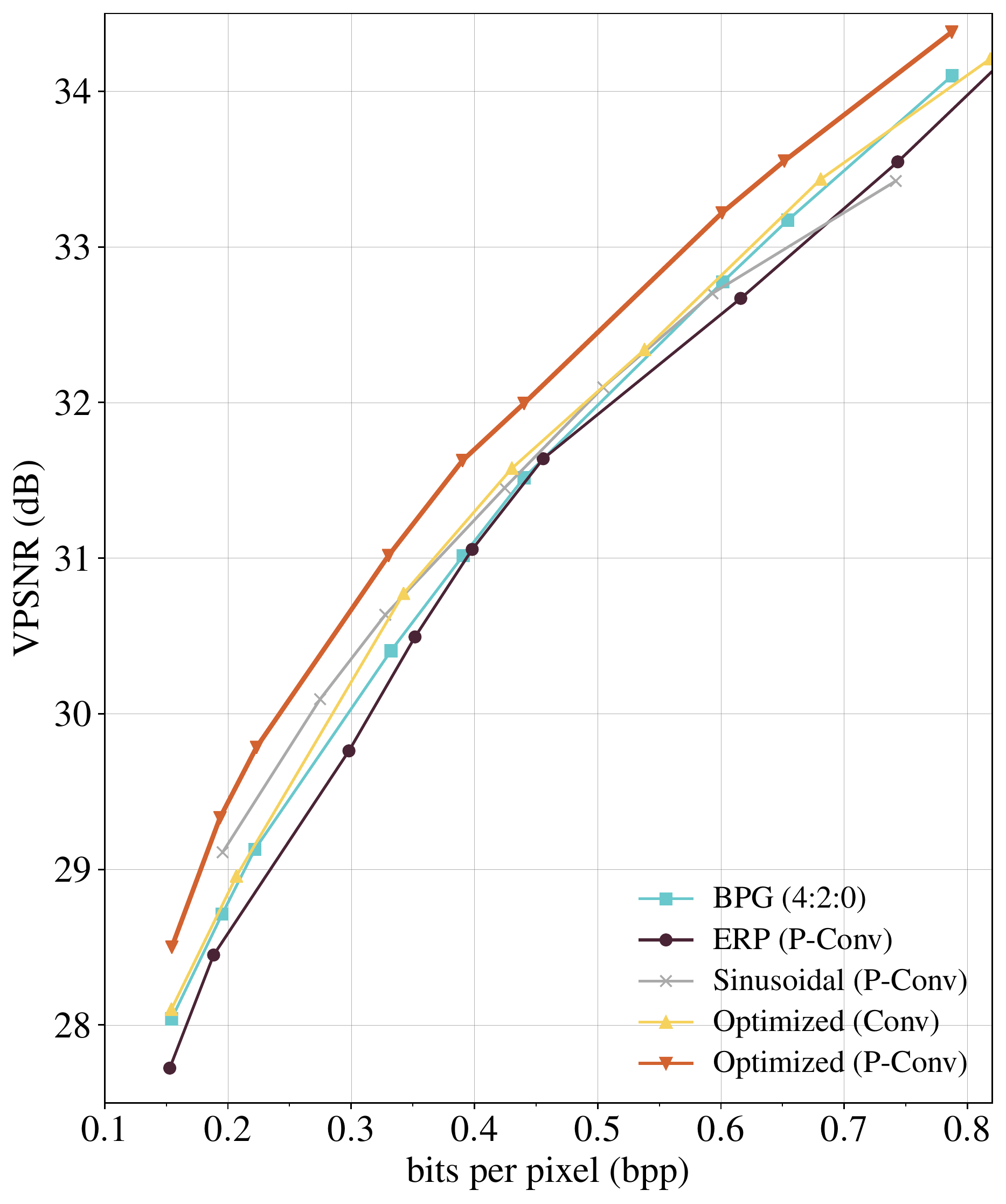}
\centering{\scriptsize{(a)}}
\end{minipage}
\begin{minipage}{0.49\linewidth}
\includegraphics[width=1.0\linewidth]{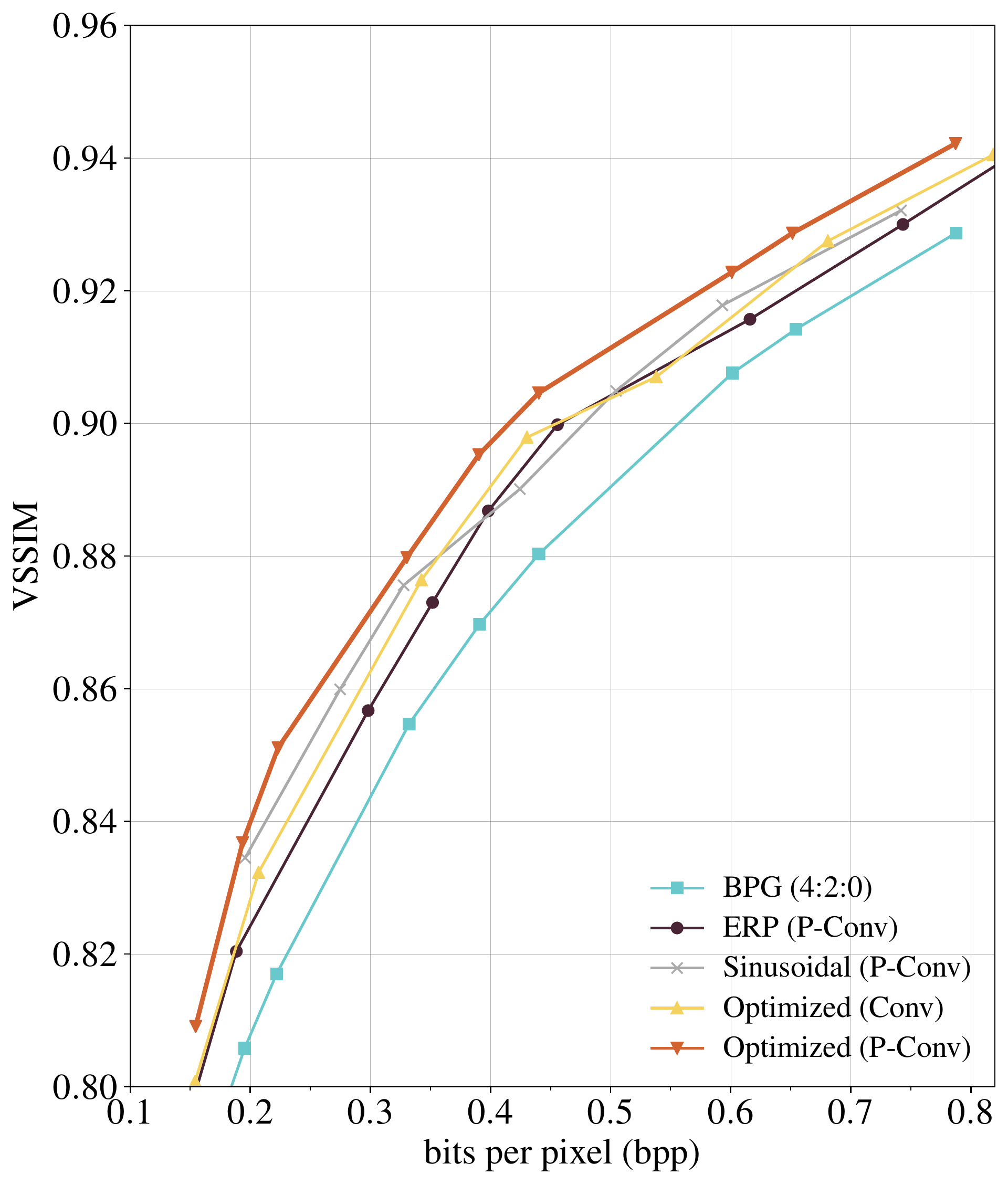}
\centering{\scriptsize{(b)}}
\end{minipage}
\caption{Rate-distortion curves of different input representations and convolutions. P-Conv: pseudocylindrical convolution.}\label{fig:ablation}
\end{figure}



\subsection{Qualitative Evaluation}
Fig.~\ref{fig:visual1} visually compares our VSSIM-optimized method against  JPEG-RWP, JPEG2000-RWP, BPG-RWP, and Minnen18 at similar bitrates, from which we have several interesting observations. First, JPEG-RWP, JPEG2000-RWP, and BPG-RWP suffer from common visual artifacts such as blocking, ringing, and blurring, which are further geometrically distorted at high latitudes. Second, \mdegr specific distortions such as the ``radiation artifacts'' in Viewports $(\frac{\pi}{2},0)$ and $(-\frac{\pi}{2},0)$ begin to emerge for all but the proposed method due to the over-sampling issue at poles even with RWP. This provides strong justifications of the proposed pseudocylindrical representation. Meanwhile, we also see different 
levels of color cast
produced by all competing methods in Viewport $(\frac{\pi}{4},\pi)$, centered at the right boundary of the ERP image (see Fig. \ref{fig:pseudo}). That is, the viewport is made up of pixels that are far apart in the ERP image, which may be compressed in substantially different ways. In contrast, our method with pseudocylindrical convolutions does not suffer from this problem. Third, compared to the three compression standards, Minnen18 appears to have better visual quality with less visible distortions, especially in flat regions. Overall, the proposed method offers the best quality, generating flat regions with little artifacts (see Viewport $(\frac{\pi}{2},0)$), reconstructing structures in a sharp way (see Viewport $(-\frac{\pi}{4},0)$), and producing plausible textures that are close to the original (see Viewport $(-\frac{\pi}{4},\frac{\pi}{2})$).

\begin{table}[t]
 \scriptsize
   \centering
   \caption{Performance comparison of different input representations and convolutions  in terms of BD-VPSNR, BD-VSSIM, and BD-BR. The BPG is the anchor method. P-Conv: pseudocylindrical convolution}\label{tab:ablation}
   \begin{tabular}{L{2.3cm} R{1.1cm} R{1.1cm} R{1.2cm} R{1.0cm}}
     \toprule
      \multirow{2}{*}{Method} & \multicolumn{2}{c}{RATE-VPSNR} & \multicolumn{2}{c}{RATE-VSSIM}\\
      \cmidrule{2-3} \cmidrule{4-5}
       & \multicolumn{1}{c}{\specialcell{BD-VPSNR \\ (dB)}} & \multicolumn{1}{c}{\specialcell{BD-BR \\ (\%)}} & \multicolumn{1}{c}{BD-VSSIM} & \multicolumn{1}{c}{\specialcell{BD-BR \\ (\%)}}\\
     \midrule
ERP (P-Conv) & $-0.161$ & $4.30$ & $0.013$ & $-14.22$\\
Sinusoidal (P-Conv) & $0.145$ & $-4.19$ & $0.018$ & $-18.34$\\
Optimized (Conv) & $0.116$ & $-3.09$ & $0.017$ & $-18.21$\\
Optimized (P-Conv) & $\mathbf{0.547}$ & $\mathbf{-13.97}$ & $\mathbf{0.025}$ & $\mathbf{-25.86}$\\
     \bottomrule
   \end{tabular}
\end{table}

\subsection{Ablation Studies}\label{sec:abl}
We conduct three sets of ablation experiments to single out two core contributions of the proposed method - the pseudocylindrical representation and convolution. First, we compare the optimized  pseudocylindrical representation (by solving Problem \eqref{eq:sop} using Alg. \ref{alg1}) with the ERP format\footnote{We may as well consider the ERP tiles by setting $W_t=W$ as input to our pseudocylindrical DNN-based compression system, which is equivalent to feeding the entire ERP image to a standard DNN for compression.} and the sinusoidal tiles by setting $W_t = \cos(\theta_t)W$ (see Fig. \ref{fig:pseudo}). For each input structure, we retrain a DNN with the same network architecture illustrated in Fig. \ref{fig:frame_analysis}, \ref{fig:frame_synthesis} and \ref{fig:frame_entropy} using the same training strategy and VMSE as the distortion measure. Next, we fix the optimized pseudocylindrical representation and compare the standard convolution with zero padding to the proposed pseudocylindrical convolution with the same network structure and training strategy.

The rate-distortion curves and the corresponding BD-BR and BD-Distortion metrics are given in Fig.~\ref{fig:ablation} and Tab.~\ref{tab:ablation}, respectively, where BPG is the anchor. We find that both sinusoidal and the optimized pseudocylindrical tiles deliver better compression performance than ERP. Moreover, the optimized representation offers more perceptual gains (and bitrate savings) than the sinusoidal tiles, validating the effectiveness of the proposed greedy search algorithm. 
Our results convey a somewhat  counterintuitive message: slightly oversampling mid-latitude regions is preferred over uniform sampling everywhere for \mdegr image compression. Meanwhile, the proposed pseudocylindrical convolution significantly outperforms the standard convolution with zero padding. Last, we compare the computational speed of the proposed pseudocylindrical convolution to the standard convolution, and report the running time of the analysis, synthesis, entropy networks in Tab.~\ref{tab:running_time}. As can be seen, the pseudocylindrical convolution has nearly the same running speed as the standard convolution. These demonstrate the promise of pseudocylindrical convolutions in modeling \mdegr images.

\begin{table}[t]
 \scriptsize
   \centering
   \caption{The running speed in seconds of the pseudocylindrical convolution and the standard convolution on the optimized pseudocylindrical representation}\label{tab:running_time}
   \begin{tabular}{C{1.6cm} C{1.8cm} C{1.8cm} C{1.8cm} }
     \toprule
      Convolution &  Analysis Network ($g_a$) & Synthesis Network ($g_s$) & Entropy Network ($g_e$)\\
     \midrule
	  Conv & $0.104$ & $0.105$ & $0.040$ \\
	   P-Conv & $0.107$ & $0.108$ & $0.045$ \\
     \bottomrule
   \end{tabular}
 \end{table}

\section{Conclusion and Discussion}
In this paper, we have introduced a new data structure for representing \mdegr images - pseudocylindrical representation. We also proposed the pseudocylindrical convolution that can be efficiently implemented by standard convolutions with pseudocylindrical padding. Relying on the proposed techniques, we implemented one of the first  DNN-based \mdegr image compression system that offers favorable perceptual gains at similar bitrates. In the future, we will try to perform joint optimization of the parameters of the front-end pseudocylindrical representation and the back-end image compression method at the image level. This may be achievable by training another DNN that takes a \mdegr image as the input for parameter estimation of the corresponding representation \cite{su2018learning}. We will also explore the possibility of combining the proposed techniques with existing video codecs (\eg, HEVC, VVC, VP9, and AV1) to improve \mdegr video compression.

The application scope of the proposed pseudocylindrical representation and convolution is far beyond \mdegr image compression. In fact, it may serve as a canonical building block  for general \mdegr image modeling, and is particular useful for \mdegr applications that expect efficiency, scalability, and transferability. For example, in \mdegr image editing and enhancement, the pseudocylindrical representation may be optimized to under-sample certain parts of the image to better account for global image context. As another example, our representation with uniform sampling density (\ie, the sinusoidal tiles) may be preferable in \mdegr computer vision tasks to localize and track objects moving from low-latitude to high-latitude places. In either \mdegr application, the proposed pseudocylindrical convolution enables reusing existing methods trained on central-perspective images, and requires only a small set of (labeled) \mdegr images for efficient adaptation.



\ifCLASSOPTIONcaptionsoff
  \newpage
\fi

\bibliographystyle{IEEEtran}
\bibliography{IEEEabrv,./egbib}

\begin{thebibliography}{10}
\providecommand{\url}[1]{#1}
\csname url@samestyle\endcsname
\providecommand{\newblock}{\relax}
\providecommand{\bibinfo}[2]{#2}
\providecommand{\BIBentrySTDinterwordspacing}{\spaceskip=0pt\relax}
\providecommand{\BIBentryALTinterwordstretchfactor}{4}
\providecommand{\BIBentryALTinterwordspacing}{\spaceskip=\fontdimen2\font plus
\BIBentryALTinterwordstretchfactor\fontdimen3\font minus
  \fontdimen4\font\relax}
\providecommand{\BIBforeignlanguage}[2]{{%
\expandafter\ifx\csname l@#1\endcsname\relax
\typeout{** WARNING: IEEEtran.bst: No hyphenation pattern has been}%
\typeout{** loaded for the language `#1'. Using the pattern for}%
\typeout{** the default language instead.}%
\else
\language=\csname l@#1\endcsname
\fi
#2}}
\providecommand{\BIBdecl}{\relax}
\BIBdecl

\bibitem{snyder1987map}
J.~P. Snyder, \emph{Map {Projections} -- A {Working} {Manual}}.\hskip 1em plus
  0.5em minus 0.4em\relax US Government Printing Office, 1987.

\bibitem{lee2017omnidirectional}
S.-H. Lee, S.-T. Kim, E.~Yip, B.-D. Choi, J.~Song, and S.-J. Ko,
  ``Omnidirectional video coding using latitude adaptive down-sampling and
  pixel rearrangement,'' \emph{Electronics Letters}, vol.~53, no.~10, pp.
  655--657, 2017.

\bibitem{boyce2017hevc}
J.~Boyce, A.~Ramasubramanian, R.~Skupin, G.~J. Sullivan, A.~Tourapis, and
  Y.~Wang, ``{HEVC} additional supplemental enhancement information (draft
  4),'' \emph{Joint Collaborative Team on Video Coding of ITU-T SG}, vol.~16,
  2017.

\bibitem{youvalari2016analysis}
R.~G. Youvalari, A.~Aminlou, and M.~M. Hannuksela, ``Analysis of regional
  down-sampling methods for coding of omnidirectional video,'' in \emph{Picture
  Coding Symposium}, 2016.

\bibitem{yu2015content}
M.~Yu, H.~Lakshman, and B.~Girod, ``Content adaptive representations of
  omnidirectional videos for cinematic virtual reality,'' in
  \emph{International Workshop on Immersive Media Experiences}, 2015, pp. 1--6.

\bibitem{li2017spherical}
Y.~Li, J.~Xu, and Z.~Chen, ``Spherical domain rate-distortion optimization for
  360-degree video coding,'' in \emph{IEEE International Conference on
  Multimedia and Expo}, 2017, pp. 709--714.

\bibitem{liu2018rate}
Y.~Liu, L.~Yang, M.~Xu, and Z.~Wang, ``Rate control schemes for panoramic video
  coding,'' \emph{Journal of Visual Communication and Image Representation},
  vol.~53, pp. 76--85, 2018.

\bibitem{tang2017optimized}
M.~Tang, Y.~Zhang, J.~Wen, and S.~Yang, ``Optimized video coding for
  omnidirectional videos,'' in \emph{IEEE International Conference on
  Multimedia and Expo}, 2017, pp. 799--804.

\bibitem{xiu2018adaptive}
X.~Xiu, Y.~He, and Y.~Ye, ``An adaptive quantization method for 360-degree
  video coding,'' in \emph{Applications of Digital Image Processing XLI}, vol.
  10752, 2018.

\bibitem{li2016novel}
J.~Li, Z.~Wen, S.~Li, Y.~Zhao, B.~Guo, and J.~Wen, ``Novel tile segmentation
  scheme for omnidirectional video,'' in \emph{IEEE International Conference on
  Image Processing}, 2016, pp. 370--374.

\bibitem{lin2019efficient}
J.-L. Lin, Y.-H. Lee, C.-H. Shih, S.-Y. Lin, H.-C. Lin, S.-K. Chang, P.~Wang,
  L.~Liu, and C.-C. Ju, ``Efficient projection and coding tools for 360
  video,'' \emph{IEEE Journal on Emerging and Selected Topics in Circuits and
  Systems}, vol.~9, no.~1, pp. 84--97, 2019.

\bibitem{he2018content}
Y.~He, X.~Xiu, P.~Hanhart, Y.~Ye, F.~Duanmu, and Y.~Wang, ``Content-adaptive
  360-degree video coding using hybrid cubemap projection,'' in \emph{Picture
  Coding Symposium}, 2018, pp. 313--317.

\bibitem{hanhart2018360}
P.~Hanhart, X.~Xiu, Y.~He, and Y.~Ye, ``360 video coding based on projection
  format adaptation and spherical neighboring relationship,'' \emph{IEEE
  Journal on Emerging and Selected Topics in Circuits and Systems}, vol.~9,
  no.~1, pp. 71--83, 2018.

\bibitem{ozcinar2017viewport}
C.~Ozcinar, A.~De~Abreu, and A.~Smolic, ``Viewport-aware adaptive 360 video
  streaming using tiles for virtual reality,'' in \emph{IEEE International
  Conference on Image Processing}, 2017, pp. 2174--2178.

\bibitem{corbillon2017viewport}
X.~Corbillon, G.~Simon, A.~Devlic, and J.~Chakareski, ``Viewport-adaptive
  navigable 360-degree video delivery,'' in \emph{IEEE International Conference
  on Communications}, 2017.

\bibitem{zelnik2005squaring}
L.~Zelnik-Manor, G.~Peters, and P.~Perona, ``Squaring the circle in
  panoramas,'' in \emph{IEEE International Conference on Computer Vision},
  vol.~1, 2005.

\bibitem{chang2013rectangling}
C.-H. Chang, M.-C. Hu, W.-H. Cheng, and Y.-Y. Chuang, ``Rectangling
  stereographic projection for wide-angle image visualization,'' in \emph{IEEE
  International Conference on Computer Vision}, 2013, pp. 2824--2831.

\bibitem{kim2017automatic}
Y.~W. Kim, C.-R. Lee, D.-Y. Cho, Y.~H. Kwon, H.-J. Choi, and K.-J. Yoon,
  ``Automatic content-aware projection for 360 videos,'' in \emph{IEEE
  International Conference on Computer Vision}, 2017, pp. 4753--4761.

\bibitem{kimerling1999comparing}
J.~A. Kimerling, K.~Sahr, D.~White, and L.~Song, ``Comparing geometrical
  properties of global grids,'' \emph{Cartography and Geographic Information
  Science}, vol.~26, no.~4, pp. 271--288, 1999.

\bibitem{eder2020tangent}
M.~Eder, M.~Shvets, J.~Lim, and J.-M. Frahm, ``Tangent images for mitigating
  spherical distortion,'' in \emph{IEEE Conference on Computer Vision and
  Pattern Recognition}, 2020, pp. 12\,426--12\,434.

\bibitem{balle2016end}
J.~Ball{\'e}, V.~Laparra, and E.~P. Simoncelli, ``End-to-end optimized image
  compression,'' in \emph{International Conference Learning Representations},
  2017.

\bibitem{theis2017lossy}
L.~Theis, W.~Shi, A.~Cunningham, and F.~Husz{\'a}r, ``Lossy image compression
  with compressive autoencoders,'' in \emph{International Conference Learning
  Representations}, 2017.

\bibitem{balle2018variational}
J.~Ball{\'e}, D.~Minnen, S.~Singh, S.~J. Hwang, and N.~Johnston, ``Variational
  image compression with a scale hyperprior,'' in \emph{International
  Conference Learning Representations}, 2018.

\bibitem{minnen2018joint}
D.~Minnen, J.~Ball\'{e}, and G.~D. Toderici, ``Joint autoregressive and
  hierarchical priors for learned image compression,'' in \emph{Neural
  Information Processing System}, 2018.

\bibitem{torfason2018towards}
R.~Torfason, F.~Mentzer, E.~Agustsson, M.~Tschannen, R.~Timofte, and
  L.~Van~Gool, ``Towards image understanding from deep compression without
  decoding,'' in \emph{International Conference Learning Representations},
  2018.

\bibitem{agustsson2019generative}
E.~Agustsson, M.~Tschannen, F.~Mentzer, R.~Timofte, and L.~V. Gool,
  ``Generative adversarial networks for extreme learned image compression,'' in
  \emph{IEEE International Conference on Computer Vision}, 2019, pp. 221--231.

\bibitem{li2020efficient}
M.~Li, K.~Ma, J.~You, D.~Zhang, and W.~Zuo, ``Efficient and effective
  context-based convolutional entropy modeling for image compression,''
  \emph{IEEE Transactions on Image Processing}, vol.~29, pp. 5900--5911, 2020.

\bibitem{zioulis2018omnidepth}
N.~Zioulis, A.~Karakottas, D.~Zarpalas, and P.~Daras, ``Omnidepth: Dense depth
  estimation for indoors spherical panoramas,'' in \emph{European Conference on
  Computer Vision}, 2018, pp. 448--465.

\bibitem{tateno2018distortion}
K.~Tateno, N.~Navab, and F.~Tombari, ``Distortion-aware convolutional filters
  for dense prediction in panoramic images,'' in \emph{European Conference on
  Computer Vision}, 2018, pp. 707--722.

\bibitem{su2017learning}
Y.-C. Su and K.~Grauman, ``Learning spherical convolution for fast features
  from 360 imagery,'' in \emph{Neural Information Processing Systems}, vol.~30,
  2017, pp. 529--539.

\bibitem{su2019kernel}
Y.-C. \vspace{0mm}Su and K.~Grauman, ``Kernel transformer networks for compact
  spherical convolution,'' in \emph{IEEE Conference on Computer Vision and
  Pattern Recognition}, 2019, pp. 9442--9451.

\bibitem{cohen2018spherical}
T.~S. Cohen, M.~Geiger, J.~K{\"o}hler, and M.~Welling, ``Spherical {CNNs},''
  \emph{International Conference Learning Representations}, 2018.

\bibitem{esteves2018learning}
C.~Esteves, C.~Allen-Blanchette, A.~Makadia, and K.~Daniilidis, ``Learning
  {SO}(3) equivariant representations with spherical {CNN}s,'' in
  \emph{European Conference on Computer Vision}, 2018, pp. 52--68.

\bibitem{perraudin2019deepsphere}
N.~Perraudin, M.~Defferrard, T.~Kacprzak, and R.~Sgier, ``Deepsphere: Efficient
  spherical convolutional neural network with healpix sampling for cosmological
  applications,'' \emph{Astronomy and Computing}, vol.~27, pp. 130--146, 2019.

\bibitem{jiang2019spherical}
C.~Jiang, J.~Huang, K.~Kashinath, P.~Marcus, M.~Niessner \emph{et~al.},
  ``Spherical cnns on unstructured grids,'' \emph{arXiv preprint
  arXiv:1901.02039}, 2019.

\bibitem{sui2021perceptual}
X.~Sui, K.~Ma, Y.~Yao, and Y.~Fang, ``Perceptual quality assessment of
  omnidirectional images as moving camera videos,'' \emph{IEEE Transactions on
  Visualization and Computer Graphics}, 2021.

\bibitem{toderici2015variable}
G.~Toderici, S.~M. O'Malley, S.~J. Hwang, D.~Vincent, D.~Minnen, S.~Baluja,
  M.~Covell, and R.~Sukthankar, ``Variable rate image compression with
  recurrent neural networks,'' \emph{arXiv:1511.06085}, 2015.

\bibitem{toderici2016full}
G.~Toderici, D.~Vincent, N.~Johnston, S.~J. Hwang, D.~Minnen, J.~Shor, and
  M.~Covell, ``Full resolution image compression with recurrent neural
  networks,'' in \emph{IEEE Conference on Computer Vision and Pattern
  Recognition}, 2017, pp. 5306--5314.

\bibitem{johnston2017improved}
N.~Johnston, D.~Vincent, D.~Minnen, M.~Covell, S.~Singh, T.~Chinen, S.~J.
  Hwang, J.~Shor, and G.~Toderici, ``Improved lossy image compression with
  priming and spatially adaptive bit rates for recurrent networks,'' in
  \emph{IEEE Conference on Computer Vision and Pattern Recognition}, 2018, pp.
  4385--4393.

\bibitem{rippel2017real}
O.~Rippel and L.~Bourdev, ``Real-time adaptive image compression,'' in
  \emph{International Conference on Machine Learning}, 2017, pp. 2922--2930.

\bibitem{li2017learning}
M.~Li, W.~Zuo, S.~Gu, D.~Zhao, and D.~Zhang, ``Learning convolutional networks
  for content-weighted image compression,'' in \emph{IEEE Conference on
  Computer Vision and Pattern Recognition}, 2018, pp. 3214--3223.

\bibitem{li2020learning}
M.~Li, W.~Zuo, S.~Gu, J.~You, and D.~Zhang, ``Learning content-weighted deep
  image compression,'' \emph{IEEE Transactions on Pattern Analysis and Machine
  Intelligence}, vol.~43, no.~10, pp. 3446--3461, 2021.

\bibitem{wang2004image}
Z.~Wang, A.~C. Bovik, H.~R. Sheikh, and E.~P. Simoncelli, ``Image quality
  assessment: {From} error visibility to structural similarity,'' \emph{IEEE
  Transactions on Image Processing}, vol.~13, no.~4, pp. 600--612, 2004.

\bibitem{wang2003multiscale}
Z.~Wang, E.~P. Simoncelli, and A.~C. Bovik, ``Multiscale structural similarity
  for image quality assessment,'' in \emph{Asilomar Conference on Signals,
  Systems, and Computers}, 2003, pp. 1398--1402.

\bibitem{goodfellow2014generative}
I.~Goodfellow, J.~Pouget-Abadie, M.~Mirza, B.~Xu, D.~Warde-Farley, S.~Ozair,
  A.~Courville, and Y.~Bengio, ``Generative adversarial nets,'' in \emph{Neural
  Information Processing System}, 2014, pp. 2672--2680.

\bibitem{ghaznavi2017comparison}
R.~Ghaznavi-Youvalari, A.~Zare, H.~Fang, A.~Aminlou, Q.~Xie, M.~M. Hannuksela,
  and M.~Gabbouj, ``Comparison of {HEVC} coding schemes for tile-based
  viewport-adaptive streaming of omnidirectional video,'' in \emph{IEEE
  International Workshop on Multimedia Signal Processing}, 2017, pp. 1--6.

\bibitem{sullivan2012overview}
G.~J. Sullivan, J.-R. Ohm, W.-J. Han, T.~Wiegand \emph{et~al.}, ``Overview of
  the high efficiency video coding({HEVC}) standard,'' \emph{IEEE Transactions
  on Circuits and Systems for Video Technology}, vol.~22, no.~12, pp.
  1649--1668, 2012.

\bibitem{xu2020state}
M.~Xu, C.~Li, S.~Zhang, and P.~Le~Callet, ``State-of-the-art in 360 video/image
  processing: {Perception}, assessment and compression,'' \emph{IEEE Journal of
  Selected Topics in Signal Processing}, vol.~14, no.~1, pp. 5--26, 2020.

\bibitem{budagavi2015360}
M.~Budagavi, J.~Furton, G.~Jin, A.~Saxena, J.~Wilkinson, and A.~Dickerson,
  ``360 degrees video coding using region adaptive smoothing,'' in \emph{IEEE
  International Conference on Image Processing}, 2015, pp. 750--754.

\bibitem{liu2017novel}
Y.~Liu, M.~Xu, C.~Li, S.~Li, and Z.~Wang, ``A novel rate control scheme for
  panoramic video coding,'' in \emph{IEEE International Conference on
  Multimedia and Expo}, 2017, pp. 691--696.

\bibitem{yu2015framework}
M.~Yu, H.~Lakshman, and B.~Girod, ``A framework to evaluate omnidirectional
  video coding schemes,'' in \emph{IEEE International Symposium on Mixed and
  Augmented Reality}, 2015, pp. 31--36.

\bibitem{sun2016ahg8}
Y.~Sun, A.~Lu, and L.~Yu, ``{AHG8}: {WS-PSNR} for 360 video objective quality
  evaluation,'' in \emph{Joint Video Exploration Team of ITU-T SG16 WP3 and
  ISO/IEC JTC1/SC29/WG11, JVET-D0040, 4th Meeting}, 2016.

\bibitem{su2018learning}
Y.-C. Su and K.~Grauman, ``Learning compressible 360$^{\circ}$ video isomers,''
  in \emph{IEEE Conference on Computer Vision and Pattern Recognition}, 2018,
  pp. 7824--7833.

\bibitem{zare2017virtual}
A.~Zare, A.~Aminlou, and M.~M. Hannuksela, ``Virtual reality content streaming:
  Viewport-dependent projection and tile-based techniques,'' in \emph{IEEE
  International Conference on Image Processing}, 2017, pp. 1432--1436.

\bibitem{ozcinar2019visual}
C.~Ozcinar, J.~Cabrera, and A.~Smolic, ``Visual attention-aware omnidirectional
  video streaming using optimal tiles for virtual reality,'' \emph{IEEE Journal
  on Emerging and Selected Topics in Circuits and Systems}, vol.~9, no.~1, pp.
  217--230, 2019.

\bibitem{hadizadeh2013saliency}
H.~Hadizadeh and I.~V. Baji{\'c}, ``Saliency-aware video compression,''
  \emph{IEEE Transactions on Image Processing}, vol.~23, no.~1, pp. 19--33,
  2013.

\bibitem{luz2017saliency}
G.~Luz, J.~Ascenso, C.~Brites, and F.~Pereira, ``Saliency-driven
  omnidirectional imaging adaptive coding: Modeling and assessment,'' in
  \emph{IEEE International Workshop on Multimedia Signal Processing}, 2017, pp.
  1--6.

\bibitem{sitzmann2018saliency}
V.~Sitzmann, A.~Serrano, A.~Pavel, M.~Agrawala, D.~Gutierrez, B.~Masia, and
  G.~Wetzstein, ``Saliency in {VR}: How do people explore virtual
  environments?'' \emph{IEEE Transactions on Visualization and Computer
  Graphics}, vol.~24, no.~4, pp. 1633--1642, 2018.

\bibitem{saff1997distributing}
E.~B. Saff and A.~B. Kuijlaars, ``Distributing many points on a sphere,''
  \emph{The Mathematical Intelligencer}, vol.~19, no.~1, pp. 5--11, 1997.

\bibitem{kostelec2008ffts}
P.~J. Kostelec and D.~N. Rockmore, ``{FFTs} on the rotation group,''
  \emph{Journal of Fourier Analysis and Applications}, vol.~14, no.~2, pp.
  145--179, 2008.

\bibitem{driscoll1994computing}
J.~R. Driscoll and D.~M. Healy, ``Computing {Fourier} transforms and
  convolutions on the 2-sphere,'' \emph{Advances in Applied Mathematics},
  vol.~15, no.~2, pp. 202--250, 1994.

\bibitem{weinstein1996groupoids}
A.~Weinstein, ``Groupoids: unifying internal and external symmetry,''
  \emph{Notices of the AMS}, vol.~43, no.~7, pp. 744--752, 1996.

\bibitem{zakharchenko2016ahg8}
V.~Zakharchenko, E.~Alshina, A.~Singh, and A.~Dsouza, ``{AHG8}: Suggested
  testing procedure for 360-degree video,'' in \emph{Joint Video Exploration
  Team of ITU-T SG16 WP3 and ISO/IEC JTC1/SC29/WG11, JVET-D0027, 4th Meeting},
  2016.

\bibitem{bjontegaard2001calculation}
G.~Bjontegaard, ``Calculation of average {PSNR} differences between
  {RD}-curves,'' \emph{VCEG-M33}, 2001.

\bibitem{He_2016_CVPR}
K.~He, X.~Zhang, S.~Ren, and J.~Sun, ``Deep residual learning for image
  recognition,'' in \emph{IEEE Conference on Computer Vision and Pattern
  Recognition}, 2016, pp. 770--778.

\bibitem{cheng2020learned}
Z.~Cheng, H.~Sun, M.~Takeuchi, and J.~Katto, ``Learned image compression with
  discretized {Gaussian} mixture likelihoods and attention modules,'' in
  \emph{IEEE Conference on Computer Vision and Pattern Recognition}, 2020, pp.
  7939--7948.

\bibitem{shi_2016_real}
W.~Shi, J.~Caballero, F.~Huszar, J.~Totz, A.~P. Aitken, R.~Bishop, D.~Rueckert,
  and Z.~Wang, ``Real-time single image and video super-resolution using an
  efficient sub-pixel convolutional neural network,'' in \emph{IEEE Conference
  on Computer Vision and Pattern Recognition}, 2016, pp. 1874--1883.

\bibitem{courbariaux2016binarized}
M.~Courbariaux, I.~Hubara, D.~Soudry, R.~El-Yaniv, and Y.~Bengio, ``Binarized
  neural networks: {Training} deep neural networks with weights and activations
  constrained to +1 or -1,'' \emph{arXiv:1602.02830}, 2016.

\bibitem{kingma2014adam}
D.~Kingma and J.~Ba, ``{Adam}: {A} method for stochastic optimization,'' in
  \emph{International Conference Learning Representations}, 2015.

\bibitem{wallace1992jpeg}
G.~K. Wallace, ``The {JPEG} still picture compression standard,'' \emph{IEEE
  Transactions on Consumer Electronics}, vol.~38, no.~1, pp. 18--34, 1992.

\bibitem{skodras2001jpeg}
A.~Skodras, C.~Christopoulos, and T.~Ebrahimi, ``The {JPEG} 2000 still image
  compression standard,'' \emph{IEEE Signal Processing Magazine}, vol.~18,
  no.~5, pp. 36--58, 2001.

\bibitem{bellard2016bpg}
\BIBentryALTinterwordspacing
F.~Bellard, ``{BPG} image format,'' 2019. [Online]. Available:
  \url{https://bellard.org/bpg/}
\BIBentrySTDinterwordspacing

\end{thebibliography}

\begin{IEEEbiography}
[{\includegraphics[width=1in,height=1.25in,clip,keepaspectratio]{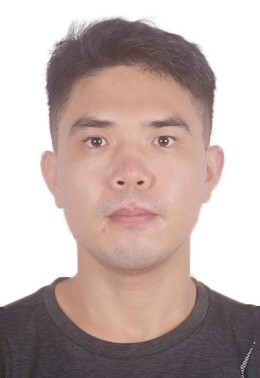}}]{Mu Li} received his BCS in Computer Science and Technology in 2015 from Harbin Institute of Technology, and the Ph.D. degree from the Department of Computing, the Hong Kong Polytechnic University, Hong Kong, China, in 2020. He is the owner of Hong Kong PhD Fellowship. He is currently a postdoctoral researcher at School of Data Science, the Chinese University of Hong Kong, Shenzhen. His research interests include deep learning, image processing, and image compression.
\end{IEEEbiography}
\begin{IEEEbiography}[{\includegraphics[width=1in,height=1.25in,clip,keepaspectratio]{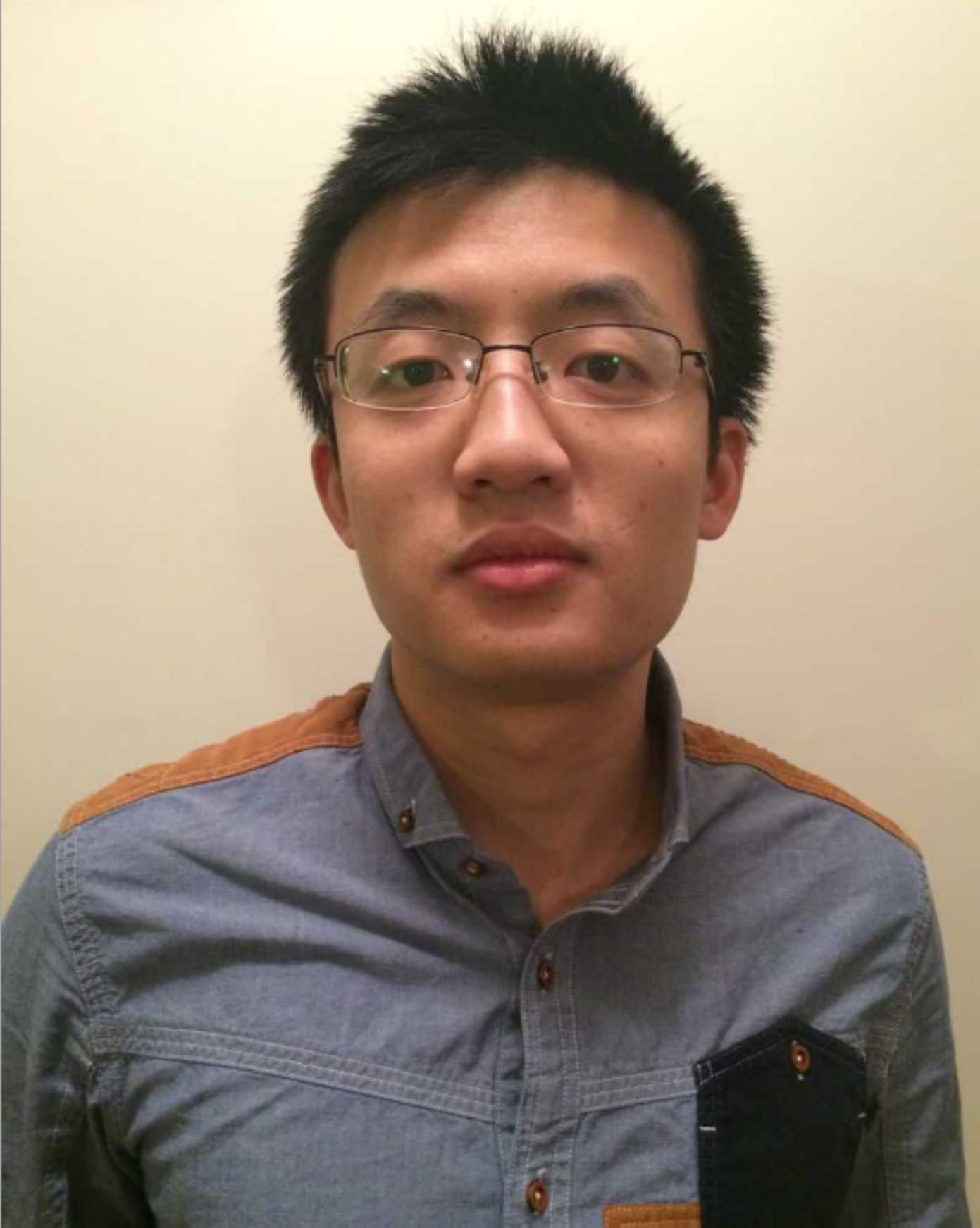}}]{Kede Ma}
(S'13-M'18) received the B.E. degree from the University of Science and Technology of China, Hefei, China, in 2012, and the M.S. and Ph.D.
degrees in electrical and computer engineering from the University of Waterloo, Waterloo, ON, Canada, in 2014 and 2017, respectively. He was a Research Associate with the Howard Hughes Medical Institute and New York University, New York, NY, USA, in 2018. He is currently an Assistant Professor with the Department of Computer Science, City University of Hong Kong. His research interests include perceptual image processing, computational vision, and computational photography.
\end{IEEEbiography}
\begin{IEEEbiography}
[{\includegraphics[width=1in,height=1.25in,clip,keepaspectratio]{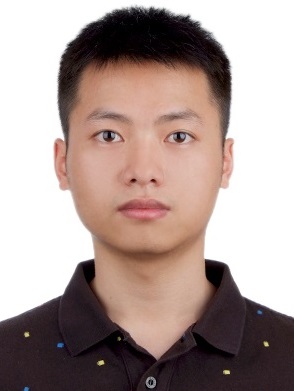}}]{Jinxing Li} received the B.Sc. degree from the department of Automation, Hangzhou Dianzi University, Hangzhou, China, in 2012, the M.Sc. degree from the department of Automation, Chongqing University, Chongqing, China, in 2015, and the PhD degree from the department of Computing, Hong Kong Polytechnic University, Hong Kong, China, in 2019. Dr. Li worked at The Chinese University of Hong Kong, Shenzhen, from 2019 to 2021. He is currently with Harbin Institute of Technology, Shenzhen, China. His research interests are pattern recognition, deep learning, medical biometrics and machine learning.
\end{IEEEbiography}
\begin{IEEEbiography}
[{\includegraphics[width=1in,height=1.25in,clip,keepaspectratio]{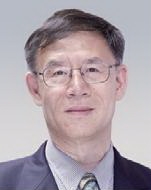}}]{David Zhang}(Life Fellow, IEEE) graduated in Computer Science from Peking University. He received his MSc in 1982 and his PhD in 1985 in both Computer Science from the Harbin Institute of Technology (HIT), respectively. From 1986 to 1988 he was a Postdoctoral Fellow at Tsinghua University and then an Associate Professor at the Academia Sinica, Beijing. In 1994 he received his second PhD in Electrical and Computer Engineering from the University of Waterloo, Ontario, Canada. He has been a Chair Professor at the Hong Kong Polytechnic University where he is the Founding Director of Biometrics Research Centre (UGC/CRC) supported by the Hong Kong SAR Government since 1998. Currently he is Presidential Chair Professor in Chinese University of Hong Kong (Shenzhen). So far, he has published over 20 monographs, 500+ international journal papers and 40+ patents from USA/Japan/HK/China. He has been continuously 8 years listed as a Global Highly Cited Researchers in Engineering by Clarivate Analytics during 2014-2021. He is also ranked about 85 with H-Index 120 at Top 1,000 Scientists for international Computer Science and Electronics. Professor Zhang is selected as a Fellow of the Royal Society of Canada. He also is a Croucher Senior Research Fellow, Distinguished Speaker of the IEEE Computer Society, and an IEEE Life Fellow and an IAPR Fellow.
\end{IEEEbiography}

\end{document}